

\magnification\magstep1
\input amssym.def
\input amssym
\input epsf
\def \bR{{\Bbb R}}

\newdimen\itemindent \itemindent=32pt
\def\textindent#1{\parindent=\itemindent\let\par=\resetpar%
\indent\llap{#1\enspace}\ignorespaces}

\let\oldpar=\par
\def\resetpar{\oldpar\parindent=20pt\let\par=\oldpar}

\font\ninerm=cmr9 \font\ninesy=cmsy9
\font\eightrm=cmr8 \font\sixrm=cmr6
\font\eighti=cmmi8 \font\sixi=cmmi6
\font\eightsy=cmsy8 \font\sixsy=cmsy6
\font\eightbf=cmbx8 \font\sixbf=cmbx6
\font\eightit=cmti8
\def\eightpoint{\def\rm{\fam0\eightrm}
  \textfont0=\eightrm \scriptfont0=\sixrm \scriptscriptfont0=\fiverm
  \textfont1=\eighti  \scriptfont1=\sixi  \scriptscriptfont1=\fivei
  \textfont2=\eightsy \scriptfont2=\sixsy \scriptscriptfont2=\fivesy
  \textfont3=\tenex   \scriptfont3=\tenex \scriptscriptfont3=\tenex
  \textfont\itfam=\eightit  \def\it{\fam\itfam\eightit}%
  \textfont\bffam=\eightbf  \scriptfont\bffam=\sixbf
  \scriptscriptfont\bffam=\fivebf  \def\bf{\fam\bffam\eightbf}%
  \normalbaselineskip=9pt
  \setbox\strutbox=\hbox{\vrule height7pt depth2pt width0pt}%
  \let\big=\eightbig  \normalbaselines\rm}
\catcode`@=11 %
\def\eightbig#1{{\hbox{$\textfont0=\ninerm\textfont2=\ninesy
  \left#1\vbox to6.5pt{}\right.\n@space$}}}
\def\vfootnote#1{\insert\footins\bgroup\eightpoint
  \interlinepenalty=\interfootnotelinepenalty
  \splittopskip=\ht\strutbox %
  \splitmaxdepth=\dp\strutbox %
  \leftskip=0pt \rightskip=0pt \spaceskip=0pt \xspaceskip=0pt
  \textindent{#1}\footstrut\futurelet\next\fo@t}
\catcode`@=12 %

\magnification\magstep1
\def\pmb#1{\setbox0=\hbox{#1}%
 \kern-.025em\copy0\kern-\wd0
 \kern.05em\copy0\kern-\wd0
 \kern-.025em\raise.0433em\box0 }
\def \olr{{\raise6pt\hbox{$\leftrightarrow  \! \! \! \! \!$}}}
\def \ollr{{\raise7pt\hbox{$\leftrightarrow  \! \! \! \! \! \!$}}}
\def\vev#1{{\langle #1  \rangle}}
\def \de{\delta}
\def \si{\sigma}
\def \Ga{\Gamma}

\def \hnab{{\hat \nabla}}

\def \pr{\partial}
\def \bx{{\bf x}}

\def \0{{\bf 0}}
\def \hX{{\hat X}}
\def \hT{{\hat T}}

\def \rO{{\rm O}}
\def \vep{\varepsilon}
\def \half{{\textstyle {1 \over 2}}}

\def \quar{{\textstyle {1 \over 4}}}
\def \hh{{1\over 2}}
\def \ts{ \textstyle}

\def \txi{{\tilde \xi}}
\def \A{{\cal A}}
\def \B{{\cal B}}
\def \C{{\cal C}}
\def \D{{\cal D}}
\def \F{{\cal F}}
\def \G{{\cal G}}
\def \H{{\cal H}}
\def \I{{\cal I}}
\def \O{{\cal O}}
\def \L{{\cal L}}
\def \R{{\cal R}}

\def \d{{\rm d}}

\def \hO{{\hat \O}}
\def \bX{{\bar {X}}}
\def \heta{{\hat \eta}}
\def \hs {{\hat {s}}}
\def \bs {{\bf {s}}}
\def \bz {{\bar z}}

\def \hC{{\hat {C}}}
\def \hcT{{\hat {{\cal T}}}}

\def \hphi {{\hat {\phi}}}

\def \ep{\epsilon}
\def \l{\langle}
\def \r{\rangle}
\def \br{{\bf r}}
\def \st{{\tilde s}}
\def \tv{{\tilde v}}
\def \tX{{\tilde X}}

\parskip 5pt
\font \bigbf=cmbx10 scaled \magstep1
{\nopagenumbers
\rightline{DAMTP/95-1}
\rightline{UBC/TP-95-002}
\rightline{cond-mat/9505127}
\vskip 2truecm
\centerline {\bigbf Conformal Field Theories}
\vskip 5pt
\centerline {\bigbf Near a Boundary in General Dimensions}
\vskip 2.0 true cm
\centerline {D.M. McAvity$^a$ and H. Osborn$^b$}
\vskip 10pt
\centerline {$^a$Department of Physics, University of British Columbia}
\centerline {6224 Agricultural Rd, Vancouver, BC, V6T 1Z2, CANADA}
\centerline {email: dmm@physics.ubc.ca}
\vskip 10pt
\centerline {$^b$Department of Applied Mathematics and Theoretical Physics,}
\centerline {Silver Street, Cambridge, CB3 9EW, UK}
\centerline {email: ho@amtp.cam.ac.uk}
\vskip 2.0 true cm
The implications of restricted conformal invariance under conformal
transformations preserving a plane boundary are discussed for general
dimensions $d$. Calculations of the universal function of a conformal
invariant $\xi$ which appears in the two point function of scalar operators
in conformally invariant theories with a plane boundary are undertaken to
first order in the $\vep=4-d$ expansion for the the operator $\phi^2$  in
$\phi^4$ theory. The form for the associated functions of $\xi$ for
the two point functions for the basic field $\phi^\alpha$
and the auxiliary field $\lambda$ in the the $N\to \infty$ limit
of the $O(N)$ non linear sigma model for any $d$ in the range $2<d<4$ are
also rederived.  These results are obtained by integrating the two point
functions over planes parallel to the boundary, defining a restricted
two point function which may be obtained more simply.
Assuming conformal invariance this transformation can be inverted to
recover the full two point function.
Consistency of the results is checked by considering the limit $d\to 4$
and also by analysis of the operator product expansions for
$\phi^\alpha\phi^\beta$ and $\lambda\lambda$. Using this method the form
of the two point function for the energy momentum tensor in the conformal
$O(N)$ model with a plane boundary is also found. General results for the
sum of the contributions of all derivative operators appearing in the operator
product expansion, and also in a corresponding boundary operator expansion,
to the two point functions are also derived making essential use of
conformal invariance.
\vfill\eject}
\pageno=1

\leftline{\bigbf 1 Introduction}
\medskip

In more than two dimensions it is not generally possible to construct
explicitly non trivial conformal field theories, with a detailed
knowledge of the spectrum of spins and scale dimensions of all operators
in the theory and further the coefficients appearing in operator product
expansions for each pair of operators,
at least to the same degree as in two dimensions.  Nevertheless for a
very large class of quantum field theories scale invariance at
possible critical points may also be extended to invariance under the
full conformal group [1] which
implies significant restrictions on the form of multi-point
correlation functions [2].  In particular the functional form of two and
three point functions is effectively unique assuming conformal covariance
since to construct conformal invariants, and hence for arbitrary functions
to be present, four or more points are required (for operators
with spin there may be more than one linearly independent conformally
covariant form for the three point function [3]).  From an experimental
viewpoint in a statistical physics context only two point functions
are mostly relevant and in this case conformal invariance  gives
little more  than just scale invariance.  On the other hand for
calculating critical exponents in the $1/N$ expansion using conformal
invariance has proved essential in obtaining results to $\rO(1/N^2)$ and
$\rO(1/N^3)$ [4].

For statistical mechanical problems involving a boundary then at a
critical point there are new critical exponents, expressing the
behaviour of physical quantities near or on the boundary, which are unrelated
(at least in any simple general fashion) to bulk critical exponents [5].
Also for any particular bulk critical point there are a variety of
possible boundary conditions with differing surface exponents.  As
Cardy first showed [6,1] there is still a residual conformal group
consisting of conformal transformations leaving the boundary
invariant.  For a plane boundary in $d$ Euclidean dimensions the
restricted conformal group is then $O(d - 1,1)$.  In this case the two
point function, involving operators at $x,x'$, depends on functions of
a single conformal invariant $\xi(x,x')$.
These functions depend on the particular theory and associated boundary
conditions, or rather on their corresponding universality class. However
conformal invariance is a significantly stronger requirement than scale
invariance as far as potential experimental implications are concerned [7].
If we define coordinates  $x_\mu = (y,\bx)$, with $y$ measuring
the perpendicular distance from the boundary, then scale invariance
by itself, with conventional translational and rotational symmetries,
only restricts the two point function to depend on functions of the
two scale invariant variables $\bs^2/y^2 , \, \bs^2/y^{\prime 2}$
where $\bs=\bx-\bx'$.

For scalar fields there is a single function of $\xi$ in the associated
two point function but for fields with spin there may be several.
As a particular illustration we consider
the energy momentum tensor $T_{\mu\nu}$, which is traceless in
the conformal limit, and the two point function then contains three
possible invariant functions.  However the conservation equation
$\partial_\mu  T_{\mu\nu} = 0$ provides two first order linear differential
equations linking these functions.  In two dimensions the number of
invariant functions is reduced to two and in this case the
differential equations have a unique solution, satisfying appropriate
boundary conditions, in accord with known results [8]. Under the
restricted conformal group it is also possible to form non zero two
point functions for fields of differing spin and scale dimensions,
such as for $T_{\mu\nu}$ and a scalar field although in this case the
functional dependence is entirely determined.

In this paper we investigate conditions on the form of two point
functions for operators in conformally invariant theories with a plane
boundary.  Previously we discussed such two point functions involving the
energy momentum tensor and derived the necessary conditions arising from
the conservation equations for $T_{\mu\nu}$ by considering a perpendicular
configuration where the two points were restricted to be on a
perpendicular to the boundary [9] (referred to subsequently as I).
Here we construct a conformally
covariant form for the general two point function $\l T_{\mu\nu} (x)
T_{\rho\sigma} (x') \r$ for arbitrary $x,x'$ by constructing in terms
of the invariant scalar $\xi(x,x')$ vectors
$X_\mu \propto \partial_\mu \xi,\,  X'{}_{\! \si} \propto
\partial'{}_{\! \si} \xi$ at $x, x'$ respectively.

In general a two point function in the presence of a boundary may be written,
expressing invariance under translations parallel to the boundary, as
$\l {\O} (x) {\O'} (x')\r = G(y,y',\bx-\bx')$.
In our discussions it is useful to
consider a transformation of $G$ obtained by integrating over planes parallel
to the boundary, $ \int \! \d^{d-1}{\bx}\, G(y,y',\bx) = G_\parallel(y,y')$,
which defines what is sometimes referred to as the
parallel susceptibility. For a conformally invariant theory, when
$\l {\O} (x) {\O'} (x')\r$ is expressible in terms of a function $g(\xi)$,
this procedure, which we refer to as parallel integration,
defines a new function $\hat{g}(\rho), \ \rho = (y-y')^2 / 4yy'$.
Crucially we are able to show that the transformation $g \rightarrow
\hat{g}$ is invertible  with $g$ expressible in terms of an integral
over ${\hat g}$, for $d=3$ the result is simply $g(\xi)= -{\hat g}'(\xi)/\pi$.
This then allows for significant simplifications
in calculations  since finding $ G_\parallel(y,y')$ is sufficient to
determine the full two point function $\l {\O} (x) {\O'} (x')\r$.

These methods are illustrated by application to the $O(N) \ \sigma$-model
in the large $N$ limit. This is in the same universality class
as an $N$ component scalar field theory with a renormalisable
$\phi^4$ interaction at the non Gaussian Wilson fixed point.
We are able to recover some results obtained some time ago for the
arbitrary function of $\xi$ associated with the two point functions of
the basic $N$ component fields $\phi$ and also the auxiliary field
$\lambda$ which is a scalar under $O(N)$ [10,11].

We also consider constraints on the invariant function $g(\xi)$ arising
from the operator product expansion in which the operators with lowest
dimension and non vanishing one point functions are relevant for the limit
$\xi \rightarrow 0$.  With conformal invariance only scalar operators, or
their derivatives, in the operator product expansion of $\O$ and $\O'$
can contribute to their two point function in the presence of a boundary.
We derive an explicit form for $g(\xi)$, in terms
of hypergeometric functions, resulting from all derivative operators
formed from any particular scalar operator occurring in the operator product
expansion. The derivative terms in the operator product expansion are
determined by the requirement that they should reproduce exactly the
appropriate full three point function for the conformal field
theory without boundary.  These results are applied to two point amplitudes
involving $T_{\mu\nu}$ and the expressions obtained automatically satisfy
the required conservation equations.

A similar boundary operator expansion, where a bulk operator ${\O}(x)$
is expanded in terms of boundary operators $\hat{\O}({\bx})$ [12],
is also investigated. This constrains the function $g(\xi)$ in the limit
$\xi\to \infty$. Again, using conformal invariance,
we are able to sum up explicitly the results for all derivative
operators formed from a given boundary operator which in general also
involves hypergeometric functions.  In this case the essential input
determining the derivative terms for a particular boundary operator $\hO$
appearing in the expansion of $\O$
is the form of the two point function for $\hO$ and also $\O$ at arbitrary
points. The boundary operator expansion of the energy momentum
tensor $T_{\mu\nu}$ defines a boundary scalar operator $\hat{T}$ which is
given by the non singular limit of $T_{\perp\perp}(x)$  as
$x \rightarrow (0, {\bx})$ and therefore has dimension $d$ [8]. However for
$d > 2$ it is necessary to also consider boundary
operators $\hcT_{ij}$, symmetric traceless tensors  whose dimension
is not constrained by general principles (this again reflects the non
uniqueness of $\l T_{\mu\nu}(x) T_{\sigma\rho} (x') \r$ for $d > 2$).

In detail in section 2 we consider the general conditions stemming
from conformal invariance for two point functions with a boundary.  In
particular we describe the notion of quasi-primary operators which have
simple properties under conformal transformations and analyse in detail
the forms of the two point amplitudes for vector
operators $V_\mu$ and also $T_{\mu \nu}$ in the presence of a boundary.
Using the vectors $X_\mu,\, X'{}_{\!\sigma}$ it is straightforward to
write down conformally covariant expressions.  In section 3 we consider
the simplest conformally invariant scalar field theories defined by free
fields and also with a quartic interaction at the non Gaussian fixed point
present in the $\vep = 4-d$ expansion. In particular we calculate the two
point function for the operator $\phi^2$ to first order in $\vep$.  In
section 4 we analyse the $O(N)$ model in the large $N$ limit and
describe in detail the transformation $g \leftrightarrow \hat{g}$
mentioned above.  Section 5 contains more calculations for the two
point functions of $V_\mu$ or $T_{\mu\nu}$ in the $O(N)$ model to
leading order in $1/N$.  The result for the two point function of the
energy momentum tensor is non trivial, involving generalised hypergeometric
functions. Sections 6 and 7 respectively contain the
details of our discussions of the consequences of the operator product
and boundary operator expansions with applications to the results
obtained in the $O(N)$ model. Various calculational details are
relegated to five appendices.
\bigskip
\leftline{\bigbf 2 Conformal Invariance with Plane Boundaries}
\medskip

In flat $d$-dimensional Euclidean space with coordinates $x_\mu\in \bR^d$ a
conformal transformation  $g$ is defined by preservation of the line element up
to a local scale factor
$$ x_\mu \to x^g{}_{\! \mu} (x)  \, ,
\quad \d x^g{}_{\! \mu} \d x^g{}_{\! \mu} = \Omega^g(x)^{-2} \d x_\mu \d x_\mu
\  \ \Rightarrow \ \ \d^d x^g = \Omega^g(x)^{-d} \d^d x \, .
\eqno (2.1) $$
Such transformations define a group, $(x^{g_1})^{g_2} = x^{g_2 g_1}$, which
is isomorphic to $O(d+1,1)$. For any conformal transformation we may define
a local orthogonal matrix belonging to $O(d)$ by
$$\R^g{}_{\!\mu \alpha}(x) = \Omega^g (x)
{\pr x^g{}_{\! \mu} \over \pr x_\alpha} \ ,
\eqno (2.2) $$
satisfying $\R^{g_2} (x^{g_1}) \R^{g_1} (x) = \R^{g_2 g_1} (x)$.
For $d>2$ arbitrary conformal transformations can be generated by combining
translations and rotations, for which $\Omega^g=1$, with inversions through
the origin, $i$, taking ${ x_\mu \to x_\mu  / x^2}$, for which
$$ \R^i{}_{\!\mu \nu} (x) =
I_{\mu \nu}(x) \equiv \de_{\mu \nu} - 2 \, {x_\mu x_\nu \over x^2} \, ,
\quad \Omega^i (x) = x^2 \ .
\eqno (2.3) $$
Under conformal transformations $I_{\mu\nu}
(x-x') \to \R_{\mu\alpha}(x)\R_{\nu\beta}(x') I_{\alpha\beta}(x-x')$ so
that it transforms as a vector at $x$ and also at $x'$. In consequence
$I_{\mu\nu}(x-x')$ may be regarded as representing a form of parallel
transport for conformal transformations and it plays a crucial role in
the construction of two point functions for vector and other tensor fields.

In a conformal field theory fields $\O^i(x)$ which form the vector space
on which some finite dimensional irreducible representation of $O(d)$ acts
are defined as quasi-primary fields
if they transform under the full conformal group according to [13]
$$ \O^i(x) \to \O^{g \, i} (x^g) =
\Omega^g (x)^{\eta}  D^i {}_{\! j} ( \R^g(x) ) \O^j (x) \, ,
\eqno (2.4) $$
where $\eta$ is the scale dimension of the fields and, for any $R\in O(d)$,
$D(R)$ is the corresponding matrix in this representation.
Since the transformation rule in (2.4) depends on $x$, derivatives of
quasi-primary fields are not in general quasi-primary but transform with extra
inhomogeneous terms. However for $V_\mu$ a vector field of dimension $d-1$ or
$T_{\mu\nu}$ a symmetric traceless tensor field of dimension $d$ it is
straightforward to show [3] that $\pr_\mu V_\mu$ and $\pr_\mu T_{\mu\nu}$ are
quasi-primary fields. Hence it follows that in a conformal field theory
a conserved vector current and the
energy momentum tensor necessarily have dimensions $d-1$ and $d$
respectively. For the full conformal group there are no invariants
which can be constructed from two or three points so that the two or three
point functions of quasi-primary fields are essentially uniquely determined,
with no arbitrary functions present.

For a flat Euclidean space with a plane boundary, as was shown by Cardy [6],
a non trivial subgroup of conformal transformations still remains. If we let
$x_\mu = (y,\bx)$ and define a plane boundary by $y=0$, so that $y$
is the perpendicular distance from $x$ to the boundary, then it is
necessary  to restrict the conformal group to those transformations
leaving $y=0$ invariant. This subgroup is then generated by $d-1$
dimensional translations and  $O(d-1)$ rotations acting on $\bx$ together
with the inversion $ x_\mu \to x_\mu  / x^2$ again and forms the group
$O(d,1)$. Under such transformations for two points $x_\mu,\, x'{}_{\!\mu}$ it
is easy to see that
$$
(x-x')^2 \to {(x-x')^2 \over \Omega(x) \Omega(x')} \ , \quad
y \to {y \over \Omega(x) } \ , \quad
y' \to {y' \over \Omega(x') } \ .
\eqno (2.5) $$
Hence we may construct invariants
$$
\xi = {(x-x')^2\over 4yy'}\, , \quad
v^2 = {(x-x')^2 \over (x - x')^2 + 4yy'} = {\xi \over \xi + 1}\ .
\eqno (2.6) $$
where $0\le \xi < \infty $ and $0\le v < 1$.

For a one point function of an operator $\O$ in
the neighbourhood of a plane boundary conformal invariance under
transformations as in (2.4) implies that this can only be non zero for
quasi-primary scalar fields, belonging to the singlet representation of
$O(d)$, when we can write
$$ \l \O (x) \r = {A_\O \over (2y)^\eta } \, .
\eqno (2.7) $$

For the two point function of quasi-primary operators the existence of the
conformal invariants in (2.5) implies that there may be an arbitrary function
present. In addition the two point function may be non zero for operators of
differing spins and scale dimension, unlike the case for conformal invariance
without a boundary. For two scalar fields we may write in general
$$
\langle \O_1(x) \O_2(x')\rangle = {1\over (2y)^{\eta_1} (2y')^{\eta_2}}\,
f_{12}(\xi) = {(2y')^{\eta_1-\eta_2}\over (x-x')^{2\eta_1}}\, F_{12}(v) \ ,
\quad \xi^{\eta_1} f_{12}(\xi) = F_{12}(v) \, .
\eqno (2.8) $$
The functions $f_{12}(\xi)$ or $F_{12}(v)$ are constrained by the operator
product expansion (OPE). If for $s=x-x'\to 0$
$$ \O_1(x) \O_2(x') \sim {C_{12}{}^3\over (s^2)^{
{1\over 2} (\eta_1 + \eta_2 - \eta_3)}} \O_3 (x') \, , \quad
\l \O_3 (x') \r = {A_3 \over (2y')^{\eta_3} } \, ,
\eqno (2.9) $$
then for $\xi \sim v^2 \to 0$ there is a contribution to the functions
$f_{12}(\xi)$ or $F_{12}(v)$ in (2.8) of the form
$$ f_{12}(\xi) \sim C_{12}{}^3 A_3 \, \xi^{-{1\over 2}
(\eta_1 + \eta_2 - \eta_3)} \, ,
\qquad F_{12}(v) \sim C_{12}{}^3 A_3 \, v^{\eta_1 - \eta_2 + \eta_3} \, .
\eqno (2.10) $$
Later, in section 6, we determine the entire contribution of non leading
derivative terms in the OPE.

There are additional constraints  on the
functions $f_{12}(\xi)$ or $F_{12}(v)$ arising
from a boundary operator expansion (BOE). For conformal field theories
with boundary it is natural to define a basis of boundary operators
$\hO_n(\bx)$, which transform irreducibly under $O(d-1)$ and have a well
defined scale dimension $\heta_n$. The precise set of such operators depends
of course on the particular conformal theory and also the precise boundary
conditions, compatible with conformal invariance imposed.  The BOE has the
form
$$ \O (x) = \sum_n {B_{\O}{}^{\hO_n}\over (2y)^{\eta - \heta_n}} \,
\hO_n (\bx) \, ,
\eqno (2.11) $$
where in (2.7) $A_\O = B_{\O}{}^1$.
The two point function of a bulk operator and a boundary operator is
determined up to an overall constant
$$ \l  \O (x) \hO_n(\bx') \r = {B_{\O \hO_n}\over (2y)^{\eta-\heta_n}
(\hs^2)^{\heta_n}} \, , \quad \hs_\mu = (y, \bx - \bx') \, .
\eqno (2.12) $$
Combining (2.11) and (2.12) it is easy to find that in the limit $\xi\to
\infty$ or $v\to 1$
$$ f_{12}(\xi) \sim B_{\O_1 \hO}B_{\O_2}{}^{\!\hO} \xi^{-\heta} \, , \qquad
F_{12}(v) \sim B_{\O_1 \hO}B_{\O_2}{}^{\!\hO} (1-v^2)^{\heta - \eta_1} \, ,
\eqno (2.13) $$
where $\heta$ is the scale dimension of the leading boundary operator $\hO$
which gives a non zero contribution when using (2.11) and (2.12) to
take the limit $y'\to 0$. Clearly there are consistency conditions arising
from considering the alternative limit $y\to 0$.
For a statistical mechanical system at a critical point the invariant
functions $f(\xi)$ or $F(v)$ appearing in two point functions are universal,
depending only on the universality class and the boundary
conditions.\footnote{*}{Note that the results (2.8) give for the limits
of the two point function perpendicular and parallel to the boundary
$\l \O_1(y,\0) \O_2(y',\0) \r \propto 1/y^{\eta_\perp}$
as $y\to \infty$ and $\l \O_1(y,\bx) \O_2(y,\bx') \r \propto
1/|\bs|^{\eta_\parallel}$ as $|\bs| \to \infty$ where the surface critical
exponents $\eta_\perp =\eta_1 + \heta$, $\eta_\parallel = 2\heta$ [10].}

For non scalar quasi-primary fields the expressions for the two point
function in the neighbourhood of a plane boundary are more complicated
to analyse. In I a conformal transformation was used
to restrict $x_\mu = (y,\0)$ and $x'{}_{\!\mu} = (y',\0)$, on a perpendicular
to the boundary, where the remaining invariance or little group is
reduced to $O(d-1)$ rotations. However we now describe an alternative
approach which gives equivalent results. This is based on
defining vectors $X_\mu, \, X'{}_{\!\mu}$, with zero scale dimension,
under restricted conformal transformations preserving the boundary at
$x, \,x'$ respectively, so that $X_\mu \to \R_{\mu\alpha}(x)X_\alpha , \,
X'{}_{\!\mu} \to \R_{\mu\alpha}(x')X'{}_{\!\beta}$. Since $\xi$ in (2.6)
is a scalar these are explicitly given by
$$ X_\mu = y {v\over \xi}\pr_\mu \xi = v \Bigl ( {2y\over s^2} s_\mu -
n_\mu \Bigl ) \, , \quad
X'{}_{\! \mu} = y' {v\over \xi}\pr'{}_{\! \mu} \xi =
v \Bigl ( - {2y'\over s^2} s_\mu - n_\mu \Bigl ) \, , \quad
n_\mu = (1,\0) \, ,
\eqno (2.14) $$
which satisfy
$$ X_\mu X_\mu = X'{}_{\! \mu} X'{}_{\! \mu} = 1 \, , \quad
X'{}_{\! \mu} = I_{\mu\nu} (s) X_\nu \, .
\eqno (2.15) $$
As $y\to 0$ $X_\mu \to - n_\mu$ and for $y'\to 0 $ $X_\mu \to
- I_{\mu\nu} (\hs) n_\nu$.

We may now easily construct invariant forms for two point amplitudes
for tensor fields. For $V_\mu$ a vector field of dimension $d-1$ then
$$ \l V_\mu (x) V_\nu (x') \r = {1\over (s^2)^{d-1}} \bigl ( I_{\mu\nu}(s)
C(v) + X_\mu  X'{}_{\! \nu} D(v) \bigl ) \, .
\eqno (2.16) $$
To impose current conservation, $\pr_\mu V_\mu = 0$, we use
$$  \eqalign {
& \pr_\mu \Bigl ( {1\over (s^2)^{d-1}} I_{\mu\nu}(s) \Bigl ){} = 0 \, , \quad
\pr_\mu \Bigl ( {1\over (s^2)^{d-1}} X_\mu \Bigl ){} = - (d-1) \,
{2y'\over (s^2)^d}\, v \, , \cr
& \pr_\mu X'{}_{\! \nu}= {2y'\over s^2} \, v
\bigl ( - I_{\mu\nu}(s) +  X_\mu  X'{}_{\! \nu} \bigl ) \, , \quad
\pr_\mu F(v) = {2y'\over s^2}\, X_\mu v^2 F'(v) \, , \cr}
\eqno (2.17) $$
to obtain
$$ v{\d \over \d v} (C+D) = (d-1) D \, .
\eqno (2.18) $$
Away from the boundary, for $v\to 0$, $D\to 0$ while $C$ is a constant
determined by the bulk conformal theory. Similarly for $T_{\mu\nu}$ the
traceless energy momentum tensor of dimension $d$ and $\O$ a scalar field
of dimension $\eta$ we may find expressions for the mixed two point functions
$$\eqalign {
\l V_\mu (x) \O (x') \r = {}&{(2y')^{d-1-\eta} \over (s^2)^{d-1}} \,
X_\mu C_{V\O}(v) \, ,\cr
\l T_{\mu\nu} (x) \O (x') \r = {}& {(2y')^{d-\eta} \over (s^2)^d} \Bigl (
X_\mu X_\nu - {1\over d} \delta_{\mu \nu} \Bigl) C_{T\O}(v) \, . \cr}
\eqno (2.19) $$
Current conservation and also $\pr_\mu T_{\mu\nu} = 0$, using (2.17)
and
$$ \pr_\mu \biggl ( {1\over (s^2)^d} \Bigl (
X_\mu X_\nu - {1\over d} \delta_{\mu \nu} \Bigl) \biggl ){} =
- (d-1) \, {2y'\over (s^2)^{d+1}} \, v X_\nu \, ,
\eqno (2.20) $$
requires
$$ v{\d \over \d v} C_{V\O} = (d-1)  C_{V\O} \, , \quad
v{\d \over \d v} C_{T\O} = d C_{T\O} \, ,
\eqno (2.21) $$
which gives a unique functional form
$$ C_{V\O}(v) = c_{V\O} v^{d-1} \, , \qquad
C_{T\O} (v) = c_{T\O} v^d \, .
\eqno (2.22) $$
The coefficients $c_{V\O}$ and $c_{T\O}$ are determined by Ward identities.
The energy momentum tensor $T_{\mu\nu}$ may be defined by the response to
arbitrary infinitesimal reparameterisations $x_\mu \to x_\mu + a_\mu(x)$
[6,14] and from conformal invariance it is possible to derive the relations
(see eqs. (E.7,8) in ref. [3])
$$ \eqalign{
\l T_{\mu\mu} (x) \O(x') \r = {}& d \, C \l \O(x') \r \de^d(s) \, , \cr
 \pr_\mu \l T_{\mu\nu} (x) \O(x') \r = {}& \Big ( {\eta\over d} + C \Big )
\l \O(x') \r \pr_\nu \de^d(s) - \l \pr_\nu \O(x') \r \de^d(s) \, , \cr}
\eqno (2.23) $$
where $C$ is undetermined (reflecting the arbitrariness of
$\l T_{\mu\nu} (x) \O(x') \r$  when regularised
up to terms ${\propto \de_{\mu\nu} \de^d (x-x') (2y')^{-\eta}}$).
The expression obtained in (2.19,22) can be shown to be compatible with (2.23)
by considering the expansion for $s\to 0$ which has the form
$$  \eqalign{ \!\!\!\!
{v^d \over (s^2)^d} \Bigl ( X_\mu X_\nu - {1\over d} \delta_{\mu \nu} \Bigl)
{} \sim {}& {1\over (2y')^d (s^2)^{\hh d}}
\bigg \{ {s_\mu s_\nu \over s^2} - {1\over d}\,\de_{\mu\nu} \cr
{}& -{1\over 2y'} \Big ( n_\mu s_\nu + n_\nu s_\mu - \de_{\mu\nu} n{\cdot s}
+ (d-2) n{\cdot s} {s_\mu s_\nu \over s^2} \Big ) \bigg \} \, . \cr}
\eqno (2.24) $$
With a suitable regularisation of the terms $\rO(s^{-d})$ we may find
\footnote{*}{In the spirit of differential regularisation [15]
we may obtain a well defined distribution for the $\rO(s^{-d})$ terms
${\displaystyle {1\over s^d}\Big({s_\mu s_\nu \over s^2} - {1\over d}\,
\de_{\mu\nu} \Big ) = {1\over d(d-2)} \Big ( \pr_\mu \pr_\nu - {1\over d}
\de_{\mu\nu} \pr^2 \Big ) {1\over s^{d-2}} + c \de_{\mu\nu} \de^d (s)}$,
where $c$ is an arbitrary constant.
The divergence and trace of this expression may be found using $-\pr^2 s^{-d+2}
= (d-2)S_d \de^d(s)$.
The trace of the $\rO(s^{-d+1})$ terms in (2.24) is zero while the
divergence is unambiguous being $\propto \de^d (s)$.}
$$
\pr_\mu {v^d \over (s^2)^d}
\Bigl ( X_\mu X_\nu - {1\over d} \delta_{\mu \nu} \Bigl){}
= - {d-1\over d^2}S_d \, {1\over (2y)^d}\pr_\nu \de^d (x-x')
+ c \, {1\over (2y')^d}\pr_\nu \de^d (x-x') \, ,
\eqno (2.25) $$
for $c$ arbitrary and $S_d = 2\pi^{\hh d}/ \Gamma (\half d)$.
Assuming the result (2.25) the regularised trace of (2.24) is then also
${dc\de^d (x-x')/(2y)^d}$. With these formulae it is easy to check that the
identities (2.23) are satisfied, using (2.7), if $CA_\O = c_{T\O} c$ and [8,7]
$$ c_{T\O} = - {d\eta\over d-1} \, {A_\O \over S_d} \, .
\eqno (2.26) $$

The two point function of the energy momentum tensor itself can also be written
in a similar fashion in terms of a basis of conformally covariant
symmetric traceless tensors at $x,x'$ in the general form
$$ \eqalign {\!\!\!\!\!\!
\l T_{\mu\nu} (x) T_{\si\rho}(x') \r = {1\over (s^2)^d} \biggl \{
& \I_{\mu\nu,\si\rho}(s) C(v) +  \Bigl (
X_\mu X_\nu - {1\over d} \delta_{\mu \nu} \Bigl)  \Bigl (
X'{}_{\!\si} X'{}_{\!\rho} - {1\over d} \delta_{\si\rho} \Bigl) A(v) \cr
{} + & \Bigl ( X_\mu X'{}_{\! \si} I_{\nu\rho}(s)
+ \mu \leftrightarrow \nu , \si \leftrightarrow \rho \cr
& {} - {4\over d}
\delta_{\mu\nu} X'{}_{\! \si} X'{}_{\!\rho} - {4\over d} \delta_{\si\rho}
X_\mu X_\nu + {4\over d^2} \delta_{\mu\nu} \delta_{\si\rho} \Bigl ) B(v)
\biggl \} \, , \cr}
\eqno (2.27) $$
where $\I_{\mu\nu,\si\rho}$ represents inversion on symmetric traceless
tensors
$$\I_{\mu \nu,\si \rho} (s) = \half \bigl ( I_{\mu \si}(s) I_{\nu \rho}
(s) + I_{\mu \rho}(s) I_{\nu \si} (s) \bigl ){} - {1\over d}\, \de_{\mu
\nu} \de_{\si \rho} \ .
\eqno (2.28) $$
If the two point function is defined on all $\bR^d$, with no boundary, then
$A, B$ are absent while $C$ is a constant $C_T$ and the
conservation equation for $T_{\mu\nu}$, giving $\pr_\mu
\l T_{\mu\nu} (x) T_{\si\rho}(x') \r = 0$, is satisfied by virtue of
$$ \pr_\mu \Bigl ( {1\over (s^2)^d} \I_{\mu \nu,\si \rho} (s) \Bigl ){}=0 \, .
\eqno (2.29) $$
In the general case the conservation equation, using (2.29) with (2.17),
(2.20) as well as
$$ \eqalign {
\pr_\mu \Bigl ({1\over (s^2)^d} \bigl ( X_\mu I_{\nu\si}(s)
+ &X_\nu I_{\mu\si}(s)  - {2\over d} \delta_{\mu\nu} X'{}_{\! \si} \bigl )
\Bigl ) \cr
& = {2y'\over (s^2)^{d+1}} \, {v\over d} \bigl ( (2-d^2) I_{\nu\si}(s)
+ (d-2) X_\nu X'{}_{\! \si} \bigl ) \, , \cr}
\eqno (2.30) $$
leads to the conditions
$$ \eqalignno {
\Bigl ( v{\d \over \d v} - d \Bigl ) (C+2B ) = {}& - {2\over d} (A+4B) -
dC \, , & (2.31a) \cr
\Bigl ( v{\d \over \d v} - d \Bigl ) \bigl ( (d-1) A + 2 (d-2) B \bigl ){}
= {} & 2A - 2(d^2 - 4) B \, . & (2.31b) \cr}
$$
These equations still leave the $A,B,C$ undetermined up to an arbitrary
function of $v$. Away from the boundary, or as $v,\xi\to 0$, the terms
involving $X_\mu, X'{}_{\!\si}$ should disappear, hence in this limit
$A,B \to 0$, while $C\to C_T$ which is the constant determining the
scale of energy momentum tensor two point function which has a simple form
in the associated conformal field theory without boundary,
$$
\l T_{\mu\nu} (x) T_{\si\rho}(x') \r_{\rm no \ boundary} = {C_T\over (s^2)^d}
\,
\I_{\mu\nu,\si\rho}(s)  \, .
\eqno (2.32) $$
On the boundary
if we assume $T_{1i} (0,\bx) = 0$, with $i$ denoting components tangential
to the boundary, then we must require $C(1) + 2 B(1) =0 $. Since
$\l T_{\mu\nu} \r =0$ with a plane boundary there are no Ward identity
relations for this two point function.

When $d=2$ there are only two linearly independent tensors instead of the
three in (2.27) so that only $A,\, C+2B$ are relevant. The differential
equations then have a unique solution obeying the boundary conditions
$$
A (v) = 4 C_T v^4 \, , \qquad C(v) + 2B(v) = C_T (1-v^4) \, .
\eqno (2.33) $$
In two dimensions with complex coordinates $z=x+iy , \bz = x-iy$,
$\delta_{z\bz}= \half$ then
$$ X_z X'{}_{\! z} = \half I_{zz}(s) = - {1\over 4}\, {\bz - \bz'\over z-z'}
\, , \quad
X_z X'{}_{\! \bz}  = {1\over 4}\, {\bz - z'\over z- \bz'} \, , \quad
I_{z\bz} ( s) = 0 \, ,
\eqno (2.34) $$
and since $v^2 = |z-z'|^2/ | z - \bz'|^2$ we thereby obtain the universal
form [1]
$$ \l T_{zz} (x) T_{zz}(z') \r = {\quar C_T \over (z-z')^4} \, , \qquad
\l T_{zz} (z) T_{\bz\bz}(\bz') \r = {\quar C_T \over (z-\bz')^4} \, .
\eqno (2.35) $$
Up to a suitable normalisation factor $C_T$ is of course the Virasoro central
charge.

For general $d$ the expression (2.27) for the energy momentum tensor two
point function may be related to the results in I, where free
field expressions were given for arbitrary $d$ and also the functional
dependence on $v$ calculated to first order in the $\vep$ expansion from
$d=4$ for the non Gaussian fixed point in $\phi^4$ field theory, by
restricting (2.27) to the perpendicular configuration, when for $y>y'$,
$X_\mu = - X'{}_{\!\mu} = n_\mu , \ v =(y-y')/(y+y')$, the general form
can be written as
$$ \eqalign {
\l T_{1j} (y, \0 ) T_{1\ell} (y',\0) \r = {}& {1\over (y-y')^{2d}} \, \gamma
(v)\delta_{j\ell} \, , \cr
\l T_{ij} (y, \0 ) T_{k\ell} (y',\0) \r = {}& {1\over (y-y')^{2d}} \bigl (
\de(v) \, \de_{ij} \de_{k\ell} + \epsilon(v) (\de_{ik} \de_{j\ell}
+ \de_{i\ell} \de_{jk} ) \bigl ) \, . \cr }
\eqno (2.36) $$
All other non zero components may be obtained trivially from (2.36) using
$T_{\mu\nu} = T_{\nu\mu}$ and $T_{\mu\mu}=0$, for instance
$$ \l T_{11} (y, \0 ) T_{11} (y',\0) \r
= {1\over (y-y')^{2d}} \, \alpha(v) \, , \qquad
\alpha = (d-1) \bigl ( (d-1)\delta + 2\epsilon \bigl ) \, .
\eqno (2.37) $$
It is straightforward to show that
$$ \alpha = {d-1\over d^2} \big ( (d-1)( A+4B) + dC \big ) \, , \quad
\gamma = -B -\half C  \, , \quad \ep = \half C \, , \quad
\delta = {1\over d^2}( A+4B-dC) \, .
\eqno (2.38) $$
Eqs. (2.31a,b) then translates into the results given in I, for instance
(2.31b) becomes more simply
$$ \Bigl ( v{\d \over \d v} - d \Bigl ) \alpha (v) = 2(d-1) \gamma (v) \, .
\eqno (2.39) $$
On the boundary itself it is easy to see that
$$ \l T_{11} (0,\bx ) T_{11} (0,\bx') \r = {1\over (\bs^2)^d} \alpha(1) \, .
\eqno (2.40) $$
\bigskip
\leftline{\bigbf 3 Scalar Field Theory}
\medskip

The simplest conformally invariant field theory is that given by free
massless scalar fields $\phi^\alpha(x), \, \alpha = 1 \dots N$ (for later
convenience we assume $N$ components). To ensure compatibility with conformal
invariance we may impose either Neumann or Dirichlet boundary conditions
which for a plane boundary at $y=0$ requires $\pr_1 \phi^\alpha(0,\bx) = 0$ or
$ \phi^\alpha(0,\bx) = 0$ respectively. If
$$ \l \phi^\alpha(x) \phi^\beta (x') \r = \delta^{\alpha\beta}G_\phi(x,x')
\, , \quad G_\phi(x,x') = {1\over s^{2\eta_\phi}} F_\phi(v) \, ,
\eqno (3.1) $$
then in the free field case it is easy to see, using the method of images, that
$$  \eta_\phi = \half d - 1 \, , \quad F_\phi(v) = A ( 1 \pm v^{d-2} ) \, ,
\eqno (3.2) $$
where the upper/lower signs correspond to Neumann/Dirichlet  boundary
conditions and
$$ A = {1\over (d-2) S_d} \, , \qquad S_d = {2\pi^{\hh d}\over \Gamma
(\half d)} \, .
\eqno (3.3) $$
For the composite operator $\phi^2$ with dimension $\eta_{\phi^2}$ then in
general
$$ \eqalign {
\l \phi^2(x) \phi^2 (x') \r & =  \l \phi^2(x) \r \l \phi^2 (x') \r
+ G_{\phi^2}(x,x') \, , \cr
G_{\phi^2}(x,x') & = {1\over s^{2\eta_{\phi^2}}} F_{\phi^2}(v) \, , \quad
\l \phi^2(x) \r = {A_{\phi^2} \over (2y)^{\eta_{\phi^2}}} \, , \cr}
\eqno (3.4) $$
and in the free field case
$$ \eta_{\phi^2} = 2 \eta_\phi = d-2 \, , \qquad F_{\phi^2}(v) = 2N
F_\phi(v)^2 \, , \quad A_{\phi^2} = \pm N A \, .
\eqno (3.5) $$

A more interesting case is the conformal field theory realised at the
non Gaussian fixed point in the theory with interaction ${1\over 24}g
(\phi^2)^2$ for which critical exponents and other universal quantities
can be calculated in the $\vep = 4 -d$ expansion. With
minimal subtraction at the fixed point $g_*/16\pi^2 = 3 \vep/
(N+8) + \rO(\vep^2)$. For
the $\l \phi \phi \r$ two point function the $\rO (\vep)$ corrections
to the free field results for the universal function $F_\phi(v)$ have
been calculated by Gompper and Wagner [16] and also in I\footnote{*}{In I the
results in (C.1,2), and also (C.3), for $G(y,y')$ should have an additional
$\pm$ sign. Other
misprints in I are that the factor multiplying $F''$ in (C.5) should
be $v^2(1-v^2)^2$, in (C.8a) the tensor structure should be
$\de_{ik}\de_{j\ell} + \de_{i\ell}\de_{jk}$ and in the last line of (C.8b)
the second factor should be $x^2+z^2-y'^2$, in (2.6) in the transformation
equation for $T_{\mu\nu}$ replace $\Omega({\tilde x})$ by
$\Omega({\tilde x})^d$, in (2.26) the correct equation is $A_{ij} =
\eta \de_{ij}/(d-1)S_d$ and in (2.29) in the formula for $\alpha(1)$
$2d\to 2^d$. Further in (A.2), and also subsequently on the same page,
$e^\mu{}_i e^\mu{}_j \to e^\mu{}_i e^\nu{}_j$.}. Since $g_*= \rO(\vep)$
it is sufficient to first order to evaluate the Feynman integrals just
for $d=4$. The results obtained at
one loop give $\eta_\phi = \half d - 1 + \rO (\vep^2)$ and
$$ F(v)^{(1)} = \mp {1\over 2}\, {N+2 \over N+8}A \, \vep \, \Bigl (
v^2 \ln {1-v^2\over v^2} \pm \ln (1-v^2) \Bigl ) \ .
\eqno (3.6) $$
In the Dirichlet case it is easy to see that adding the result (3.6) to
the free result (3.2) is compatible to the order calculated with the form
$$ F_\phi(v)_{\rm ord} = C_{\phi} \bigl ( 1 - v^{\eta_{\phi^2}}
\bigl )^{\heta_n - \eta_\phi} \, ,
\eqno (3.7) $$
where $C_\phi = A (1+\rO(\vep))$ and $\heta_n$ is the dimension of the
surface operator $\hphi_n(\bx) = \pr_1\phi (0,\bx)$. The expression (3.6)
implies the one loop corrections
$$
\eta_{\phi^2} = d - 2 + {N+2\over N+8} \vep + \dots \, , \quad
\heta_n = \half d - \half \, {N+2\over N+8} \vep + \dots \, ,
\eqno (3.8) $$
in accord with long established results.
The subscript $\rm ord$ in (3.7) denotes that this function is appropriate
to the ordinary transition in a statistical mechanical context.
It is straightforward then to verify that the functional form in (3.7)
is consistent with the limiting behaviours given by (2.10), since $\phi^2$
is here the lowest dimension operator appearing in the operator product
$\phi\phi$ beyond the identity, and also with (2.13) since in the Dirichlet
case $\hphi_n$ is the leading boundary operator
appearing in the BOE of $\phi$.

For the Neumann case, which is appropriate for the so called special
transition,
we may write
$$ F_\phi(v)_{\rm sp} = C_{\phi} \bigl ( 1 + v^{\eta_{\phi^2}}
\bigl ) ( 1 - v^{\eta_{\phi^2}} \bigl )^{\heta - \eta_\phi} \, ,
\eqno (3.9) $$
where
$$ \heta = \half d -1 - \half \, {N+2\over N+8} \vep + \dots \, ,
\eqno (3.10) $$
is the scale dimension of the the boundary operator $\hphi(\bx)=\phi(0,\bx)$.
(3.9) is again in accord with the usual OPE for $\phi\phi$ and as $y'\to 0$
or $v\to 1$ it is easy to see, with (3.1),
$$
\l \phi^\alpha(x) \phi^\beta (x') \r_{\rm sp}  \sim \delta^{\alpha\beta}\,
{2C_\phi \over \hs^{2\eta_\phi}}\Bigl ( {4yy'\over \hs^2}
\Bigl)^{\heta - \eta_\phi}\Bigl ( 1 + \bigl ( 2\eta_\phi - \half \eta_{\phi^2}
- \heta \bigl ) {2yy'\over \hs^2} \Bigl ) \, ,
\eqno (3.11) $$
where $\hs^2$ is given in (2.12). It is crucial that, with the above results
for scale dimensions, the next to leading corrections vanish, corresponding
to the boundary operator $\hphi_n =0$.

As $v\to 0$ the particular boundary conditions are irrelevant so we must
require $ F_\phi(0)_{\rm sp} =  F_\phi(0)_{\rm ord}$, as exemplified in
(3.7.9).
By considering the terms $\propto v^{\eta_{\phi^2}}$ we easily see,
following (2.10), that
$$ {A_{\phi^2,{\rm sp}}\over A_{\phi^2,{\rm ord}}} = - 1 - {N+2\over N+8}\,
\vep + \rO(\vep^2) \, ,
\eqno (3.12) $$
which is independent of the normalisation of the operator $\phi^2$.

We have also calculated the leading $\vep$ corrections for the two point
function of the operator $\phi^2$, which represents the energy density in
a statistical physics context. There are two Feynman graphs shown in
figs. (1a,b), fig. (1a) corresponding to the one loop correction to
$G_\phi(x,x')$.

\epsfbox{1loop.eps}

Details of the calculation are given in appendix A.
After removing divergences the result is
$$
F_{\phi^2}(v)^{(1)} = 2A^2N {N+2\over N+8} \vep \Bigl ( 1 + v^4 -
(1-v^2)^2 \ln (1-v^2)
+ 2 v^4 {2- v^2 \over 1- v^2} \ln v^2 \pm v^2 \ln v^2 \Bigl ) \, .
\eqno (3.13) $$
In the Dirichlet case we therefore have to this order, combining with (3.2,5),
$$ F_{\phi^2}(v)_{\rm ord} = C_{\phi^2}\bigg (\big ( 1- v^{\eta_{\phi^2}}\big )
^{d-\eta_{\phi^2}} + {N+2\over N+8}\, \vep \Bigl ( 2 v^2 +
v^4 {3- v^2 \over 1- v^2} \ln v^2 \Bigl ) \bigg ) \, ,
\eqno (3.14) $$
for $C_{\phi^2} = 2A^2N(1+\rO (\vep))$. In the limits $v\to 0,1$ this
behaves as
$$ \eqalignno {
F_{\phi^2}(v)_{\rm ord} \sim {}& C_{\phi^2} - 2C_{\phi^2}
\bigg ( 1 - {3\over 2}\,{N+2\over N+8}\, \vep \bigg ) \,
v^{\eta_{\phi^2}} \, , & (3.15a)\cr
F_{\phi^2}(v)_{\rm ord} \sim {}&  C_{\phi^2}
\bigg ( 1 + {11\over 6}\,{N+2\over N+8}\, \vep - \vep \bigg )
\big (1-v^2\big )^{d-\eta_{\phi^2}} \, . & (3.15b)\cr}
$$
The first case corresponds to the leading contributions of the identity
and $\phi^2$ in the OPE as expected.  For $v\to 1$ the result corresponds to
the boundary operator $\hT(\bx)=T_{11}(0,\bx)$
whose scale dimension is $d$ at any conformal fixed point. In the Neumann
case the results to this order from (3.13) can be simply expressed by
$$   F_{\phi^2}(v)_{\rm sp} = 4C_{\phi^2}\bigg ( 1 - {N+2\over N+8}\, \vep
\bigg ) \, v^{\eta_{\phi^2}} + F_{\phi^2}(v)_{\rm ord} \, .
\eqno (3.16) $$
For $v\to 1$ the first term is a constant and so corresponds to the
contribution of a boundary operator $\hphi^2$ with dimension
$\heta_{\hphi^2} = \eta_{\phi^2} + \rO (\vep^2) = 2+\rO(\vep)$ which is
now present in addition to the operator $\hT$.
For $ v\to 0$ using $F_{\phi^2}(v) \sim C_{\phi^2} +
C_{\phi^2\phi^2}{}^{\phi^2}\! A_{\phi^2} v^{\eta_{\phi^2}}$ it is easy to
see that (3.15a,16) are in accord with (3.12).
\bigskip
\leftline{\bigbf 4 O(N) Model for Large N}
\medskip

Besides the $\vep$ expansion it is also possible to obtain non trivial
fixed points in quantum field theories by considering expansions in $1/N$
for theories with $N$-component fields [17]. Such large $N$ methods have the
virtue of allowing calculations of critical exponents at conformally
invariant fixed points for any dimension, at least in the range $2<d<4$.
The simplest example is the non linear $O(N)$ $\si$ model for  fields
$\phi^\alpha\in S^{N-1}$. In a lagrangian formalism it is classically
equivalent to remove the non linear constraint on $\phi$ but
introduce an auxiliary field  $\lambda$,
without any kinetic term and with an interaction $\L_I = \half \lambda \phi^2$,
whose field equation then
enforces the condition $\phi^2 = {\rm const}.$. The $1/N$ expansion
for the corresponding quantum field theory
may naturally be written in terms of propagators for $\phi,\lambda$
with a single vertex given by $\L_I$. At the scale
invariant fixed point Vasil'ev, Pis'mak and Khonkonen [4] formulated
the $\si$ model in terms of a skeleton graph expansion for the two point
functions for $\phi,\lambda$ which can be solved self consistently for the
scaling dimensions $\eta_\phi , \, \eta_\lambda$ by imposing conformal
invariance in a tractable $1/N$ expansion. Using these methods
$\eta_\phi , \, \eta_\lambda$, which determine all other bulk critical
indices, have been determined to $\rO ( N^{-3}),\rO ( N^{-2})$ respectively.
To leading order  $\eta_\phi= \half d -1,
\, \eta_\lambda = 2$ and results are consistent with those from the $\vep$
expansion for the renormalisable ${1\over 24}g (\phi^2)^2$, as considered
in the previous section, with the identification $\eta_\lambda =
\eta_{\phi^2}$.
In recent years the $O(N)$ $\si$ model has been extensively investigated
at its conformal fixed point so that the spectrum of operators present and
their scaling dimensions are well understood [17,14].

Writing the two point functions for the basic fields $\phi^\alpha,\lambda$ as
$$ \l \phi^\alpha(x) \phi^\beta (x') \r = \delta^{\alpha\beta}G_\phi(x,x')
\, , \quad \l \lambda(x) \lambda (x') \r =  \l \lambda(x) \r
\l \lambda (x') \r + G_\lambda(x,x') \, ,
\eqno (4.1) $$
since we assume manifest $O(N)$ invariance and $\l \phi^\alpha \r =0$.
To zeroth order in $1/N$ [4] the basic equations determining these are
equivalent to
$$ \eqalignno {
\bigl ( - \nabla^2 +  \l \lambda(x) \r \bigl ) G_\phi(x,x')& = \de^d (x-x')\, ,
& (4.2a) \cr
\int \! \d^d x'' \, G_\phi(x,x'')^2 \, G_\lambda(x'',x')& = - {2\over N}
\de^d (x-x')\, . & (4.2b) \cr}
$$
Alternatively in the renormalisable $(\phi^2)^2$ theory,
with the identification $\lambda = {1\over 6}g\phi^2$,
$G_\lambda$ represents the leading contribution at large $N$ due to summing
all bubble diagrams. With no boundary present (4.1a,b) are trivial to
solve at a fixed point.  If, as required by conformal invariance,
$$ G_\phi(x,x') = {A\over s^{2\eta_\phi}} \, , \quad
G_\lambda(x,x') = {B \over s^{2\eta_\lambda}} \, ,
\eqno (4.3) $$
then (4.2b) requires in general that
$$\eta_\lambda = d - 2 \eta_\phi  \, , \quad
A^2 B = - {2\over N}f(2\eta_\phi) \,, \quad f(\alpha) =
 {1\over \pi^d} \, {\Gamma(d-\alpha)
\Gamma(\alpha)\over \Gamma(\alpha - \half d)
\Gamma(\half d-\alpha)} \, ,
\eqno (4.4) $$
while (4.2a) gives $\eta_\phi = \half d - 1, \, \l \lambda \r =0$,
with $A$ as in (3.3), and hence $\eta_\lambda =2$ for $N\to \infty$.
The relation between $\eta_\lambda$ and $\eta_\phi$ in (4.4) is necessary
for the integral in (4.2b) to be conformally invariant, using the
transformation of the measure in (2.1). It is also useful to
note that for $\eta_\phi=\half d - 1$ $B>0$ for $2<d<4$ and as $d\to 4$
$B\sim 8\vep^2/N$.
In general the normalisations of the fields $\phi,\lambda$ are arbitrary but
with the above choices the scale of the vertex function is determined.
Without a boundary, when $\l\lambda\r=0$, conformal invariance dictates
the general form
$$ \l  \phi^\alpha(x_1) \phi^\beta (x_2) \lambda (x_3) \r = \de^{\alpha\beta}
\, {C_{\phi\phi\lambda}\over (x_{12}^{\, 2})^{\eta_\phi-\hh \eta_\lambda}
(x_{13}^{\, 2} x_{23}^{\, 2})^{\hh \eta_\lambda}} \, , \quad
x_{ij} = x_i - x_j \, .
\eqno (4.5) $$
To leading order in $1/N$,\footnote{*}{This may be obtained by
calculating the lowest order contribution, using the conformal star
triangle relation,
$\l  \phi^\alpha(x_1) \phi^\beta (x_2) \lambda (x_3) \r
= - \de^{\alpha\beta}\int \! \d^d x \, {A^2B\over (
(x_1-x)^2 (x_2-x)^2)^{{d/ 2} -1} (x_3 - x)^4}$
but the result is also in accord with a more systematic
analysis based on requiring cancellation of shadow fields in the OPE [14].}
$$ {C_{\phi\phi\lambda}\over A B} = {1\over 2\vep}\, .
\eqno (4.6) $$
The normalisation independent vertex coupling is then
$C_{\phi\phi\lambda}/(A^2B)^\hh = \rO
(N^{-\hh})$ which implies formally that the $1/N$ expansion is consistent.

The aim here is to extend these results to the situation for a plane
boundary. The essential results in the large $N$ limit have been
obtained by Bray and Moore [10] and also Ohno and Okabe [11] some time ago but
we differ in making essential use of conformal invariance which leads to a
a more direct and perhaps simpler derivation of the $\lambda$ two point
function by solving (4.2b) (in both treatments Dirichlet boundary conditions,
or the ordinary transition, is more straightforward than Neumann boundary
conditions, as appropriate for the special transition).
Following (2.4) if we write
$$ G_\phi (x,x') = {1\over (4yy')^{\hh d - 1}} f_\phi(\xi) =
{1\over (s^2)^{\hh d - 1}} F_\phi (v) \, , \quad F_\phi (v) = \xi^{\hh d -1}
f_\phi(\xi) \, ,
\eqno (4.7) $$
then imposing (4.2a) requires $F_\phi(0)=A$ and also, since $\eta_\lambda = 2$,
if we take
$$ \l \lambda (x) \r = {A_\lambda \over 4y^2} \, ,
\eqno (4.8) $$
$F_\phi$ satisfies the differential equation
$$ \xi (1+\xi) {\d^2\over \d \xi^2} F_\phi + (\half \vep + 2\xi )
{\d \over \d \xi} F_\phi - {\ts {1\over 4}}A_\lambda \, F_\phi = 0 \, .
\eqno (4.9) $$
This is easily seen to be soluble in general in terms of hypergeometric
functions, the appropriate solution with the relevant behaviour as $\xi \to
0$ being
$$A F(a,b,;\half\vep;-\xi) = A (1-v^2)^a F(a,\half \vep-b;\half\vep;v^2) \, ,
\quad a+b=1 \, , \quad ab = - {\ts{1\over 4}} A_\lambda \, .
\eqno (4.10) $$
For the particular applications of interest here we assume that
the leading term for $G_\phi$ in the large $N$ limit is given by the
$N\to \infty$ results of the previous section. It is important to note that
the first order in $\vep$ results for the scaling dimensions are exact in
this limit\footnote{${}^\dagger$}{The two loop results [12,19,20,21]
for the surface exponents $\heta,\heta_n$ vanish as $N\to \infty$.}.
Hence in the Dirichlet case we obtain
$$  f_\phi(\xi)_{\rm ord} = A\bigl ( \xi (1+\xi)\bigl )^{-\hh d + 1}\, , \qquad
F_\phi ( v)_{\rm ord} = A \bigl ( 1 - v^2 \bigl )^{\hh d - 1} \, .
\eqno (4.11) $$
This solution is in exact agreement with the general form (4.10)
for $a=\half d - 1 , \, b = \half \vep$. Hence in (4.8) in this case
$$ A_{\lambda,{\rm ord}} = - (4-d)(d-2) \, .
\eqno (4.12) $$
The consistency of this result may be verified by considering the OPE for
$\phi\phi$ where from (4.5) and (4.1,3) in the limit $s=x-x'\to 0$
$$ \phi^\alpha(x) \phi^\beta (x') \sim {A\over (s^2)^{\hh d-1}}\,
\de^{\alpha\beta}
+ {C_{\phi\phi}{}^\lambda\over (s^2)^{- \hh \vep}}\,
\de^{\alpha\beta}\lambda (x') \, , \quad  C_{\phi\phi}{}^\lambda =
{C_{\phi\phi\lambda}\over B} \, .
\eqno (4.13) $$
Applying this to the two point function (4.1) with the general form (4.7)
then we must require as $v\to 0$ $F_\phi(v) \sim A + C_{\phi\phi}{}^\lambda
A_\lambda v^2$. Using the explicit expression (4.11) in conjunction with (4.6)
is easily seen to recover (4.12).

Given an explicit formula for $G_\phi$ then $G_\lambda$ is determined by
solving (4.2b). Given that $\eta_\lambda = 2$ to leading order in $1/N$ then
conformal invariance dictates the general form
$$ G_\lambda (x,x') = {1\over (4yy')^2} f_\lambda(\xi) =
{1\over (s^2)^2} F_\lambda (v) \, , \quad
F_\lambda (v) = \xi^2f_\lambda(\xi)\,.
\eqno (4.14) $$

We first discuss the essential mathematical
problem of inverting an operator represented by a kernel $G(x,x')$ which
transforms as a scalar of dimension $\alpha$ at $x,x'$ on $\bR_+^d = \{
(y,\bx);y>0\}$. Representing the inverse by a kernel $H(x,x')$ we require
$$ \int_{\bR_+^d} \!\!\!\! \d^d x \, G(x_1,x) \, H(x,x_2) =
\de^d (x_1-x_2)\, ,\quad x_1,x_2 \in \bR_+^d \, .
\eqno (4.15) $$
Under a conformal transformation as in (2.1)
$\de^d(x_1-x_2)\to {\Omega^g(x_1)^d\de^d(x_1-x_2)}$. Hence
the integral (4.15) is compatible with restricted conformal transformations
preserving the boundary if we take
$$ G (x,x') = {1\over (4yy')^\alpha} g(\xi) \, , \qquad
H (x,x') = {1\over (4yy')^{d-\alpha}} h(\xi) \, .
\eqno (4.16) $$
Hence we may expect to reduce (4.15) to an equation for the single variable
functions $g,h$. To achieve this
it is convenient to define a transform, $g\to {\hat g}$,  by integrating $G$
over planes parallel to the boundary\footnote{*}{This is analogous to the
radon transform [22].}
$$ \int \! \d^{d-1} \bx \, G (x,x') = {1\over (4yy')^{\alpha-\lambda}}
{\hat g}(\rho) \, , \quad \rho = {(y-y')^2\over 4yy'} \, , \quad
\lambda = \half (d-1) \, ,
\eqno (4.17) $$
where
$$
{\hat g}(\rho) = {\pi^\lambda \over \Gamma (\lambda)} \int_0^\infty \!\!\!\!
\d u \, u^{\lambda -1}g(u+\rho) \, .
\eqno (4.18) $$
This transform may be inverted, ${\hat g} \to g$, by
$$ g(\xi) = {1\over \pi^\lambda \Gamma (-\lambda)} \int_0^\infty \!\!\!\!
\d \rho \, \rho^{-\lambda - 1}{\hat g}(\rho + \xi) \, ,
\eqno (4.19) $$
where the integral of $\rho^{-\lambda - 1}$ is singular for $d$ of interest
here but may be defined by analytic continuation in $\lambda$ from
${\sl Re}(\lambda)<0$.
The inversion formula may be verified by using
$$ \int \! \d u \, (\rho -u)_+^{\mu-1} u_+^{\lambda-1} = B(\mu,\lambda)\,
\rho_+^{\mu+\lambda-1} \sim \Gamma (-\lambda) \Gamma (\lambda) \de (\rho)
\ \ \hbox{as} \ \ \mu\to -\lambda \, .
\eqno (4.20) $$
For $d=3 \Rightarrow \lambda=1$ we use
$$ {\rho_+^{-\lambda -1}\over \Gamma(-\lambda)} \sim  \de'(\rho) \ \ \hbox{for}
\ \ \lambda \to 1 \, ,
\eqno (4.21) $$
to reduce (4.19) to the simple form
$$ g(\xi) = - {1\over \pi} \, {\hat g}'(\xi) \, ,
\eqno (4.22) $$
which is easy to check directly from (4.18).

With the definition (4.17) for $\hat g$, and correspondingly for $\hat h$,
it is easy to see, by integrating over $\bx_1$, that (4.15) reduces to
$$ \int_0^\infty \!\!\! {\d y \over y}\, {\hat g}(\rho_1) \, {\hat h}
(\rho_2) = 4y_1 \de(y_1-y_2) \, , \quad \rho_i = {(y-y_i)^2\over 4yy_i} \, .
\eqno (4.23) $$
If $y=e^{2\theta}, \, y_i = e^{2\theta_i}$ this may be simplified to
$$ \int_{-\infty}^\infty \!\!\!\! \d \theta \, {\hat g}
\bigl (\sinh^2(\theta-\theta_1)\bigl ) \, {\hat h}
\bigl (\sinh^2(\theta-\theta_2)\bigl ) = \de (\theta_1 - \theta_2) \, .
\eqno (4.24) $$
This is then straightforward to solve by Fourier transforms
$$ {\tilde {\hat g}}(k) = \int \! \d \theta \, e^{ik\theta} \,
{\hat g}(\sinh^2\theta) \, ,
\eqno (4.25) $$
which finally gives
$$ {\tilde {\hat g}}(k) \, {\tilde {\hat h}}(k) = 1 \, .
\eqno (4.26) $$
Hence ${\tilde {\hat h}}$  may be determined and by taking the inverse Fourier
transform and applying (4.19) it is possible to find $h$.

For the problem of interest here then, given the result (4.11) for $G_\phi$,
we should take $\alpha=d-2$ and, with $f_\phi(\xi)_{\rm ord}^{\, 2} = A^2
g(\xi)$,
$$ g(\xi) = \bigl [ \xi(\xi+1)\bigl ]^{-d+2} \, .
\eqno (4.27) $$
For simplicity we describe here the application of the above method for the
physically interesting case of $d=3$ when the integrals are also relatively
standard. General $d$ is discussed in appendix B. For $d=3$ we easily obtain
$$ {\hat g}(\rho) = \pi \ln {\rho + 1\over \rho}\, , \qquad
{\hat g}(\sinh^2 \theta) = - 2\pi \ln |\tanh \theta | \, ,
\eqno (4.28) $$
and therefore
$$ {\tilde {\hat g}}(k) = {2\pi^2 \over k} \tanh \quar \pi k \, .
\eqno (4.29) $$
Using (4.26) to give ${\tilde {\hat h}}(k)$ we then get
$$ {\hat h}(\rho) = - {1\over 2\pi^3} \, {1\over \rho(\rho+1)} \, , \quad
h(\xi) = - {1\over 2\pi^4} \biggl ( {1\over \xi^2} - {1\over (\xi+1)^2}
\biggl ) \, ,
\eqno (4.30) $$
with ${\hat h}\to h$ given by (4.22). Hence for $d=3$, when $A=1/4\pi$, the
solution of (4.2b) gives in (4.14)
$$ f_\lambda (\xi)_{\rm ord} = {16\over \pi^2 N}{1+2\xi\over \xi^2(1+\xi)^2}
\, , \qquad
F_\lambda (v)_{\rm ord} = {16\over \pi^2 N} (1-v^4) \, .
\eqno (4.31) $$

For general $d$ the results of appendix B lead to
$$ \eqalign {
f_\lambda (\xi)_{\rm ord} = {}& B \, {\Gamma(d) \Gamma(d-2)\over \Gamma(2d-4)}
\, \xi^{-d} F\bigl (d-2,d;2d-4;-{1\over \xi}\bigl ){}
= 2B \, {Q_{d-3}^2(1+2\xi)\over \xi(1+\xi)} \, , \cr
& B = {16\over N}\, {\vep\, \Gamma(d-2)\over \Gamma(2-\half d)
\Gamma(\half d - 1)^3} \, , \cr}
\eqno (4.32) $$
where $Q_{d-3}^2$ is an associated Legendre function and, given $A$ in (3.2),
$B$ is in accord with (4.4) for $\eta_\phi= \half d -1$.
For $d=3$, using $F(1,3;2;z)=\half ((1-z)^{-2} + (1-z)^{-1})$,
this is easily seen to be in agreement with (4.31).  It is also of interest
to consider the limit as $d\to 4$ when
$$\eqalign {
\xi^{-d+2} F\bigl (d-2 & ,d;2d-4;-{1\over \xi}\bigl ){}
= {\xi^\vep\over(1+\xi)^2} F\bigl (d-2,-\vep;2d-4;-{1\over \xi}\bigl ) \cr
&\sim {1\over(1+\xi)^2}\biggl ( 1 +\vep \ln \xi -6 \vep\sum_{n=1}^\infty
{(-1)^n\over n(n+2)(n+3)} \, {1\over \xi^n} \biggl ) \cr
& = (1-v^2)^2\bigl(1-\vep \ln (1-v^2)\bigl){}
+ \vep \biggl\{ 2v^2 + v^4 {3-v^2\over 1-v^2} \ln v^2 \biggl \} \, . \cr}
\eqno (4.33) $$
With $B\sim 8\vep^2/N$ this agrees exactly with the results in (3.4) and
(3.14) in the $N\to \infty$ limit for $\lambda \to {1\over 6}g_* \phi^2
\sim 8\pi^2\vep  \phi^2/N$.

Manifestly from (4.32) $f_\lambda (\xi)_{\rm ord}$ may be expanded in powers
$\xi^{-d-n}$ or $(2\xi+1)^{-d-2n}$ for $n=0,1,\dots$ which, from later results,
represents the contributions of boundary operators with dimensions $d+2n$ in
the BOE. For $\xi\to \infty$ $f_\lambda(\xi)_{\rm ord} \propto
\xi^{-d}$ as expected for the leading operator appearing in the BOE for
$\lambda(x)$ being $T_{11}(0,\bx)\equiv \hT(\bx)$.
An alternative expansion of $f_\lambda (\xi)_{\rm ord}$ valid for small
$\xi$ also verifies the consistency of the the leading $1/N$ expression for
the $\lambda$ two point function with the OPE for $\lambda(x)\lambda(x')$.
{}From (B.8) in the limit  $\xi\to 0$ we get
$$ f_\lambda (\xi)_{\rm ord} = B \Bigl ( {1\over \xi^2}\big (1 - (d-2)(d-3) \xi
\big ) +\half (d-1)(d-2)(d-3)(d-4) \ln{1\over \xi} + \dots \Bigl ) \,.
\eqno (4.34) $$
The leading operator, apart from the identity, appearing in the OPE for
$\lambda\lambda$ is $\lambda$ itself and the relevant coefficient is
determined from the three point function
$$ \l  \lambda(x_1) \lambda (x_2) \lambda (x_3) \r =
{C_{\lambda\lambda\lambda}\over (x_{12}^{\, 2}
x_{13}^{\, 2} x_{23}^{\, 2})^{\hh \eta_\lambda}} \, .
\eqno (4.35) $$
To leading order in $1/N$ [14], with our normalisations,
$$ C_{\lambda\lambda\lambda} = B^2\, {d-3\over 4-d} \,
\eqno (4.36) $$
and, with $\eta_\lambda = 2$, we can write
$$ \lambda(x)\lambda(x') \sim {B\over (x-x')^4} + {C_{\lambda\lambda}{}^\lambda
\over (x-x')^2}\lambda(x') \, , \quad C_{\lambda\lambda}{}^\lambda
= {C_{\lambda\lambda\lambda}\over B} \, ,
\eqno (4.37) $$
which implies $F_\lambda (v) \sim B + C_{\lambda\lambda}{}^\lambda A_\lambda
v^2$, using (4.8) for $\l \lambda \r$.  It is easy to see from this that (4.34)
is compatible with the result (4.12) in this case.
Note that $C_{\lambda\lambda\lambda}=0$
for $d=3$ which is reflected by the absence of a $v^2$ term in (4.31).

In the Neumann case if we take the $N\to \infty$ limit in (3.9) we get
$$  f_\phi(\xi)_{\rm sp} = A\, {1+2\xi \over
\bigl(\xi (1+\xi)\bigl )^{\hh d-1}} \, , \qquad
F_\phi ( v)_{\rm sp} = A (1+v^2) ( 1 - v^2  )^{-\hh \vep} \, ,
\eqno (4.38)  $$
where the leading surface operator has dimension $d-3$ which agrees with
the $N\to \infty$ limit of (3.10). The expression given in (4.38) corresponds
to the general solution (4.10) if $a=-\half\vep,\, b= 1+\half\vep$ so
that in this case
$$ A_{\lambda,{\rm sp}} = (4-d)(6-d)=\vep(2+\vep) \, .
\eqno (4.39) $$
{}From (4.12,39) $A_{\lambda,{\rm sp}}/A_{\lambda,{\rm ord}} = - (2+\vep)/
(2-\vep)$ which is in accord with the large $N$ limit of (3.12).
The solution of (4.2b) to find $G_\lambda$ is more involved and is undertaken
in appendix B. The conformal invariant function $f_\lambda$ in (4.14) is given
by
$$ \eqalign{ \!\!\!\!\!\!
f_\lambda (\xi)_{\rm sp} = B \bigg \{ &
{1\over 3} \, {6-d\over d-2}\, {\Ga(d)\Ga(d -2)\over \Ga (2d-5)}\cr
& \times {\xi+\half\over\big [\xi(1+\xi)\big ]^{\hh (d+1)}}\,
{}_3F_2\big ( \half d+ \half,\half d-{\textstyle{3\over2}}
,{\textstyle{3\over2}};
d-{\textstyle{5\over2}},{\textstyle{5\over2}}; -{1 \over 4\xi(1+\xi)}\big ) \cr
{}+{} &{\pi\,\Ga(\half d-1)^2\over \Ga (d-3)\Ga(\half d -
{\textstyle{3\over2}})
\Ga ({\textstyle{7\over2}}-\half d)}\,
{8\over
(1+2\xi)^2} F\big ( {\textstyle{3\over 2}},1;{\textstyle{7\over2}}-\half d;
{1\over (1+2\xi)^2}\big )
\bigg \} \, . \cr}
\eqno(4.40) $$
The two terms correspond in the BOE to two classes of boundary operators with
dimensions $d+2n$ and $2+2n$, $n=0,1,\dots$ respectively.
Corresponding to (4.34) we also find for the singular behaviour as $\xi\to 0$
$$ f_\lambda (\xi)_{\rm sp} = B \bigg ( {1\over \xi^2}\big (1 +(d-3)(6-d) \xi
\big ) + {(d-1)(d-3)(d-4)(d-6)^2\over 2(d-2)} \ln{1\over \xi} + \dots \bigg )
\, ,
\eqno (4.41) $$
in which the first two terms may also be easily seen to be compatible with
the OPE (4.37) together with (4.39).

In both (4.34) and (4.41) there appear $\ln \xi$ terms which are naively
unexpected on the basis of the OPE. However these may be understood by
considering the role of an operator, denoted as $\lambda^2$, which has
dimension $\eta_{\lambda^2} = 4 + \rO(N^{-1})$.
This may be defined in terms of the OPE for $\lambda\lambda$ where we take
$C_{\lambda\lambda}{}^{\lambda^2} = 1$ and we assume that, with this
normalisation, $A_{\lambda^2} = A_\lambda{}^{\! 2}( 1 + \rO(N^{-1}))$.
According to (2.10) we then expect from the OPE  a contribution to the
$\lambda$ two point function given by
$$ f_{\lambda\lambda}(\xi)_{\lambda^2} \sim A_\lambda{}^{\! 2} \Big (
1 + \half \big ( 2\eta_{\lambda,1} - \eta_{\lambda^2,1}\big ) {1\over N}
\ln {1\over \xi} + \dots \Big ) \, ,
\eqno (4.42) $$
where $\eta_{\lambda,1}, \, \eta_{\lambda^2,1}$ are the coefficients of
the $1/N$ terms in the large $N$ expansion of the scale dimensions of
$\lambda, \, \lambda^2$. The
first term on the r.h.s. of (4.42) clearly represents the disconnected pieces
$\l\lambda \r \l\lambda \r$ in (4.1), which are $\rO(1)$ as $ N\to \infty$,
while the second $\rO(N^{-1})$ term corresponds exactly to the $\ln \xi$
terms in (4.34,41), with (4.12,39), if
$$  \big ( 2\eta_{\lambda,1} - \eta_{\lambda^2,1} \big ){1\over N} =
{(d-1)(d-3)\over (d-2)(d-4)} B \, ,
\eqno (4.43) $$
which agrees with direct calculations.\footnote{*}{$\eta_{\lambda,1}$ and
$\eta_{\lambda^2,1}$  are given by (5.29) and (5.30) in the fifth paper
listed in ref. [18] where $\eta_1(S)\equiv\eta_{\phi,1}$ with
$B=4d(d-2)\eta_{\phi,1}/N$.}
\bigskip
\vfill\eject
\noindent{\bigbf 5 Two Point Functions for the Conserved Current and Energy
Momentum Tensor in the Large N Limit}
\medskip

In the $O(N)$ model described in the previous section other operators may
be defined in terms of the basic fields $\phi^\alpha$ through the OPE.
Here we focus on the conserved current $J_\mu^{\alpha\beta} = -
J_\mu^{\beta\alpha}$, whose charges generate the $O(N)$ symmetry, and the
energy momentum tensor $T_{\mu\nu}$ in this model. These have scale
dimension $d-1$ and $d$ respectively. The former may be
defined through the OPE
$$ \phi^{[\alpha}(x) \phi^{\beta]}(x') \sim {A\over S_d C_J}\,
{1\over (s^2)^{\eta_\phi - \hh d + 1}}\, s_\mu J_\mu^{\alpha\beta}(x') \, ,
\eqno (5.1) $$
where the coefficient is determined by $O(N)$ Ward identities [14] with $C_J$
setting the scale of the two point function for $J_\mu^{\alpha\beta}$. This
may be written as
$$ \l J_\mu^{\alpha\beta}(x) J_\nu^{\gamma\delta}(x') \r =
( \de^{\alpha\gamma}\de^{\beta\delta} -  \de^{\alpha\delta}\de^{\beta\gamma})
G_{J\, \mu\nu}(x,x') \, ,
\eqno (5.2) $$
where,  as in (2.16), conformal invariance dictates
$$ G_{J\, \mu\nu}(x,x') = {1\over (s^2)^{d-1}} \bigl ( I_{\mu\nu}(s)
C_J(v) + X_\mu  X'{}_{\! \nu} D_J(v) \bigl ) \, , \quad C_J = C_J(0) \, .
\eqno (5.3) $$
In the $N\to \infty$ limit $\eta_\phi = \half d -1$ and $A$ is given by
(3.3). The leading contributions to the two point function are then simply
$$ G_{J\, \mu\nu}(x,x') = \Big ( {S_d C_J\over A} \Big )^{\! 2} \half \big (
\pr_\mu G_\phi(x,x') {\overleftarrow \pr}{}'{}_{\!\! \nu} G_\phi(x,x')
- \pr_\mu G_\phi(x,x')
\, G_\phi(x,x') {\overleftarrow \pr}{}'{}_{\!\! \nu} \big ) \, .
\eqno (5.4) $$
Substituting the conformally invariant form for $G_\phi$ in (3.1) then gives
$$ \eqalign {
C_J(v) = {}& \Big ( {S_d C_J\over A} \Big )^{\! 2} \half \big ( (d-2)
F_\phi(v)^2 - (1-v^2) v F_\phi(v) F_\phi {}^{\! \prime} (v) \big ) \, , \cr
D_J(v) = {}& \Big ( {S_d C_J\over A} \Big )^{\! 2} \half \Big (
F_\phi(v) v {\d\over \d v} \big ( (1-v^2) v F_\phi {}^{\! \prime} (v) \big )
- v^2 (1-v^2 ) F_\phi {}^{\! \prime} (v)^2 \Big ) \, , \cr}
\eqno (5.5) $$
which leads to $C_J = 2/(d-2)S_d^2$. Using (4.9) we may verify that the
general results (5.5) satisfy the conservation condition (2.18).
For the Dirichlet case\footnote{*}{For a free complex
scalar field which has a conserved current $V_\mu = i \phi^* \olr \pr_\mu \phi$
then in (2.16) using (3.2) gives the corresponding results, if
${C_V = \Gamma(\half d)^2 / (2\pi^d (d-2))}$,
${C(v) = C_V \bigl ( 1 \pm v^{d-2} (1+v^2) + v^{2d-2} \bigl )}$ and
$ C(v)+D(v) = {C_V \bigl ( 1 \pm (d-1) v^{d-2} (1-v^2) - v^{2d-2} \bigl )}$.}
using (4.11) we have
$$
C_J(v)_{\rm ord} = C_J(1+v^2)(1-v^2)^{d-2} \, , \quad
C_J(v)_{\rm ord} + D_J(v)_{\rm ord} =  C_J (1-v^2)^{d-1} \, .
\eqno (5.6) $$

The calculation of two point functions involving the energy momentum
tensor $T_{\mu\nu}$ is more involved. In a non trivial conformal field
theory the natural definition of $T_{\mu\nu}$ is through its appearance in
OPEs. In the simplest case we may extend (4.13) to give generally
$$ \eqalign {
\phi^\alpha (x) \phi^\alpha (x') \sim {} & {NC_\phi\over s^{2\eta_\phi}}
+ {N C_{\phi\phi}{}^\lambda\over (s^2)^{\eta_\phi - \hh \eta_\lambda}}
C^{\eta_\lambda,0}(s,\pr_{x'}) \lambda (x') \cr
& {} -  {d\eta_\phi\over d-1}\, {N C_\phi\over C_T S_d} \,
{1\over (s^2)^{\eta_\phi - \hh d +1}} s_\mu s_\nu T_{\mu\nu}(x') +\dots \,
,\cr}
\eqno (5.7) $$
where $C_\phi$ is the coefficient of the bulk two point function for $\phi$,
in (3.1) $C_\phi = F_\phi(0)$ while $C_T$ is given by the scale of the two
point function for the energy momentum tensor, as in (2.32). The
coefficient of the energy momentum tensor term in OPEs is determined through
Ward identities [3,14]. In the term containing $\lambda$ in (5.7)
$C^{\eta_\lambda,0}(s,\pr)= 1 + \rO(s)$ is introduced to include all
derivatives of $\lambda$ in the above OPE. It is determined by the
requirement of reproducing the three point function (4.5) where in
general $C_{\phi\phi\lambda}=C_{\phi\phi}{}^\lambda C_\lambda$. It is
necessary to consider such derivative operators, unlike in (5.1), since
$T_{\mu\nu}$, which has dimension $d$, is not the lowest dimension operator
appearing in this OPE. Explicitly
$$ \eqalign {
C^{a,b}(s,\pr) = {1\over B(a_+,a_-)}& \int_0^1\!\! \d \alpha \, \alpha^{a_+-1}
(1-\alpha)^{a_--1} \cr
&\times \sum_{m=0} {1\over m!}\, {1\over(a+1-\half d)_m}
\, \bigl [ -\quar s^2 \alpha (1-\alpha) \pr^2 \bigl ]^m e^{\alpha s{\cdot \pr}}
\, , \cr}
\eqno (5.8) $$
for $a_{\pm} = \half (a\pm b)$ with the Pochammer symbol
$(a)_m = \Gamma(a+m)/\Gamma(a)$ and we assume $[s,\pr]=0$ to move all
derivatives to the right. For our purposes we need to expand this to $\rO(s^2)$
$$ C^{a,0}(s,\pr) \sim 1 + \half s {\cdot \pr} + {1\over 8(a+1)} \Big (
(a+2) s_\mu s_\nu \pr_\mu \pr_\nu - {\half a \over a+1-\half d} \, s^2 \pr^2
\Big ) \, .
\eqno (5.9) $$

Initially we focus on determining the two point function $\l T_{\mu\nu} (x)
\lambda(x') \r$ which as a consequence of (2.19,22) has a unique functional
form in the conformally invariant limit. In order to use the OPE (5.7) to
find this we need to determine an expression for the $\l \phi \phi \lambda \r$
three point function. To leading order in $1/N$ the connected three point
function in the $O(N)$ $\sigma$-model with a plane boundary discussed
previously is simply given by
$$\l  \phi^\alpha(x_1) \phi^\beta (x_2) \lambda (x_3) \r_{\rm conn}
= - \de^{\alpha\beta}\int_{\bR_+^d} \!\!\!\!  \d^d r \,
G_\phi (x_1,r) G_\phi (x_2,r) G_\lambda (r,x_3) \, ,
\eqno (5.10) $$
where $G_\phi, G_\lambda$ are given by the results of the section 4
for the ordinary and special critical points. The application of the OPE to
this is based on the equation (verified in appendix C)
$$ \eqalign {
{1\over x_{1}^{\, 2\eta_1} x_{2}^{\, 2\eta_2}} =  {}&
C^{\eta,\eta_1-\eta_2}( x_{12}, \pr_{x_2}) {1\over x_{2}^{\, 2\eta}}
+ \rho \,(x_{12}^{\, 2})^{\hh d -\eta}
C^{d-\eta,\eta_2-\eta_1} (x_{12}, \pr_{x_2}) \de^d (x_{2})\, , \cr
\eta =  \eta_1{}&  +\eta_2 \, , \qquad
\rho = \pi^{\hh d} { \Ga (\hh d - \eta_1) \Ga (\hh d - \eta_2)
\Ga (\eta - \hh d) \over \Ga (\eta_1) \Ga (\eta_2) \Ga( d-\eta )} \, . \cr}
\eqno (5.11) $$
Neglecting the $\de$ function term, this result is the essential condition
which determines the derivative operator $C^{a,b}(s,\pr)$ in the OPE for
scalar fields in conformal theories. For our purpose we take
$\eta_1=\eta_2 = \half d -1$ and generalise this to write
$$ \eqalignno {
G_\phi (x_1,r) G_\phi (x_2,r) \sim {}& C^{d-2,0}( x_{12}, \pr_{x_2})
G_\phi (x_2,r)^2 + \rho A^2 (x_{12}^{\, 2})^{\hh \vep}
C^{2,0}( x_{12}, \pr_{x_2}) \de^d (x_2 - r) \cr
&  - \half (x_{12})_\mu (x_{12})_\nu t_{\mu\nu}(x_2, r) +  \dots \, , & (5.12)
\cr}
$$
where other terms are less singular and irrelevant here (for free fields
without boundary (5.11) of course shows that $t_{\mu\nu}$
and other higher order contributions in the expansion (5.12) vanish). Using
(5.12) in (5.10) shows that it is compatible with the OPE (5.7), taking
$\eta_\phi=\half d -1, \eta_\lambda =2$ and $C_\phi = A$, since the
$G_\phi (x_2,r)^2$ term is absent due to (4.2b). In this case it is necessary
that $\rho A^2 = - A/(2\vep) = - C_{\phi\phi}{}^\lambda$ to leading order in
$1/N$, which is in accord with (4.6).
If we use the result for $C_T$ when $N\to \infty$, which is the same as
for free scalar fields [3,14],
$$ C_T = {Nd \over (d-1) S_d^{\, 2}} \, ,
\eqno (5.13) $$
we may also obtain
$$ \l T_{\mu\nu} (x) \lambda(x') \r = - N \int_{\bR_+^d} \!\!\!\!  \d^d r \,
t_{\mu\nu}(x,r) \, G_\lambda (r,x') \, .
\eqno (5.14) $$
Using (5.9,11) and the basic Green function equation (4.2a), with (4.8),
gives
$$ \eqalign {
t_{\mu\nu}(x,x') = {}& {\hat t}_{\mu\nu}(x,x')
- {1\over d}\de_{\mu\nu} \, {A_\lambda \over (2y)^2} G_\phi(x,x')^2 \, , \cr
{\hat t}_{\mu\nu}= {}& - G_\phi  \D_{\mu\nu} G_\phi
+{ d\over 4(d-1)} \D_{\mu\nu} \big ( G_\phi{}^{\! 2} \big ) \, ,
\quad \D_{\mu\nu} = \pr_\mu \pr_\nu - {1\over d} \de_{\mu\nu} \pr^2 \, , \cr}
\eqno (5.15) $$
Using the form (4.7) for $G_\phi$ ${\hat t}_{\mu\nu}$ becomes
$$ \eqalign {
{\hat t}_{\mu\nu}(x,x') = {}& {(2y')^2 \over s^{2d}} \Bigl (
X_\mu X_\nu - {1\over d} \delta_{\mu \nu} \Bigl) f(v)  \, , \cr
f(v) = {}& - {2\over d-1} \xi(\xi+1) \bigg ( (d-2) F_\phi {\d \over \d \xi}
\Big ( \xi^2 {\d \over \d \xi}F_\phi \Big ) - d \, \xi^2 \Big (
{\d \over \d \xi}F_\phi \Big )^2 \bigg ) \, , \cr}
\eqno (5.16) $$
where $X_\mu$ is defined as in (2.14). Under conformal transformations
$t_{\mu\nu}(x,x')$ transforms as a scalar with scale
dimension $d-2$ at $x'$, and also as a symmetric tensor of dimension $d$ at
$x$,
so that the integration in (5.14) preserves conformal invariance.
{}From (5.15), or from (5.16) and using (4.9), we may also find
$$
\Big ( (d-2) n_\nu + y \pr_\nu \Big ) G_\phi(x,x'){}^{2} \, ,
\eqno (5.17) $$
which in turn gives
$$ \pr_\mu t_{\mu\nu}(x,x') = n_\nu \, {2A_\lambda\over (2y)^3}
G_\phi (x,x')^2 \, ,
\eqno (5.18) $$
where it is important to note that since $f(v)\propto v^4$ as $v\to 0$
and hence $t_{\mu\nu}(x,x') = \rO(s^{4-2d})$ there are no $\de$ function
contributions. Since $t_{\mu\mu}$ is easily found from (5.15) or (5.16)
and using (5.18), and applying the basic equation (4.2b),
we therefore can derive
$$  \l T_{\mu\mu} (x) \lambda(x') \r = - {2A_\lambda \over (2y)^2}
\, \de^d(s) \,, \quad \pr_\mu \l T_{\mu\nu} (x) \lambda(x') \r =  n_\nu\,
{4A_\lambda \over (2y)^3} \, \de^d(s) \, .
\eqno (5.19) $$
This is in exact agreement with (2.23) for $\O \to \lambda, \, \eta\to 2$
if we take $C= -2/d$.
As a consequence of the general conformal invariance results given in
(2.19,22,26) the integral (5.14) is completely determined therefore by
$$ \l T_{\mu\nu} (x) \lambda(x') \r = -{2dA_\lambda\over (d-1)S_d}\,
{(2y')^{d-2} \over s^{2d}}
\Bigl ( X_\mu X_\nu - {1\over d} \delta_{\mu \nu} \Bigl) v^d \, .
\eqno (5.20) $$

The same procedures may also be applied to find the energy momentum tensor
two point function if we start from $\l \phi^\alpha(x_1) \phi^\alpha( x_2)
\phi^\beta(x_3) \phi^\beta (x_4) \r$ and consider the OPEs for
$x_{12}, x_{34} \sim 0$. The relevant contributions for large $N$ are
$$ \eqalign {
& N \big ( G_\phi (x_1,x_3) G_\phi (x_2,x_4) + G_\phi (x_1,x_4)
G_\phi (x_2, x_3) \big ) \cr
& + N^2 \int_{\bR_+^d} \!\!\!\!  \d^d r \int_{\bR_+^d} \!\!\!\!  \d^d r' \,
G_\phi (x_1,r) G_\phi (x_2,r) G_\phi (x_3,r) G_\phi (x_4,r) \,G_\lambda (r,r')
\, . \cr}
\eqno (5.21) $$
For $x_{12}=x_{34}=0$ this vanishes due to (4.2b). This is crucial in
cancelling the $\rO(1)$ contribution as $x_{12}\sim x_{34} \to 0$, which would
correspond to  a `shadow operator' of dimension $d-2$ in the OPE
(for free fields this is represented by the operator $\phi^2$)
so that the leading behaviour is $\propto (x_{12}^2)^{\hh \vep},
(x_{34}^2)^{\hh \vep}$ which is appropriate for $\lambda$, with dimension $2$,
being the lowest dimension operator apart from the identity.
Following a similar discussion to the above we may now obtain
the large $N$ expression for $\l T_{\mu\nu}(x) T_{\si\rho}(x')\r $ as
$$ \eqalign{
& G^f{}_{\! \mu\nu\si\rho}(x,x')
- N {2\over d} \Big ( \de_{\mu\nu}\,{A_\lambda \over (2y')^2} t_{\si\rho}(x',x)
+ \de_{\si\rho}  \, {A_\lambda \over (2y)^2} t_{\mu\nu}(x,x') \Big ) \cr
& - N {2\over d^2} \de_{\mu\nu}\de_{\si\rho} {A_\lambda {}^{\! 2} \over
(4yy')^2} G_\phi(x,x')^2
+ N^2 \! \int_{\bR_+^d} \!\!\!\!  \d^d r \! \int_{\bR_+^d} \!\!\!\!  \d^d r' \,
t_{\mu\nu} (x,r) t_{\si\rho}(x',r') \, G_\lambda (r,r') \, , \cr}
\eqno (5.22) $$
where, with the operator $\D_{\mu\nu}$ defined in (5.15), the `free field'
part $G^f$ is given by
$$ \eqalign{
G^f{}_{\! \mu\nu\si\rho} = N \Big \{ {}& G_\phi \, \D_{\mu\nu}
\D'{}_{\! \si\rho}
G_\phi +  \D_{\mu\nu}G_\phi \, \D'{}_{\! \si\rho} G_\phi \cr
& - {d\over 2(d-1)} \Big ( \D_{\mu\nu} \big ( G_\phi \D'{}_{\! \si\rho}
G_\phi \big ) {} + \D'{}_{\! \si\rho} \big ( G_\phi D_{\mu\nu} G_\phi \big )
\Big ) \cr
& + {d^2\over 8(d-1)^2} \, \D_{\mu\nu} \D'{}_{\! \si\rho} \big (
G_\phi {}^{\! 2} \big ) \Big \} \, . \cr}
\eqno (5.23) $$
It is easy to verify that the trace on $\mu\nu$ and $\si\rho$ in (5.22)
vanishes and hence the terms involving $A_\lambda$ explicitly in (5.22)
and also contained in $t_{\mu\nu}, \, t_{\si\rho}$ can be dropped. Hence
we can write
$$ \l T_{\mu\nu}(x) T_{\si\rho}(x')\r =
G^f{}_{\! \mu\nu\si\rho}(x,x') + G^\lambda{}_{\! \mu\nu\si\rho} (x,x') \, ,
\eqno (5.24) $$
where
$$
G^\lambda{}_{\! \mu\nu\si\rho} (x,x') =
N^2 \! \int_{\bR_+^d} \!\!\!\!  \d^d r \! \int_{\bR_+^d} \!\!\!\!  \d^d r' \,
{\hat t}_{\mu\nu} (x,r) {\hat t}_{\si\rho}(x',r') \, G_\lambda (r,r') \, .
\eqno (5.25) $$
Writing $r_\mu = (z,\br)$ and using the result (5.14,20),
$$ G^\lambda{}_{\! \mu\nu\si\rho}(x,x')
= N  {2d A_\lambda \over (d-1)S_d} \int_0^\infty \!\!\!\!\! \d z \! \int
\! \d^{d-1} \br \Big ( {2z \tv \over \st^2 \, \st^{\prime 2}} \Big )^{\! d}
f(\tv') \Big ( \tX_\mu \tX_\nu - {1\over d} \de_{\mu\nu} \Big )
\Big ( \tX'{}_{\!\si} \tX'{}_{\! \rho} - {1\over d} \de_{\si\rho} \Big )   ,
\eqno (5.26) $$
for $\st^2 = (x-r)^2, \, \tv^2 = \st^2/(\st^2 + 4yz)$, $\st^{\prime 2},
\, \tv^{\prime 2}$ similarly defined with $x\to x'$, and $\tX_\mu(x,r), \,
\tX'{}_{\!\si}(x',r)$ vectors formed as in (2.14).

To verify the conservation equations we may use the definition (5.23)
along with (4.2a,8)
to find
$$
\pr_\mu G^f{}_{\! \mu\nu\si\rho}(x,x') = {N\over d}\, {4A_\lambda\over (2y)^3}
\Big ( (d-2) n_\nu + y \pr_\nu \Big ) {\hat t}_{\si\rho}(x',x) \, ,
\eqno (5.27) $$
and using (5.17) in (5.25) it is easy to verify that this is cancelled by
$\pr_\mu G^\lambda{}_{\! \mu\nu\si\rho}$. By virtue of conformal invariance
$ G^f{}_{\!\mu\nu\si\rho}(x,x')$ may be expressed in the form (2.27) but
(5.27) now gives
$$ \eqalign{
\Big ( v{\d\over \d v} - d \Big ) \gamma^f(v) = {}& {d\over (d-1)^2}
\alpha^f(v)
+ {(d+1)(d-2)\over d-1} \ep^f(v)  + 2NA_\lambda {1\over d} f(v) \, , \cr
\Big ( v{\d\over \d v} - d \Big )\alpha^f(v) =  {}& 2(d-1) \gamma^f(v)
+ 2NA_\lambda {d-1\over d^2} \Big ( vf'(v) - {2d\over 1-v^2} f(v) \Big ) \, .
\cr}
\eqno (5.28) $$

By explicit calculation we find using the results in (4.11) or (4.38)
$$ \eqalignno {
f(v)_{\rm ord} = {}& 2A^2 {(d-2)^2\over d-1}\, v^4 \big (1-v^2\big )^{d-4} \, ,
&(5.29a) \cr
f(v)_{\rm sp} = {}& {2A^2 \over d-1} \, v^4 \big (1-v^2\big )^{d-6} \Big \{
(d-3)(d-4) \big (1+v^2\big )^2 - d \big (1-v^2\big )^2 \Big \} \, . & (5.29b)
\cr}
$$
For the ordinary transition then (5.23), in conjunction with (2.27,38), gives
$$ \eqalign {
\alpha^f(v)_{\rm ord} = {}& {N\over S_d^{\,2}} \bigg \{ \big (1 - v^2\big )^d
+ {2\over d-1} \big (1 - v^2\big )^{d-2} v^2 +
8 {(d-2)^2 \over d^2} \big (1 - v^2\big )^{d-4} v^4 \bigg \} \, , \cr
\gamma^f(v)_{\rm ord} = {}& - {N\over S_d^{\,2}} \big (1+v^2\big )
\bigg \{ {d\over 2(d-1)} \big (1 - v^2\big )^{d-1}
+ {d-2 \over (d-1)^2}\big (1 - v^2\big )^{d-3} v^2 \bigg \} \, , \cr
\ep^f(v)_{\rm ord} = {}& {N\over S_d^{\,2}} \big (1- v^2\big )^{d-2}  \bigg \{
{d\over 2(d-1)}\big (1+v^2\big )^2 - {d-2 \over (d-1)^2} v^ 2 \bigg \} \, .
\cr}
\eqno (5.30) $$
It is easy to check that (5.29a) and (5.30) satisfy (5.28).
The evaluation of the integral in (5.26) is discussed in appendix D. The
results can be similarly expressed in the basis given by (2.27).

{}From (D.33.34) we obtain
$$ \eqalign {
\alpha^\lambda(v)_{\rm ord} = {}& -{N\over S_d^{\,2}}{8\over d^2}(d-2)^2
v^4 \big ( 1-v^2\big )^{d-4} \cr
{}& -{N\over S_d^{\,2}} {2 \over d-1} v^2\big ( 1-v^2\big )^{d-2}
{}_3F_2\big (1,d-1,\half d-2;d-{\textstyle {3\over 2}},\half d;
-{(1-v^2)^2\over 4v^2} \big ) \cr
{}& + \alpha(1)_{\rm ord}\, v^d \Big ( 1 + {d\over d-1} {(1-v^2)^2\over 4v^2}
\Big ) \, , \cr
\gamma^\lambda(v)_{\rm ord} = {}& {N\over S_d^{\,2}} {d-2\over (d-1)^2}
v^2(1+v^2)\big ( 1-v^2\big )^{d-3} \cr
&\qquad \qquad \quad \times
{}_3F_2\big (1,d-1,\half d-2;d-{\textstyle {3\over 2}},\half d-1;
-{(1-v^2)^2\over 4v^2} \big ) \cr
{}& - {d\over 4(d-1)^2}\alpha(1)_{\rm ord}\, v^{d-2}(1-v^4) \, , \cr
\epsilon^\lambda(v)_{\rm ord} = {}& -{N\over S_d^{\,2}} {d-2\over
(d-1)^2(2d-3)}
v^2\big ( 1-v^2\big )^{d-2} \cr
&\qquad \qquad \quad \times
{}_3F_2\big (1,d-1,\half d-2;d-\half ,\half d-1;
-{(1-v^2)^2\over 4v^2} \big ) \cr
{}& + {d\over 4(d-1)^2(d+1)}\alpha(1)_{\rm ord}\, v^{d-2}(1-v^2)^2 \, , \cr}
\eqno (5.31) $$
where
$$
\alpha(1)_{\rm ord} = {N\over S_d^{\,2}} 2^{d-2}
{\Ga(\half d)^2\Ga(3-\half d)\Ga(d-{\textstyle {3\over 2}})
\over \Ga(d-1) \Ga(\half d + \half )} \, .
\eqno (5.32) $$
These contributions vanish if $d=4$ when $\alpha(1)_{\rm ord}=2N/S_4^{\, 2}$
as expected since the integral (5.26) contains a factor $4-d$ in $A_\lambda$
and in agreement with the conformal invariant theory becoming then
of course just a free scalar field theory. For the non trivial case when
$2<d<4$ it is crucial that
adding (5.31) to (5.30) cancels the $(1-v^2)^{d-4},(1-v^2)^{d-2}$
terms in $\alpha^f(v)$ and also the  $(1-v^2)^{d-3}$ term in $\gamma^f(v)$ so
that it vanishes as $v\to 1$. The final result for $\alpha$ can be expressed
more compactly as
$$ \eqalignno { \!\!
\alpha(v)_{\rm ord} = {}&{N\over S_d^{\,2}}\big ( 1-v^2\big )^d
\bigg \{ 1 + {d-4\over d(2d-3)}{}_3F_2\big (1,d, \half d-1;d-\half,\half d
+ 1;-{(1-v^2)^2\over 4v^2} \big ) \bigg \} \cr
{}& + \alpha(1)_{\rm ord}\, v^d \Big ( 1 + {d\over d-1} {(1-v^2)^2\over 4v^2}
\Big ) \, . & (5.33) \cr }
$$
We have verified that expanding this to $\rO(\vep)$ gives the same
expressions as the $\vep$ expansion results in I in the large $N$ limit.
When $d=2$ the first term in (5.33) vanishes leaving only the contribution
proportional to $\alpha(1)_{\rm ord}= 2N/S_2^{\, 2}$.

In order to consider how this result for $\alpha(v)_{\rm ord}$
behaves in the limit $v\to 0$, which is
appropriate for the OPE, we may use the inversion formula for generalised
hypergeometric functions [28] to write this alternately as
$$ \eqalignno { \!\!
\alpha(v)_{\rm ord} = {}&{N\over S_d^{\,2}}\big ( 1-v^2\big )^d \cr
&\times \bigg \{ 1 + {1\over 2(d-1)}{4v^2\over(1-v^2)^2} \,
{}_3F_2\big (1,{\ts {5\over 2}}-d,1-\half d;2-d,3-\half d;-{4v^2\over(1-v^2)^2}
\big )\bigg \} \cr
{}& - {N\over S_d^{\,2}}{\Ga(2d-3)\Ga(3-d)\over \Ga(d+1)}{d-4\over d+2}
\, {4v^d\over (1-v^2)^d} F\big ({\ts{3\over 2}},\half d;2+\half d;
-{4v^2\over(1-v^2)^2} \big ) \, . & (5.34) \cr}
$$
\bigskip
\leftline{\bigbf 6 Operator Product Expansions}
\medskip

As described in section 2 there are constraints on the two point functions
in the presence of a boundary, which in the conformal limit depend on a
single variable function of $\xi$ or $v$, arising from the OPE. The
coefficients appearing in this are a property of the conformal theory on
$\bR^d$ without any boundary and dependence on boundary conditions arises
solely through the one point functions of those operators appearing in the
OPE. Since, as remarked earlier, only scalar
operators have non zero one point functions, which have the simple form (2.7),
we need only consider for our purposes the contributions of such operators
to the OPE. In this section we determine the functional forms for the two
point function in the neighbourhood of a boundary arising from all
derivative operators formed from a given quasi-primary operator.

We initially consider the simplest case of just scalar operators when the
general conformally
invariant three point function without any boundary can be written as
$$ \l \O_1 (x_1) \O_2(x_2) \O_3(x_3) \r_{\rm no \ boundary}
= {C_{123}\over (x_{12}^{\, 2})^{\hh(\eta_1+\eta_2-\eta_3)}
(x_{23}^{\, 2})^{\hh(\eta_2+\eta_3-\eta_1)}
(x_{31}^{\, 2})^{\hh(\eta_3+\eta_1-\eta_2)} } \, .
\eqno (6.1) $$
If $C_3$ denotes the normalisation factor for the two point function of
the operator $\O_3$, so that in general $\l \O_i (x) \O_j(x')\r = \delta_{ij}
C_i / s^{2\eta_i}$, and $C_{12}{}^3 = C_{123}/C_3$ we may rewrite the
OPE (2.9) to include all derivatives of $\O_3$ as
$$ \O_1(x) \O_2(x') = {C_{12}{}^3\over (s^2)^{
{1\over 2} (\eta_1 + \eta_2 - \eta_3)}}
C^{\eta_3,\eta_-}(s,\pr_{x'})\O_3 (x') \, ,
\quad s= x-x' \, , \ \ \eta_- = \eta_1 -\eta_2 \, ,
\eqno (6.2) $$
where $C^{a,b}(s,\pr)$ is the derivative operator defined in (5.8).
By virtue of (5.11) the OPE (6.2) exactly reproduces the three point function
(6.1) for non coincident points. Of course the OPE of $\O_1$ and
$\O_2$ in general contains contributions from infinitely many quasi-primary
operators but such a summation is left implicit here.

For subsequent use it is convenient to generalise the expression (5.8) to
$$ \eqalign {
C^{a,b}_n(s,\pr) = {1\over B(a_+,a_-)}& \int_0^1\!\! \d \alpha \,
\alpha^{a_+-1}
(1-\alpha)^{a_--1} \cr
&\times \sum_{m=0} {1\over m!}\, {1\over(a+1-n-\half d)_m}
\, \bigl [ -\quar s^2 \alpha (1-\alpha) \pr^2 \bigl ]^m e^{\alpha s{\cdot \pr}}
\, , \cr}
\eqno (6.3) $$
which reduces to (5.8) when $n=0$. The essential result for application of
the OPE in the presence of a boundary is
$$ C^{a,b}_n(s,\pr_{x'}){1\over (2y')^a} = {1\over (2y)^{a_+} (2y')^{a_-}}
F(a_+, a_- ; a+1-n - \half d ; -\xi) \, ,
\eqno (6.4) $$
which is easily obtained by application of the explicit form (6.3).

Applying (6.2) to $\l \O_1(x)\O_2(x')\r$ then gives in (2.8) the equivalent
results for the contribution of all derivative operators generated from
$\O_3$
$$ \eqalign {
f_{12}(\xi)_{\O_3}& = C_{12}{}^3 A_3 \, \xi^{-{1\over 2}
(\eta_1 + \eta_2 - \eta_3)}
F\bigl(\half(\eta_3+\eta_-), \half (\eta_3-\eta_-); \eta_3 + 1 -\half d;-\xi
\bigl )  \, , \cr
F_{12}(v)_{\O_3}& = C_{12}{}^3 A_3 \, v^{\eta_3+\eta_-}
F\bigl(\half(\eta_3+\eta_-),
\half (\eta_3+\eta_-) + 1 -\half d; \eta_3 + 1 -\half d; v^2 \bigl ) \, . \cr}
\eqno (6.5) $$
The leading singular piece as $\xi,v\to 0$ of course coincides with (2.10).
When $\eta_-=0$ and $\eta_3 = d-2$ then $F_{12}(v)_{\O_3}=1$. This is
appropriate  for free scalar fields with $\O_3 \to \phi^2$. In this case
the above results show that the free field expression for
$\l \phi^\alpha (x) \phi^\beta(x') \r$, as given by (3.1,2,3) for either
Dirichlet or Neumann boundary conditions, is reproduced solely by the
contribution in the OPE of the identity and the operator $\phi^2$ and its
derivatives, taking $C_{\phi\phi}{}^{\phi^2} = 1/N$ and with $A_{\phi^2}$
as in (3.5).

We may also derive the corresponding formulae for two point functions involving
the energy momentum tensor $T_{\mu\nu}$. The OPE is determined by knowledge
of the three point function and detailed expressions in the relevant cases
for conformal field theories in any dimension $d$ were obtained by one of us
recently. The construction given depends essentially on the result
that for three points $x_1,x_2,x_3$
it is possible to construct vectors $X_i,\, i=1,2,3$ which transform
homogeneously under conformal transformations, ${X_{i\mu} \to \Omega(x_i)
\R_{\mu\alpha}(x_i) X_{i\alpha}}$, at each $x_i$. For $X_1$ we have
$$ X_{1\mu} = {x_{12\mu}\over x_{12}^{\, 2}} - {x_{13\mu}\over x_{13}^{\, 2}}
\, , \quad
X_{1}^{\, 2} = {x_{23}^{\, 2}\over x_{12}^{\, 2}x_{13}^{\, 2}} \, ,
\eqno (6.6) $$
while $X_2, X_3$ are obtained by cyclic permutation.\footnote{*}{It is
perhaps worth noting that if $x_1\to x,x_2\to x', x_3\to {\bar x}$, for
${\bar x}_\mu = (-y,\bx)$, then ${X_1 \to X /(2yv)}$ and $X_2 \to
- (1-v^2)X'/(2y'v)$, with $X,X'$ as given in (2.14).}
A crucial result is that under `parallel transport' by the inversion
transformation, as defined in (2.3),
$$ I_{\mu\alpha}(x_{ij}) X_{j\alpha} = - {x_{ik}^{\, 2} \over x_{jk}^{\, 2}}
X_{i\mu} \, , \quad k\ne i,j \,.
\eqno (6.7) $$

For the three point function of the energy momentum tensor with two scalar
fields $\O$ of dimension $\eta$ we may then write [3]
$$ \l T_{\mu\nu}(x_1)\O(x_2) \O(x_3)\r_{\rm no \ boundary}
= - {\eta d\over d-1}\, {C_\O \over S_d}
\biggl ( {X_{1\mu} X_{1\nu}\over X_1^{\, 2}} - {1\over d} \delta_{\mu\nu}
\biggl ) {(X_1^{\, 2})^{\hh d} \over x_{23}^{\, 2\eta}} \, ,
\eqno (6.8) $$
where the normalisation is determined by Ward identities which relate this
three point function to the two point function for $\O$. The result (6.8)
determines the form of the OPE of $T_{\mu\nu}$ and $\O$ which can be
written as
$$ T_{\mu\nu}(x) \O(x') = - {\eta d\over d-1}\, {1\over S_d} \,
{1\over (s^2)^{\hh d}} A_{\mu\nu} (s,\pr_{x'}) \O(x') \, ,
\eqno (6.9) $$
where it is easy from (6.8) in the limit $x_{12}\to 0$ to see that
$$ A_{\mu\nu}(s,\pr) =
{s_\mu s_\nu\over s^2} - {1\over d} \delta_{\mu\nu}  + \rO(s) \, .
\eqno (6.10) $$
Using the full result (6.8), with the definition (6.6) for $X_1$, we may find
after some work, in terms of $C^{a,b}_n(s,\pr)$ given by (6.3), that this
may be extended to include derivative terms in the form
$$
A_{\mu\nu}(s,\pr) = {1\over \eta(\eta+1)}\, \quar s^2 \pr_\mu\pr_\nu
C_2^{\eta+2,d-\eta}(s,\pr) -{1\over d}\, \delta_{\mu\nu}
C_1^{\eta,d-\eta}(s,\pr) + \hbox{terms} \propto s_\mu \, \hbox{or} \, s_\nu \,.
\eqno (6.11) $$
In the presence of a boundary the OPE (6.9) gives
$$ \l T_{\mu\nu} (x) \O (x') \r = - {\eta d\over d-1}\, {A_\O\over S_d} \,
{1\over (s^2)^{\hh d}} A_{\mu\nu} (s,\pr_{x'}) {1\over (2y')^\eta} \, ,
\eqno (6.12) $$
and applying (6.4) with the explicit expression (6.11) leads to
$$ \eqalign {
A_{\mu\nu}(s,\pr_{x'})& {1\over (2y')^\eta} \cr
={}& {1\over (2y)^{\hh d}(2y')^{\eta-\hh d}}\biggl \{ n_\mu n_\nu \xi
F\bigl (\half d+1,\eta+1-\half d;\eta+1-\half d;-\xi \bigl)\cr
&{}\qquad \qquad \qquad \ -{1\over d}\, \delta_{\mu\nu}
F\bigl (\half d,\eta-\half d;\eta-\half d ;-\xi \bigl)
{} + \hbox{terms} \propto s_\mu \, \hbox{or} \, s_\nu \biggl \} \cr
={}& {(2y')^{d-\eta}\over (s^2)^{\hh d}} \Bigl (X_\mu X_\nu
- {1\over d} \delta_{\mu \nu} \Bigl) v^d \, , \cr}
\eqno (6.13) $$
where in the last line we have completed the $s_\mu,s_\nu$ terms
to achieve the desired general expression dictated by conformal
invariance in accord with (2.19). Clearly this result agrees exactly with
with the general expression obtained in (2.19,22) with also the result (2.26)
for the coefficient $c_{T\O}$. Of course getting the
correct form is a necessary consistency check of the above treatment.

In order to determine the OPE for two energy momentum tensors involving
a scalar operator $\O$ we need the three point function [3]
$$ \!\!\!\eqalign {\!\!\!\!\!\!\!\!\!\!\!\!\!\!
\l &  T_{\mu\nu} (x_1) T_{\si\rho}(x_2)\O(x_3) \r_{\rm no \ boundary}  \cr
{}& = {1\over (x_{12}^{\, 2})^d}
\Bigl ( {x_{12}^{\, 2}\over x_{13}^{\, 2}x_{23}^{\, 2}}\Bigl )^{\!\hh \eta}
\biggl \{ \I_{\mu\nu,\si\rho}(x_{12})\, \C_\O +  \Bigl (
{X_{1\mu} X_{1\nu} \over X_1^{\, 2}}- {1\over d} \delta_{\mu \nu} \Bigl)
\Bigl ({X_{2\si} X_{2\rho}\over X_2^{\, 2}} -
{1\over d} \delta_{\si\rho} \Bigl) \A_\O \cr
& \qquad \qquad \qquad \qquad \qquad {} + \Bigl ( x_{12}^{\, 2}
\bigl ( X_{1\mu} X_{2\si} I_{\nu\rho}(x_{12})
+ \mu \leftrightarrow \nu , \si \leftrightarrow \rho \bigl ) \cr
& \qquad \qquad \qquad \qquad \qquad \qquad {} + {4\over d} \,
\delta_{\mu\nu} {X_{2\si} X_{2\rho}\over X_2^{\, 2}}
+ {4\over d} \, \delta_{\si\rho} {X_{1\mu} X_{1\nu}\over X_1^{\, 2}}
- {4\over d^2} \delta_{\mu\nu} \delta_{\si\rho} \Bigl ) \B_\O
\biggl \} \, , \cr}
\eqno (6.14) $$
where $\A_\O,\B_\O,\C_\O$ are constants and $\I_{\mu\nu,\si\rho}$ is defined
previously in (2.28).
The three terms in (6.14) are separately conformally invariant. However
imposing the conservation equation for $T_{\mu\nu}$, using (2.29) and also
$$ \eqalign{
\pr_{1\mu} X_{2\sigma} = {}& {1\over x_{12}^{\, 2}} I_{\mu\si}(x_{12})\, ,\quad
\pr_{1\mu} \biggl( {1\over (x_{12}^{\, 2})^d}
\Bigl ({X_{1\mu} X_{1\nu}\over X_1^{\, 2}}
- {1\over d} \delta_{\mu \nu} \Bigl) \biggl ) {}= - (d-1) {X_{1\nu}\over
(x_{12}^{\, 2})^d} \, , \cr
\pr_{1\mu} \biggl( {1\over (x_{12}^{\, 2})^d} & \Bigl ( x_{12}^{\, 2} \bigl (
X_{1\mu} I_{\nu\si}(x_{12}) + X_{1\nu} I_{\mu\si}(x_{12}) \bigl ) {}
+ {2\over d}\, \de_{\mu\nu} {X_{2\si}\over X_2^{\, 2}} \Bigl ) \biggl ) \cr
= {}& - {x_{23}^{\, 2} \over (x_{12}^{\, 2})^d x_{13}^{\, 2}}\, {1\over d}
(d-2) \Bigl ( (d+1) I_{\nu\si}(x_{12}) + 2x_{12}^{\, 2}X_{1\nu} X_{2\si}
\Bigl ) \, , \cr}
\eqno (6.15) $$
gives rise to the conditions
$$ \eqalign {
\half \eta d \, \C_\O + \A_\O + (d^2-4-\eta d) \B_\O = {}& 0 \, , \cr
\bigl ( 1 + \half (d-\eta)(d-1) \bigl ) \A_\O + (\eta+2)(d-2)\B_\O
= {}& 0 \, . \cr}
\eqno (6.16) $$
In consequence there remains a single constant parameterising the three point
function (6.14) but there are no Ward identities which relate this to any two
point amplitude in this case.

{}From the three point function (6.14) we may derive the contribution of the
operator $\O$ to the OPE for two energy momentum tensors which can be written
as
$$ T_{\mu\nu}(x) T_{\si\rho}(x') = {1\over (s^2)^{d-\hh \eta}}\,
A_{\mu\nu\si\rho}(s,\pr_{x'}) \O(x') \, ,
\eqno (6.17) $$
where $A_{\mu\nu\si\rho}(s,\pr)$ is determined by requiring it to
reproduce the full expression on the r.h.s. of (6.14). In the presence
of a boundary the operator $\O$ then gives rise to an expression for the
two point function $\l T_{\mu\nu}(x) T_{\si\rho}(x')\r$ using the result
(2.7) for $\l\O(x')\r$. Following a similar, albeit more tedious, procedure
to that outlined above for  $\l T_{\mu\nu}(x) \O(x')\r$  the result is
compatible with the form (2.26) where the corresponding form for the
invariant functions $A,B,C$ due to the operator $\O$ in the OPE is given by
$$ \eqalign {
A(v)_\O ={}& {A_\O\over C_\O} \A_\O \, \xi^{\hh \eta} \Bigl ( 1 +
4{d+2\over \eta^2} \xi {\d\over \d\xi} + 4 {(d+2)(d+4)\over \eta^2
(\eta+2)^2}\xi^2 {\d^2\over \d \xi^2} \Bigl ) f(\xi) \, , \cr
B(v)_\O ={}& - {A_\O\over C_\O} \xi^{\hh \eta} \biggl ( \B_\O
\Bigl ( 1 + {2d\over \eta^2} \xi {\d\over \d\xi} \Bigl )\cr
& \qquad \qquad \quad \
+ \A_\O {2\over \eta^2} \Bigl ( \xi {\d\over \d\xi} + 2
{d+2\over (\eta+2)^2} \xi^2{\d^2\over \d \xi^2} \Bigl ) \biggl )
f(\xi) \, , \cr
C(v)_\O ={}& {A_\O\over C_\O} \xi^{\hh \eta} \Bigl ( \C_\O +
\B_\O {8\over \eta^2} \xi{\d\over \d\xi} + \A_\O
{8\over \eta^2(\eta+2)^2} \xi^2{\d^2\over \d \xi^2} \Bigl ) f(\xi) \, , \cr
& \qquad \qquad \quad \ f(\xi) =
F(\half \eta,\half \eta; \eta + 1 - \half d; -\xi) \, . \cr}
\eqno (6.18) $$
We may solve (6.16) by writing
$$ \eqalign {
\A_\O & = C_{TT\O} {d-2\over d-1}\,\eta(\eta+2) \, ,  \cr
\B_\O & = C_{TT\O}\Bigl ( {d-2\over d-1} - \half (d+2-\eta)\Bigl ) \eta \, ,\cr
\C_\O & = C_{TT\O}\Bigl ( - {2d\over d-1} + (d-\eta)^2 \Bigl ) \, , \cr}
\eqno (6.19) $$
and then, using various properties of hypergeometric functions and with
the results in (2.36,37,38), we may obtain, with $ C_{TT}{}^{\O}
=  C_{TT\O}/C_\O$,
$$ \eqalign {
\alpha(v)_\O &  = C_{TT}{}^{\O}A_\O(d-2)(d+1)v^\eta(1-v^2) \cr
& \qquad \quad \quad {}\times \biggl \{ \Bigl ( 1 -{dv^2
\over \eta+2} \Bigl ) F(v^2) + {2v^2\over \eta(\eta+2)} \bigl (
2+d(1-v^2) \bigl ) F'(v^2) \biggl \} \, , \cr
\gamma(v)_\O & = C_{TT}{}^{\O}A_\O{d-2\over d-1}(d+1) v^\eta(1-v^2) \cr
& \qquad \quad \quad {}\times \biggl \{ \Bigl (
\half (\eta-d)  - {dv^2\over \eta+2} \Bigl ) F(v^2) + {v^2\over \eta+2}
\Bigl ( \eta-d - {2dv^2\over \eta} \Bigl )  F'(v^2) \biggl \} \, , \cr
&\qquad\qquad F(v^2) = F(\half\eta,\half\eta+1-\half d; \eta+1-\half d;v^2)
\, . \cr}
\eqno (6.20) $$
Although these expressions are somewhat lengthy they satisfy differential
equations equivalent to the conservation conditions (2.31a,b).

As a check on (6.20) we may consider the trivial case when $\O\to 1$, the
identity
operator, with $\eta=0$. In this case in (6.14) $\C_1 = C_T$, as defined in
(2.32), and $A_1 = 1$ and from (6.19) $C_T = C_{TT1}d(d-2)(d+1)/(d-1)$. Using
$F(v^2) \sim 1 -\half \eta \ln(1-v^2)$ for $\eta \sim 0$ then taking the limit
$\eta\to 0$ in (6.20) and expressing the coefficient in terms of $C_T$ gives
the expected results for $\alpha_1 ,\gamma_1$ independent of $v$.

If $\eta=d-2$ then $F(v^2)=1$. As discussed earlier this case is relevant
for the operator $\phi^2$ in free field theory when now
$C_{TT}{}^{\phi^2} = \quar d(d-2)^2 A /(d-1),
\, A_{\phi^2} = \pm AN$, with $A$ given in (3.3). In this case (6.20) becomes
$$ \eqalign {
\alpha(v)_{\phi^2} = {}&\pm {N\,d\over4(d-1)S_d^{\, 2}}(d-2)(d+1)(1-v^2)^2
v^{d-2}\, , \cr
\gamma(v)_{\phi^2} = {}&\mp {N\,d\over4(d-1)^2S_d^{\, 2}}(d-2)(d+1)(1-v^4)
v^{d-2}\, , \cr}
\eqno (6.21) $$
which is identical to the results calculated in I for free scalar field theory
for the terms in $\l T_{\mu\nu}(x) T_{\si\rho}(x')\r$ dependent on the
boundary conditions.
In general however the expression obtained from a single operator $\O$ in the
OPE, as given by (6.20), fails to satisfy the required behaviour
as $y,y'\to 0$, in particular boundary
conditions such as $\gamma(1)=0$.\footnote{*}{In two dimensions the
contribution of general scalar operators from (6.20) is zero. Nevertheless
the two point function (2.34) is reproduced by the OPE
$ T_{zz}(z) T_{\bz\bz}(\bz') = e^{(z-z')\pr_z}\O(x')$ in terms of the
quasi-primary operator $\O(x) = T_{zz}(z) T_{\bz\bz}(\bz)$, which has
dimension 4 and $A_\O = \quar C_T$.}

In the $O(N)$ model for large $N$ the result for $\alpha(v)$ in (5.34)
shows that the scalar operators appearing in the OPE of two energy momentum
tensors have dimensions $2+2n$ and also $2d+2n$ for $n=0,1,\dots$.
\bigskip
\leftline{\bigbf{7 Boundary Operator Expansion}}
\medskip

In the previous section we described how the contribution of all derivative
operators formed from a quasi-primary operator and appearing in the OPE
for operators $\O_1, \O_2$ to the two point function $\l \O_1(x) \O_2(x')\r$
in the presence of a boundary may be explicitly calculated. In this section
we show how this can also be achieved for all derivative operators formed
from a particular boundary operator $\hO$ appearing in the BOE of $\O_1$
or $\O_2$. We assume a basis of boundary operators which have a well defined
spin under $O(d-1)$ rotations and scale dimension under scale transformations.
For $\hO$ a scalar operator of scale dimension $\heta$, so that the
corresponding derivative operators have dimension $\heta+n, \, n=1,2,\dots$,
the two point function on the boundary has the form
$$ \l \hO(\bx) \hO(\bx')\r = {\hC_\hO\over \bs^{2\heta}}
\, , \quad \bs = \bx - \bx' \, .
\eqno (7.1) $$
The contribution to the BOE in (2.11) for a scalar operator $\O$ of dimension
$\eta$ involving derivatives of the operator $\hO$ may then written as
$$ \O (x) = {B_{\O}{}^{\hO}\over (2y)^{\eta - \heta}} \,
D^{\heta}(y^2 \hnab^2) \hO (\bx) \, ,
\eqno (7.2) $$
where the differential operator $D^{\heta}(y^2 \hnab^2)$, $\hnab_i = \pr_i$,
is determined by consistency with (2.12). Defining
$B_{\O}{}^{\hO} = B_{\O\hO}/\hC_\hO$ this requires
$$ D^{\heta}(y^2 \hnab^2) {1\over \bs^{2\heta}} = {1\over(\bs^2+y^2)^\heta}
= \sum_{m=0} {1\over m!} (\heta)_m {(-y^2)^m\over (\bs^2)^{\heta+m}} \, .
\eqno (7.3) $$
It is easy to see that this is satisfied if we take
$$ D^{\heta}(y^2 \hnab^2) = \sum_{m=0} {1\over m!} \, {1\over
(\heta + {\ts{3\over 2}}-\half d)_m} \big ( - \quar y^2 \hnab^2 \big )^m \, .
\eqno (7.4) $$

For application to the BOE of two operators at points away from the boundary
we extend the result (7.3) to
$$ \eqalignno {
D^{\heta}(y^2 \hnab^2) & D^{\heta}(y^{\prime \, 2} \hnab^{\prime \, 2})
{1\over \bs^{2\heta}} = \sum_{m,n} {1\over m!n!} \,
{(\heta)_{m+n} (\heta + {\ts{3\over 2}}-\half d)_{m+n} \over
(\heta + {\ts{3\over 2}}-\half d)_m (\heta + {\ts{3\over 2}}-\half d)_n}
{(-y^2)^m (-y^{\prime 2})^n \over (\bs^2)^{\heta+m+n}} \cr
& = {1\over (\bs^2 + y^2 + y^{\prime 2} )^\heta} F\big (\half \heta,\half \heta
+ \half ; \heta + {\ts{3\over 2}}-\half d ; {1\over (2\xi+1)^2} \big ) \, .
& (7.5) \cr}
$$
Applying this to $\l \O_1(x)\O_2(x')\r$ we can extend the leading order result
(2.13) to give
$$ f_{12}(\xi)_\hO = B_{\O_1 \hO}B_{\O_2}{}^{\hO}\,{1\over \xi^\heta}
F\big (\heta,\heta+1-\half d; 2\heta+2-d;-{1\over \xi} \big ) \, ,
\eqno (7.6) $$
which therefore includes all derivatives of $\hO$ in the BOE of $\O_1$ and
$\O_2$.

As a special case we may consider $\heta = \half d -1$ when
$F(\heta,\lambda;2\lambda;z) \to \half \big ( (1-z)^{-\heta} + 1\big ) $
as $\lambda\to 0$. This is relevant for free scalar field theory
with Neumann boundary conditions for the surface operator $\hphi(\bx) =
\phi (0,\bx)$ when $B_\phi {}^\hphi = 1 $ and $\hC_\hphi = 2A$ so that
(7.6) gives
$$ f_\phi(\xi)_{{\rm sp},\hphi} = A \Big ( {1\over \xi^{\hh d -1}} +
{1\over (1+\xi)^{\hh d -1}}\Big ) \, ,
\eqno (7.7) $$
which is the exact result in this case. Alternatively if $\heta = \half d$
then the hypergeometric function simplifies to
$F(\heta,1;2;z) =  \big ( (1-z)^{1-\heta} - 1\big )/(\heta - 1)z$.
This is appropriate for free scalar field theory with Dirichlet boundary
conditions for the surface operator $\hphi_n(\bx) = \pr_1 \phi(0,\bx)$
when now $B_\phi {}^{\hphi_n} = \half $ and $\hC_{\hphi_n} = 2(d-2)A$
so that (7.6) becomes
$$ f_\phi(\xi)_{{\rm ord},\hphi_n} = A \Big ( {1\over \xi^{\hh d -1}} -
{1\over (1+\xi)^{\hh d -1}}\Big ) \, ,
\eqno (7.8) $$
which is again of course the exact result.

Another case when simple formulae are obtained is when $\heta = d-2$.
This is relevant for the boundary operator $\hphi^2(\bx) = \phi(0,\bx)^2$
in free field theory with Neumann boundary conditions. Applying this
to the two point function of $\phi^2$, with $B_{\phi^2}{}^{\! \hphi^2}=1$
and $\hC_{\hphi^2} = 8NA^2 $, we then obtain
$$ F_{\phi^2}(v)_{{\rm sp}, \hphi^2} = 8N A^2  \, v^{d-2} \, , \quad
F_{\phi^2}(v)_{\rm sp} = F_{\phi^2}(v)_{{\rm sp}, \hphi^2} + 2N  A^2
(1- v^{d-2})^2 \, ,
\eqno (7.9) $$
where $F_{\phi^2}(v)_{\rm sp}$ is given by (3.2,5). The remaining part, after
subtraction of the contribution arising from $\hphi^2$, is identical with
the result in the Dirichlet case for which the leading behaviour as $v\to 1$
is produced by an operator of dimension $d$.

We may also extend this treatment to the BOE for operators with spin
although we confine our attention here to the energy momentum tensor
$T_{\mu\nu}$. Following (2.14) for $x_\mu = (y,\bx)$ and $\bx'$ defining
a point on the boundary we may define a vector under conformal
transformations by
$$ \hX_\mu = {2y\over \hs^2} \hs_\mu - n_\mu = - I_{\mu\nu}(\hs) n_\nu
\, , \quad \hX^2 = 1 \, \qquad \hs_\mu = (y, \bx-\bx') \, .
\eqno (7.10) $$
Using
$$ \pr_\mu \biggl ( {(2y)^{\heta-d}\over (\hs^2)^\heta} \Bigl (
\hX_\mu \hX_\nu - {1\over d} \delta_{\mu \nu} \Bigl) \biggl ){} =
(d-1)\Big ( 1 - {\heta\over d} \Big ){1\over y}  {(2y)^{\heta-d}
\over (\hs^2)^\heta} \, \hX_\nu \, ,
\eqno (7.11) $$
it is easy that $T_{\mu\nu}$ can only have a non zero two point function,
invariant under conformal transformations and satisfying the conservation
equation, with a scalar surface operator provided that this operator
has dimension $d$. In any conformal theory there is such
a surface operator $\hT(\bx) = T_{11}(0,\bx)$, since $T_{11}$ is non
singular as the boundary is approached, and it is natural to assume
that this is unique. The surface two point function for $\hT$ has
the form (7.1) with a coefficient $\hC_\hT$. From the general result (2.27)
or specifically from (2.40) we may then show that
$$ \hC_\hT =  {d-1\over d}C(1) + {(d-1)^2\over d^2}\big ( A(1) + 4B(1)\big )
= \alpha (1) \, ,
\eqno (7.12) $$
with $\alpha(v)$ defined in (2.37,38). The two point function of $T_{\mu\nu}$
and $\hT$ can then be written
$$ \l T_{\mu\nu}(x) \hT (\bx') \r = {d\over d-1}\hC_\hT\, {1\over \hs^{2d}}
 \Bigl ( \hX_\mu \hX_\nu - {1\over d} \delta_{\mu \nu} \Bigl) \, ,
\eqno (7.13) $$
where we have taken $B_T{}^\hT = 1$.

{}From (7.13) we may deduce the form of the BOE for $T_{\mu\nu}$ involving
$\hT$,
$$ T_{\mu\nu} (x) = D_{\mu\nu} (y , \hnab) \hT (\bx) \, ,
\eqno (7.14) $$
where we must require
$$
D_{\mu\nu} (y , \hnab) {1\over \bs^{2d}} = {d\over d-1}\, {1\over \hs^{2d}}
 \Bigl ( \hX_\mu \hX_\nu - {1\over d} \delta_{\mu \nu} \Bigl) \, .
\eqno (7.15) $$
Writing
$$ D_{\mu\nu} (y , \hnab) = {\hat D}_{\mu\nu} (y , \pr) D^d (y^2 \hnab^2)\, ,
\eqno (7.16) $$
with $D^d (y^2 \hnab^2)$ defined by (7.4), then we may take\footnote{*}{Since
$(y^2 \pr^2 + (d+2) y n{\cdot \pr} ) 1/\hs^{2d} = 0 $ there is some
ambiguity in the definition of ${\hat D}_{\mu\nu} (y , \pr)$ but this does
not affect the final results in the BOE.}
$$ \eqalign{
{\hat D}_{ij}(y , \pr ) = {}& {1\over (d-1)(d+1)}\Big (
y^2 \pr_i \pr_j - \de_{ij} y{\pr\over \pr y} - (d+1) \de_{ij} \Big ) \,,\cr
{\hat D}_{i1}(y , \hnab) = {}& {1\over (d-1)(d+1)} \, y \Big (d+1
+y{\pr\over \pr y} \Big ) \pr_i \, , \quad {\hat D}_{11}(y , \pr) = -
{\hat D}_{ii}(y , \pr) \, . \cr}
\eqno (7.17) $$
With these results we may calculate the two point function for the
energy momentum tensor $T_{\mu\nu}$ and a scalar operator $\O$ of dimension
$\eta$ by using the BOE involving $\hT$ which gives
$$ \eqalign {\!\!\!\!\!\!
\l T_{\mu\nu}(x) \O(x') \r = {}& B_\O{}^\hT \hC_\hT
(2y')^{d-\eta}D_{\mu\nu} (y , \hnab)
D^d (y^{\prime\, 2} \hnab^{\prime \, 2}) {1\over \bs^{2d}} \cr
= {}& B_\O{}^\hT \hC_\hT (2y')^{d-\eta} {\hat D}_{\mu\nu} (y , \pr)
{1\over (4yy')^d} F(\xi) \cr
= {}& B_\O{}^\hT \hC_\hT {(2y')^{-\eta}\over (2y)^d}\,{1\over(d-1)(d+1)}
\bigg \{ \Big ( X_\mu X_\nu - {1\over d}\de_{\mu\nu} \Bigl ) \xi(\xi+1)
{\d^2\over \d \xi^2} F(\xi) \cr
&\qquad \quad {} - \Big ( n_\mu n_\nu - {1\over d}\de_{\mu\nu} \Big ) \Big (
\xi(\xi+1) {\d^2\over \d \xi^2} + d (\xi+\half) {\d \over \d \xi} - d \Big )
F(\xi) \bigg \} \cr
= {}& {d\over d-1}B_\O{}^\hT \hC_\hT {(2y')^{-\eta}\over (2y)^d}
\Big ( {v\over \xi} \Big )^{\! d} \Big ( X_\mu X_\nu - {1\over d}\de_{\mu\nu}
\Bigl ) \, , \cr
& \qquad \hbox{for} \ \ F(\xi) = {1\over \xi^d} F\big ( d,\half d+1;d+2;
-{1\over \xi} \big ) \, . \cr}
\eqno (7.18) $$
It is easy to see that this is in agreement with the general form given by
(2.19,22) and comparing with the result (2.26) shows that we must have
$$ B_{\O\hT}=B_\O{}^\hT \hC_\hT = - \eta \, {A_\O \over S_d} \, ,
\eqno (7.19) $$
which was first obtained by Cardy [7] for consistency of the OPE and BOE.

We also consider the contribution of the boundary operator $\hT$ to the
two point function of the energy momentum tensor where using (7.14) we
obtain
$$ \eqalign { \!\!\!\!\!
\l T_{\mu\nu}(x) T_{\si\rho}(x')\r = {}& \hC_\hT \, D_{\mu\nu}(y,\hnab)
D_{\si\rho}(y',\hnab') {1\over \bs^{2d}} \cr
= {}& {d\over d-1} \hC_\hT \, {\hat D}_{\si\rho}(y',\pr') \bigg (
{1\over (4yy')^d} \Big ( X_\mu X_\nu - {1\over d}\de_{\mu\nu} \Big )
\big [ \xi(\xi+1) \big ]^{-\hh d} \bigg ) \, , \cr}
\eqno (7.20) $$
using the result already obtained in (7.19). The evaluation of (7.20)
is straightforward albeit tedious. Some details are given in appendix E.
The final expression is of the necessary form shown in (2.27) where the
coefficients are given by
$$ \eqalign {
C(v)_\hT = {}& {d\over (d-1)^2 (d+1)}\hC_\hT \, \half v^{d-2}(1-v^2)^2 \, , \cr
2B(v)_\hT + C(v)_\hT = {}& {d\over (d-1)^2}\hC_\hT \, \half v^{d-2}(1 - v^4)
\, , \cr
A(v)_\hT+4B(v)_\hT = {}& {d^2\over (d-1)^2 (d+1)}\hC_\hT \, \quar v^{d-2}
\big ( (d+2) ( 1+v^4) + 2d v^2 \big ) \, . \cr}
\eqno (7.21) $$
These results satisfy the conservation equations (2.31a,b). Further for $d=2$
they are exact since clearly only scalar operators on the boundary need be
considered and in this case (7.21) coincides with (2.33) if $C_T =\hC_\hT$.
{}From (2.37,38) we may also write
$$ \alpha(v)_\hT = {1\over d-1}\hC_\hT \,  \quar v^{d-2}\big ( d(1+v^4)
+2(d-2) v^2 \big ) \, , \quad \alpha (1) = \hC_\hT \, .
\eqno (7.22) $$

For application to a non trivial conformal field theory we may consider first
results to first order in $\vep$ at the non Gaussian fixed point in scalar
field theory. In ref. [23] and in I $\hC_\hT$ was calculated to this order
for Neumann/Dirichlet boundary conditions giving
$$ \hC_\hT = {N\over S_d^{\, 2}} \Big ( 2 \pm {5\over 3} \, {N+2\over
N+8} \vep \Big ) \, .
\eqno (7.23) $$
The coefficient appearing in (3.15b) determines $(B_{\phi^2}{}^{\hT}
{}_{\! \rm ord})^2 \hC_{\hT,\rm ord}$ and hence using (7.19) with
$\O\to \phi^2$, together with (3.8) for $\eta_{\phi^2}$, we find
$$ {(A_{\phi^2, \rm ord})^2 \over C_{\phi^2}} = {N\over 2} \big (
1 + \rO(\vep^2) \big ) \, .
\eqno (7.24) $$
It is also of interest to compare the expressions obtained in (7.21,22)
with the results of the $\rO(\vep)$ calculations for
$\l T_{\mu\nu}(x) T_{\si\rho}(x')\r$ in I. In the Neumann case for $v\to 1$
$$ \eqalign {
\alpha(v)_{\rm sp} - \alpha(v)_\hT \sim {}& {N\over S_d^{\, 2}} \,
{\ts {20\over 3}} \big (1-v^2\big )^{2- {N+2\over N+8} \vep} \, , \cr
\gamma(v)_{\rm sp} - \gamma(v)_\hT  \sim {}& - {N\over S_d^{\, 2}} \,
{\ts {40\over 9}} \big (1-v^2\big )^{1- {N+2\over N+8} \vep} \, , \cr
\epsilon(v)_{\rm sp} - \epsilon(v)_\hT \sim {}& {N\over S_d^{\, 2}} \,
{\ts {8\over 3}} \big (1-v^2\big )^{- {N+2\over N+8} \vep} \, , \cr}
\eqno (7.25) $$
in terms of the functions defined in (2.36,37,38). These results
correspond to the contributions expected to arise from a boundary operator
which is a symmetric traceless tensor
$\hcT_{ij}$ of dimension $\heta_\hcT = d - {N+2\over N+8} \vep +
{\rO}(\vep^2)$. For such an operator we may define a two point function
with the energy momentum tensor
$$ \l T_{\mu\nu}(x) \hcT_{ij} (\bx')\r = B_{T\hcT} {(2y)^{\heta_\hcT -d} \over
\hs^{2\heta}} \bigg ( \I_{\mu\nu,ij}(\hs) +  {1\over d-1}
\Bigl ( \hX_\mu \hX_\nu - {1\over d} \delta_{\mu \nu} \Bigl) \de_{ij} \bigg )
\, , \eqno (7.26) $$
using $\I$ which is defined in (2.28). This satisfies the required conformal
transformation properties for $T_{\mu\nu}$ and obeys the necessary conservation
equation for arbitrary $\heta_\hcT$. In the Neumann case the boundary
operator $\hcT_{ij}(\bx)$ relevant for the limiting behaviour in (7.24)
appears as the leading term in the BOE of the operator formed
from the traceless part of $T_{ij}(x)$.
The above analysis shows that the coefficient, $\propto (2y)^{-d+\heta_\hcT}$,
is singular for $y\to 0$ unlike the situation for
$T_{11}(x)=-T_{ii}(x)$ which tends smoothly
to $\hT(\bx)$. $\hcT_{ij}$ cannot contribute to the BOE of scalar operators
so its role is not apparent in previous discussions. In the Dirichlet case,
corresponding to (7.24), we have
$$\eqalign{
\alpha(v)_{\rm ord}-\alpha(v)_\hT \sim {}&  {N\over S_d^{\, 2}}
\big (1-v^2\big )^{4-{N+2\over N+8}\vep}\,  ,\cr
\gamma(v)_{\rm ord}-\gamma(v)_\hT \sim {}& - {N\over S_d^{\, 2}}
{\ts {4\over 3}} \big (1-v^2\big )^{3-{N+2\over N+8}\vep}\, , \cr
\epsilon(v)_{\rm ord}-\epsilon(v)_\hT \sim {}& {N\over S_d^{\, 2}}
{\ts {12\over 5}} \big (1-v^2\big )^{2-{N+2\over N+8}\vep } \, , \cr}
\eqno (7.27) $$
which represents the contribution for an operator $\hcT_{ij}$ with dimension
$d+2- {N+2\over N+8}\vep$. This is as expected since such an operator
should be constructed from the field $\phi$ in terms of expressions of the
form $\pr_i\pr_1 \phi \pr_j \pr_1 \phi$.

In the critical $O(N)$ model in the $N\to \infty$ limit we may also use
the results obtained in sections 4 and 5 to verify the consequences of the
BOE. By considering the limit $\xi\to \infty$ or $v\to 1$ of the $\lambda$
two point function we may identify the contribution of the boundary operator
$\hT$ of dimension $d$. In the ordinary case, from the $\rO(\xi^{-d})$ term
in (4.32) we may identify
$$ \big ( B_\lambda{}^\hT{}_{\!{\rm ord}}\big )^2 \hC_{\hT,{\rm ord}}
= B{\Ga(d)\Ga(d-2)\over \Ga(2d-4)} \, .
\eqno (7.28) $$
Using now (7.19), with $\eta=2$ and $A_{\lambda,{\rm ord}}$ given by (4.12),
we then find
$$
{\hat C}_{{\hat T}{\rm ord}} = {2N\over S_d^{\, 2}}\,
{\Ga(2d-3) \Ga (3-\half d) \Ga(\half d)^3\over \Ga(d) \Ga(d-1)^2} \, .
\eqno (7.29) $$
Since $\alpha(1)={\hat C}_{{\hat T}}$ this is identical with the result
obtained by direct calculation in (5.32). The results for the energy
momentum tensor two point function given in (5.30,31) and (5.33) exhibit
explicitly the full contributions of the boundary operator $\hT$, as given
in (7.21,22), while the remaining parts involving ${}_3F_2$ functions
arise from non scalar boundary operators with dimension $2d-2+2n$. For the
special case (4.40) gives
$$ \big ( B_\lambda{}^\hT{}_{\!{\rm sp}}\big )^2 \hC_{\hT,{\rm sp}}
= B{1\over 3} {6-d\over d-2}{\Ga(d)\Ga(d-2)\over \Ga(2d-5)} \, ,
\eqno (7.30) $$
which then implies
$$
{\hat C}_{{\hat T}{\rm sp}} = {2N\over S_d^{\, 2}}\, 6(6-d)
{\Ga(2d-5) \Ga (3-\half d) \Ga(\half d)^3\over \Ga(d) \Ga(d-1)^2} \, .
\eqno (7.31) $$
It is easy to check that (7.29,31) are in accord with the $\vep$ expansion
results in (7.23).
\bigskip
\leftline{\bigbf{8 Conclusion}}
\medskip

Although the critical behaviour of statistical systems with a boundary is
relatively unexplored experimental investigation is feasible.
In this paper we have discussed theoretically the form of the functional
behaviour of two point functions at a conformal invariant crtical point in
dimensions $d>2$. In particular the $O(N)$ model, for $N\to \infty$, provides
a tractable non trivial example which may be analysed for ${2<d\le4}$. The
results of calculations in this model, which have been here extended to
include correlation functions involving the energy momentum tensor, are
complementary to the $\vep$ expansion and provide a limiting case for other
approximations.

We may perhaps note that the functions of the invariants $v$ or $\xi$ are
initially defined on the physical region $0\le v\le 1$ or $0\le \xi <\infty$
but may be analytically continued to the whole complex plane. The singularities
as $v\to 0$ and $v\to 1$ are well understood in terms of the OPE and BOE. It
would be interesting to understand more directly the form of the singular
behaviour as $v\to \infty$ or $\xi\to -1$. In this context we may note that
our results have simple transformation properties under $v\to v^{-1}$ or
$\xi\to - 1 -\xi$ (which is equivalent to taking $y\to -y$ or $y'\to -y'$)
which may merit further investigation.

\vfill\eject
\leftline{\bigbf Appendix A}
\medskip

Here we present some details of the perturbative calculation of the
two point function of $\phi^2$ to first order in $\varepsilon=4-d$
assuming the renormalisable interaction
$\L_{\rm int}={1\over 24}g(\phi^2)^2$ with $\phi^\alpha$ an $N$-component
scalar field. The coupling is restricted to the non-gaussian fixed point
$g=g_* = 48\pi^2\vep/(N+8)$ to lowest order.
To simplify the calculation we choose a
perpendicular configuration where $x_\mu=(y,\0)$ and
$x'{}_{\!\mu}=(y',\0)$, with $y>y'$, so that $v=(y-y')/(y+y')$.
For the free field case the basic propagator from (3.1,2) in this case
is just
$$
G(y,y') ={A\over (y-y')^{d-2}}\big(1\pm v^{d-2} \big )\, ,
\quad \quad A={1\over (d-2) S_d} \, .
\eqno (A.1)
$$
To first order in $g$ we may write
$$
\vev{\phi^2(y,\0)\phi^2(y',\0)}= 2NG(y,y')^2 \mp {\textstyle{{2\over3}}}
N(N+2)g_* G(y,y') \, I_a -{\textstyle {{1\over3}}} N(N+2)g_* \, I_b \, ,
\eqno (A.2)
$$
where $I_a,I_b$ are integrals corresponding to fig. (1a,b). To this order
we may restrict these to $d=4$ since $g_* \propto \vep$ and $I_a$ then
has the form
$$
I_a =A^3\int_0^\infty\!\!\! \d z\,\int \! \d^3x \,
\Big ({1\over X^2} \pm {1\over \bX^2} \Big ) \Big ({1\over X'{}^2} \pm
{1\over \bX'{}^2} \Big ){1\over 4z^2}\, .
\eqno (A.3)
$$
with $X^2=\bx^2+(y-z)^2$, $X'^2=\bx^2+(y'-z)^2$, $\bX^2=\bx^2+(y+z)^2$,
and  $\bX'^2=\bx^2+(y'+z)^2$. For finiteness we exclude from the
$z$-integral an $\epsilon$ neighbourhood of the boundary $z=0$. In
the Dirichlet case the result is finite for $\epsilon \to 0$, but in
the Neumann case the divergence must be removed by the addition of a
surface counter term $\propto \phi^2$. This calculation is equivalent
to finding the one loop correction to $G_\phi(x,x')$ which was
calculated previously [9,14]. Consequently we easily obtain the result
$$
F_{\phi^2}(v)_a = {2N(N+2)A^2\vep\over(N+8)}\Big ( v^4 \ln{(1-v^2)\over v^2}
+\ln{(1-v^2)} \pm v^2 \ln{(1-v^2)^2 \over v^2}\Big ) \, .
\eqno (A.4)
$$
For $I_b$ when $d=4$ we have
$$
I_b=A^4\int_0^\infty\!\!\! \d z\,\int \! \d^3x \,
\Big ({1\over X^2} \pm {1\over \bX^2} \Big )^2 \Big ({1\over X'{}^2} \pm
{1\over \bX'{}^2} \Big )^2 \, .
\eqno (A.5)
$$
The integrals involved here are more difficult than the previous case.
They contain the usual short distance divergence arising from the singular
behaviour of $G_\phi(x,x')^2$ when $d=4$ but this is removed
by hand by restricting the integral over the vertex point $(z,\bx)$ to exclude
$z$ from $\epsilon$ neighbourhoods of $y,y'$. Evaluating the integral then
gives
$$\eqalign{
F_{\phi^2}(v)_b ={}& {2N(N+2)A^2\vep\over(N+8)}\Big ( 1+v^4 + v^2
\ln{ (1-v^2)^2} +  v^4 \ln {v^2} +{v^4 \over 1-v^2}\ln {v^4} \cr
& \hskip 2.0 truein \pm v^2 \ln{(1-v^2)^2} + (1\pm v^2)^2 \ln
{4\epsilon^2 \over (y-y')^2} \Big )\, .\cr }
\eqno (A.6)
$$
The divergent term as $\epsilon \to 0$ is proportional to $G(y,y')^2
\ln{\big({(y-y')^2/\epsilon^2}\big )}$ which of course reflects
the modification of the singularity as $y\to y'$ from its free value
arising from the one loop correction to the scaling dimension of $\phi^2$.
The complete result to this order is now
$$ F_{\phi^2}(v) = 2N A^2 \bigl ( 1 \pm v^{d-2} \bigl )^2{} + F_{\phi^2}(v)_a
+ F_{\phi^2}(v)_b \, .
\eqno (A.7) $$
Dropping the divergent piece then gives the result quoted in (3.13).
\bigskip
\leftline{\bigbf{Appendix B}}
\medskip

In order discuss the evaluation of the integrals in section 4 for general
$d$ when $g(\xi)$ is given by (4.27) it is convenient to work backwards
and consider an expression for $\tilde{\hat g}(k)$ of the form
$$ \tilde{\hat g}_{a,b}(k) = {\Gamma (a-{\ts{i\over 4}}k)
\Gamma (a+{\ts{i\over 4}}k) \over \Gamma (b-{\ts{i\over 4}}k)
\Gamma (b+{\ts{i\over 4}}k) } \, .
\eqno (B.1) $$
Having found the corresponding $g_{a,b}(\xi)$ it is then trivial to find the
associated $h(\xi)$, which determines the inverse kernel $H(x,x')$, since
it is obviously equal to $g_{b,a}(\xi)$.
The inverse Fourier transform can be found as a sum of residues of poles,
at ${i\over 4}k= a + n $ for $\theta>0$, giving a hypergeometric function
$$ \eqalign { \!\!\!\!\!\!\!
{\hat g}_{a,b}(\sinh^2 \theta)& =
{1\over 2\pi}\int \! \d k \, e^{-ik\theta} \, \tilde{\hat g}(k) \cr
&= {4\Gamma(2a)\over \Gamma(b-a)\Gamma(b+a)}\,  e^{-4a|\theta|}
F \bigl ( 2a,a-b+1;a+b; e^{-4|\theta|} \bigl ) \, . \cr
&= {4\Gamma(2a)\over \Gamma(b-a)\Gamma(b+a)}\,
{1\over (4\cosh^2 \theta)^{2a}}
F \bigl ( 2a,a+b-\half; 2a+2b-1;{1\over \cosh^2\theta}\bigl ) \, . \cr}
\eqno (B.2) $$
The transform ${\hat g}\to g$ is then straightforward since we can use
$$ {1 \over \Gamma (\lambda)} \int_0^\infty \!\!\!\!
\d u \, u^{\lambda -1} \, {\Gamma(p+\lambda)\over (1+\rho+u)^{p+\lambda}} =
{\Gamma(p)\over (1+\rho)^{p}} \, .
\eqno (B.3) $$
Applying this term by term in (B.2),with $p=2a+n$, we get
$$ \eqalignno{
g_{a,b}(\xi)& =
{\Gamma(2a+\lambda)\over 4^{2a-1}\pi^\lambda\Gamma(b-a)\Gamma(b+a)}\,
{1\over (1+\xi)^{2a+\lambda}}\,
F \bigl ( 2a+\lambda,a+b-\half; 2a+2b-1;{1\over 1+\xi}\bigl ) \cr
& ={\Gamma(2a+\lambda)\over 4^{2a-1}\pi^\lambda\Gamma(b-a)\Gamma(b+a)}\,
{1\over \xi^{2a+\lambda}}\,
F \bigl ( 2a+\lambda,a+b-\half; 2a+2b-1;-{1\over \xi}\bigl ) \, . &(B.4)\cr}
$$

For application in the Dirichlet case we take
$$ g(\xi) = {1\over \xi^\alpha(1+\xi)^\alpha} =
{\pi^{\lambda+1}\over 4^\lambda}\, {\Gamma(2\alpha)\Gamma(\half d -\alpha)
\over \Gamma(\alpha+\half)\Gamma(\alpha)^2}\, g_{\alpha-\hh \lambda,
\hh(\lambda+1)} (\xi) \, ,
\eqno (B.5) $$
and hence in this case
$$
h(\xi) = f(\alpha) {\Gamma(d)\Gamma(\alpha)\over \Gamma(2\alpha)
\Gamma(d-\alpha)}\, {1\over\xi^d} F\bigl ( d,\alpha;2\alpha;-{1\over \xi}\bigl)
\, ,
\eqno (B.6) $$
where $f(\alpha)$ is as in (4.4). Alternatively (B.6) can be written
as
$$ \eqalign {
\xi^d h(\xi) = f(\alpha) \Bigl ( &v^{2\alpha}
F (\alpha, 2\alpha-d; \alpha-d+1; v^2 ) \cr
& + { \Gamma(d)\Gamma(\alpha-d)\over \Gamma(2\alpha-d) \Gamma(d-\alpha) } \,
v^{2d} F (d,\alpha; d-\alpha+1; v^2 ) \Bigl ) \, , \cr}
\eqno (B.7) $$
so that $h(\xi)\sim f(\alpha)\, \xi^{\alpha-d}$ as $\xi\to 0$, assuming
$\alpha<d$. When $\alpha=d-2$, as required to leading order in $1/N$, then
$$ \eqalign {
h(\xi) = {}& f(d-2) \bigg \{ {1\over \xi^2}\big ( 1-(d-2)(3-d)\xi \big ) {}+
(d-4)_4 \Big ( \ln {1\over \xi} \, \half F(d,5-d;3;\xi) + {\tilde h}(\xi) \Big
)
\bigg \} \, , \cr
&{\tilde h}(\xi) = \sum_{n=0} \xi^n \, {(d)_n (5-d)_n \over n! (n+2)!} \Big (
\psi(n+3) + \psi(n+1) - \psi (d+n) - \psi (d-4-n) \Big ) \, . \cr}
\eqno (B.8) $$

For the Neumann case we consider functions of the form
$$ g(\xi) = {(1+2\xi)^2\over \xi^\alpha(1+\xi)^\alpha} \, .
\eqno (B.9) $$
Writing $(1+2\xi)^2 = 1 + 4\xi(1+\xi)$ it is easy to see from the above
discussion that the corresponding transform becomes
$$ \tilde{\hat g} (k) = {\pi^{\lambda+1}\over 4^{\lambda+1}}\,
{\Gamma(2\alpha)\Gamma(\half d -\alpha)
\over \Gamma(\alpha+\half)\Gamma(\alpha)^2}\,
\big ( \mu^2 + \quar k^2 \big ) \,
{\tilde{\hat g}}_{\alpha-1-\hh \lambda, \hh(\lambda+1)} (k) \, , \quad
\mu^2 = \lambda^2 - 2(\alpha-1) \, .
\eqno (B.10) $$
Defining, as in (4.26), ${\tilde{\hat h}}(k) = 1/ {\tilde{\hat g}}(k)$ then
two approaches are possible in order to obtain $h(\xi)$. Firstly finding
the inverse Fourier transform may be simplified to the previous discussion
by removing the pole at $k^2 = - 4 \mu^2$ which is equivalent to
expressing $\hat h$ in terms of a differential equation,
$$ \bigg ( \mu^2 - \quar {\d^2 \over \d \theta^2} \bigg )
{\hat h} (\sinh^2 \theta ) = { 4^{\lambda+1} \over \pi^{\lambda+1}}\,
{ \Gamma(\alpha+\half)\Gamma(\alpha)^2 \over
\Gamma(2\alpha)\Gamma(\half d -\alpha)} \, {\hat g}_{\hh(\lambda+1),
\alpha-1-\hh \lambda} (\sinh^2 \theta) \, ,
\eqno (B.11) $$
which can further be rewritten as
$$ \eqalign { {}\!\!\!\!\!\!\!\!
\bigg ( \rho(1+\rho) {\d^2 \over \d \rho^2}
+ (\rho + \half ) {\d \over \d \rho} - \mu^2 \bigg ) {\hat h}(\rho)
& = - \pi^\lambda f(\alpha) (\alpha-\half d -1) {\Gamma(\alpha-1)
\Gamma(\lambda+1)\over \Gamma(2\alpha-2) \Gamma(d-\alpha)} \cr
& \, {}\times {1\over (1+\rho)^{\lambda+1}}
F \big(\lambda+1,\alpha-1;2\alpha-2;{1\over 1+\rho}\big ) \, . \cr}
\eqno (B.12) $$
By considering the definition of the transform (4.18), or its inverse (4.19),
we can then derive an equivalent differential equation for $h$
$$ \eqalign {
\bigg ( \xi(1 & +\xi) {\d^2 \over \d \xi^2} + d(\xi + \half ) {\d \over \d \xi}
+ \lambda^2 - \mu^2 \bigg )  h(\xi) \cr
& = - f(\alpha)  (\alpha- \half d -1) {\Gamma(\alpha-1)
\Gamma(d)\over \Gamma(2\alpha-2) \Gamma(d-\alpha)} \, {1\over \xi^d}
F \big(d,\alpha-1;2\alpha-2;-{1\over \xi}\big ) \, , \cr}
\eqno (B.13) $$
which is essentially identical with an equation obtained, by very different
means, by Ohno and Okabe [11] when $\alpha = d-2$ which is appropriate for
discussion of the large $N$ limit in section 4. The homogeneous equation is
easily solved in terms of standard hypergeometric functions. The relevant
boundary conditions, which ensure a unique solution, are that as $\xi\to
\infty$, taking $\mu>0$, $h(\xi) \sim \xi^{-\mu-\lambda}, \xi^{-d}$
(with no $\xi^{\mu-\lambda}$ behaviour) and as $\xi\to 0$ then
$h(\xi) \sim 1, \xi^{\alpha-d}$ (with no terms $\propto \xi^{1-\hh d}$).

However a more direct approach is to take the inverse  Fourier transform
${\tilde{\hat{h}}} \to {\hat{h}}$ when we obtain  by contour integration
$$ \eqalign{
{\hat{h}}(\sinh^2{\theta})={4^{\lambda+1}\over \pi^{\lambda +1}}
{\Gamma(\alpha+\half)\Ga(\alpha)^2\over\Ga(2\alpha)\Ga(\half d-\alpha)}
{}& \Bigg({4 \Ga(\lambda+1) \over\Ga(\alpha-\half)\Ga(\alpha-\lambda-
{\textstyle{3\over2}})(\mu^2-(\lambda+1)^2)} G(\sinh^2 \theta)\cr
{} + &{\Ga(\half(\lambda+\mu+1))\Ga(\half(\lambda-\mu+1))\over\Ga(\alpha-1-
\half(\lambda+\mu))\Ga(\alpha-1-\half (\lambda -\mu))}{1\over\mu}
e^{-2\mu |{\theta}|} \Bigg )
\cr }
\eqno(B.14)
$$
with
$$
G(\sinh^2{\theta})=e^{-2(\lambda+1)|{\theta}|} {}_4F_3
\big (\lambda+1,B,C_+,C_-;\alpha-\half,C_+ +1,C_- +1;e^{-4|\theta|}\big ) \, ,
\eqno(B.15)
$$
where
$$
B=\lambda+{\textstyle{5\over 2}}-\alpha \, , \qquad
C_\pm =\half(\lambda+1\pm\mu)\, .
\eqno(B.16)
$$
The first term on the r.h.s of equation (B.14) represents a particular
inhomogeneous solution to (B.11) while the second term is a
solution to the homogeneous equation.
It is possible to take the inverse transform
${\hat {h}}(\rho)\to h(\xi)$ by noting that with
$\rho=\sinh^2{\theta}$ then
$$
e^{-2a|{\theta}|}=
\big(\sqrt{\rho^{\vphantom x}}+\sqrt{1+\rho}\big)^{-2a}={1\over
4^a(1+\rho)^a}F\big (a,a+\half;2a+1;{1\over 1+\rho}\big ) \, ,
\eqno(B.17)
$$
so that using (B.3) gives
$$\eqalign {
{\Ga (a)\over \Ga(a+\lambda)}&{1\over \Ga(-\lambda)}\int_\xi^\infty \!\!\!
\d\rho \, (\rho-\xi)^{-\lambda-1}
\, {1\over (1+\rho)^a}F\big (a,a+\half;2a+1;{1\over 1+\rho}\big ) \cr
={}&  {1\over (1+\xi)^{a+\lambda}}
F\big (a+\lambda,a+\half;2a+1;{1\over 1+\xi}\big ) \cr
={}& {\xi+\half\over \big [ \xi(\xi+1)\big ]^{\hh (a+\lambda+1)}}
F\big (\half(a+\lambda+1),\half(a-\lambda)+1;a+1;-{1\over 4\xi(\xi+1)}
\big ) \, . \cr}
\eqno(B.18)$$
This formula can be used for both terms present in (B.14) if
$G(\sinh^2{\theta})$, as given by (B.15), is expanded in a series
of exponentials in $\theta$.  The part of the final result for $h(\xi)$
corresponding to the $e^{-2\mu|\theta|}$ term in (B.16)
is then quite simple,
$$\eqalign {
 {4^{2C_-}\over \pi^d}
{\Gamma(\alpha+\half)\Ga(\alpha)^2\over\Ga(2\alpha)\Ga(\half d-\alpha)} &
{\Ga(C_+) \Ga(C_-) \over \Ga (\alpha-\half - C_+)\Ga (\alpha-\half - C_-)}
{\Ga(\lambda+\mu)\over \Ga(\mu+1)} \cr
&{} \times {\xi+\half\over \big [ \xi(\xi+1)\big ]^{C_+}}
F\big ( C_+, \mu+{\textstyle{3\over2}}-C_+;\mu+1 ;-{1\over 4\xi(\xi+1)}
\big ) \, , \cr}
\eqno (B.19) $$
but finding a nice form for the transform of the
piece involving the ${}_4F_3$ generalised hypergeometric function
poses some challenges, a resolution  of which is illustrated below.

After expanding ${}_4F_3$ in (B.15) and using (B.18) for $a\to \lambda+1+2n$
we may then define
$$\eqalign {
F(\xi) = {}& {\Ga(\lambda+1)\over \Ga(d)}
{1\over \Ga(-\lambda)}\int_0^\infty \!\!\! \d\rho \,
\rho^{-\lambda-1} G(\rho+\xi) \cr
= {}& {\xi+\half\over(4\xi(1+\xi))^{\lambda+1}}
\sum_{n=0}^\infty{1\over n!}
{(\lambda+1)_n(B)_n(C_+)_n(C_-)_n \over
(\alpha-\half)_n(C_++1)_n (C_-+1)_n}{(d)_{2n} \over (\lambda+1)_{2n}}\cr
& \quad \quad \quad \times {1\over 4^n (4\xi(1+\xi))^n}
\, {}_2F_1 \big (\lambda+1+n,{\textstyle{3\over2}}+n;\lambda+2+2n;-
{ 1\over 4\xi(1+\xi)}\big )\, ,\cr }
\eqno(B.20) $$
so that, in addition to (B.19), the remaining part of $h(\xi)$ becomes
$$
- {4^{\lambda+1}\over \pi^d}
{\Gamma(\alpha+\half)\Ga(\alpha)^2\over\Ga(2\alpha)\Ga(\half d-\alpha)}
{\Ga(d) \over \Ga(\alpha-\half)\Ga(\alpha-\lambda- {\textstyle{3\over2}})}
{1\over C_+ C_-} F(\xi) \, .
\eqno (B.21) $$
The result (B.20) for $F(\xi)$ may be rewritten by expanding the hypergeometric
function as
$$
F(\xi)= {\xi+\half\over(4\xi(1+\xi))^{\lambda+1}}
\sum_{N=0}^\infty {1\over N!} h_N \Big ({- {1\over 4\xi(1+\xi)}}\Big )^{\!N}\,
,
\eqno(B.22)$$
where $h_N$ is given by the finite sum
$$ \eqalign{
h_N={}& \sum_{n=0}^N{(-1)^{n}\over 4^n} \Big ( {N\atop n} \Big )
{(\lambda+1)_n(B)_n(C_+)_n(C_-)_n(d)_{2n}
(\lambda+1+n)_{N-n}({\textstyle{3\over2}}+n)_{N-n} \over (\alpha-\half)_n
(C_++1)_n (C_-+1)_n(\lambda+1)_{2n}(\lambda+2+2n)_{N-n} } \, .\cr }
\eqno(B.23)
$$
By using identities for the $\Ga$ function, such as
$(d)_{2n}=4^n (\half d)_n (\lambda+1)_n$,  and (B.16) we can
write this in the compact form
$$\eqalign{
h_N={({\textstyle{3\over 2}})_N (\lambda+1)_N \over
(\lambda+2)_N }\, {}_7F_6\big (\lambda+1,& B,C_+,C_-,
\lambda+\half, \half\lambda+\textstyle{3\over2},-N;\cr
&  \alpha-\half ,C_+ + 1,C_-+1,\half\lambda+\half,{\textstyle{3\over2}},
\lambda+2+N; 1\big ) \, . \cr }
\eqno(B.24)
$$
All that remains is to find a closed expression for the sum of the
terminating series represented by
the ${}_7F_6$  generalized hypergeometric function. For the present
this remains elusive for general $\alpha$.
However, if we take $\alpha=d-2$, as required for section 4, then
$\mu=\hh(5- d)$ so that $C_+={\textstyle{3\over2}}$ and $C_-=\lambda-\half$.
Consequently the ${}_7F_6$ series reduces to a finite
${}_5F_4$ series which is summable by a special case of Dougall's theorem [27]
$$ \eqalign {
{}_5F_4\big (\lambda+1,{\textstyle{7\over2}}-\lambda,\lambda-\half,
\half\lambda+\textstyle{3\over2},& -N;
d-{\textstyle{5\over2}},{\textstyle{5\over2}},\half\lambda+\half,
\lambda+2+N ; 1\big )\cr
& {} = {(\lambda+2)_N (\lambda-\half)_N \over (d-{\textstyle{5\over2}})_N
({\textstyle{5\over2}})_N}\, , \cr}
\eqno(B.25)
$$
so that in this case
$$
F(\xi)= {\xi+\half\over\big [4\xi(1+\xi)\big ]^{\lambda+1}}\,
{}_3F_2\big ( \lambda+1,\lambda-1,{\textstyle{3\over2}};
d-{\textstyle{5\over2}},{\textstyle{5\over2}}; -{1 \over 4\xi(1+\xi)}\big ) \,
{}.
\eqno(B.26)
$$
Hence, with $f$ defined in (4.4), the final result becomes
$$ \eqalign{
 h(\xi)={}f(d-2)&\Ga(\half d-1) \bigg \{
{32\over 3} (6-d){\Ga(\lambda+1)\over \Ga (d-{\textstyle{5\over 2}})}\,
F(\xi) \cr
&{}+ {\sin \pi \lambda\over \lambda-2} \, {\Ga(\lambda-\half)\over \Ga (d-3)}\,
{8\over
(1+2\xi)^2} F\big ( {\textstyle{3\over 2}},1;3-\lambda;{1\over (1+2\xi)^2}\big
)
\bigg \} \, . \cr}
\eqno(B.27) $$

The form (B.27), with (B.26), for $h(\xi)$ is not appropriate for considering
the limit $\xi \to 0$. However by using the relation\footnote{*}{This does not
seem to appear in standard references but related results were found long
ago [28]. The particular result may be derived by multiplying by
$(-z)^\rho$ and differentiating when it reduces to a well know result for
standard hypergeometric functions, the constant of integration is determined
by taking $z=1$.}
$$\eqalign{
{}_3F_2\big (&a,b, \rho;c,\rho+1;- z\big ) \cr
= {}& \Ga(\rho+1){\Ga(a-\rho)\Ga(b-\rho) \Ga(c) \over
\Ga(a) \Ga(b) \Ga(c-\rho)}\, {1\over z^\rho} \cr
{}&- {\Ga(c)\Ga(b-a)\over \Ga(b)\Ga(c-a)}{\rho\over a-\rho}\, z^{-a}
{}_3F_2\big (a,a+1-c,a-\rho;a+1-b,a+1-\rho;-{1\over z}\big )\cr
{}&- {\Ga(c)\Ga(a-b)\over\Ga(a)\Ga(c-b)} {\rho\over b-\rho}\, z^{-b}
{}_3F_2\big (b,b+1-c,b-\rho;b+1-a,b+1-\rho;-{1\over z}\big )  \, ,\cr }
\eqno(B.28) $$
we may find an alternative expression for $F(\xi)$,
$$\eqalign{
F(\xi)& =\Ga({\textstyle{5\over2}}){\Ga(d -{\textstyle{5\over2}})
\over \Ga(d-4)}
{\Ga(\lambda-\half) \Ga(\lambda-{\textstyle{5\over2}}) \over
\Ga(\lambda-1)\Ga(\lambda+1)}{\xi+\half\over\big [4\xi(1+\xi)
\big ]^{\lambda-\hh}}\cr
&\ + {\Ga(d -{\textstyle{5\over2}})\over \Ga (\lambda+1) \Ga(\half d-1)}\,
{3\over 6-d}\, {\xi+\half\over\big [4\xi(1+\xi)\big ]^2}\Big ( 1 +
(d-3)(6-d) \xi(\xi+1) \Big ) \cr
&\ + {\Ga(d -{\textstyle{5\over2}})\over \Ga (\lambda-1) \Ga(\half d-3)}
\, {3\over d-2}\, \half (\xi+\half) \cr
&\ \ \ \times \Big ( \ln 4\xi(\xi+1) \,
{}_3F_2\big (\lambda+1,4-\half d,\half d-1;3,\half d; - 4\xi(\xi+1) \big )
{} + {\tilde F}(\xi) \Big )  \, , \cr}
\eqno(B.29) $$
where ${\tilde F}(\xi)$ is given by a power series in $\xi$, analogous to
${\tilde h}(\xi)$ in (B.8). With this form the behaviour for $\xi\sim 0$ is
manifest. It can also be readily shown that
$$\eqalign{
{1\over (2\xi+1)^2}F\big({\textstyle{3\over2}},1;3-\lambda;
{1\over (2\xi+1)^2}\big)={}& {2\sqrt \pi\over \sin \pi \lambda}
{\Ga(\lambda-\half)\over \Ga(\lambda-2)} \,
{2\xi+1\over\big [4\xi(1+\xi)\big ]^{\lambda-\hh}}  \cr
& {}+ {\lambda-2\over \lambda-\half}
F\big (\lambda-1,1;\lambda+\half;-4\xi(\xi+1)\big ) \, . \cr}
\eqno(B.30)
$$
Hence in (B.27) the first terms on the r.h.s.'s of (B.29,30) cancel and
$$\eqalign{
h(\xi)\sim {}& f(d-2) {2\xi+1\over \big [\xi(\xi+1)\big]^2} \Big ( 1 +
(d-3)(6-d) \xi(\xi+1) \Big ) \cr
& + f(d-2) {(d-1)(d-3)(d-4)(d-6)^2\over d-2}(\xi+\half)
\ln\big [4\xi(\xi+1)\big ]^{-1} \, . \cr}
\eqno(B.31)
$$
\bigskip
\leftline{\bigbf Appendix C}
\medskip

In order to justify (5.11), and also determine the appropriate form for
the derivative operator which is exhibited in (5.8) and plays a crucial
role in the OPE, we consider first the Fourier transform
$$ \eqalign {
\int \! \d^d r \, {1\over (r^2 + \mu^2)^\eta} \, e^{i p{\cdot r}} = {}&
{2 \pi^{\hh d} \over \Ga (\eta)} \, \Big ( {|p| \over 2\mu }
\Big )^{\! \eta - \hh d} K_{\eta - \hh d} \big (\mu |p|\big ) \cr
={}& { \pi^{\hh d} \over \Ga (\eta)} \bigg ( \Ga (\half d -\eta)
\big (\quar p^2\big )^{\eta - \hh d}
F_1\big (\eta - \half d + 1 ; \quar \mu^2 p^2 \big ) \cr
&\qquad \quad + \Ga ( \eta - \half d ) \big (\mu^2\big )^{\hh d -\eta}
F_1\big (\half d -\eta + 1 ; \quar \mu^2 p^2 \big )\bigg ) \, , \cr}
\eqno (C.1) $$
where $F_1(\alpha;z) = \sum_n z^n/(n! (\alpha)_n)$ is defined by series
expansion. Hence we may write
$$ \eqalign { \!\!\!\!\!
\int \! \d &^d r \,{1\over (x-r)^{2\eta_1} (x'-r)^{2\eta_2}}
e^{i p{\cdot r}} \cr
= {}&{1\over B(\eta_1,\eta_2)} \int_0^1 \!\!\! \d \alpha \, \alpha^{\eta_1-1}
(1-\alpha)^{\eta_2-1}  e^{i p{\cdot (\alpha x + (1-\alpha) x')}} \!
\int \! \d^d r \, {1\over \big (r^2 + \alpha(1-\alpha) s^2 \big )^\eta}
e^{i p{\cdot r}} \cr
={}& {1\over B(\eta_1,\eta_2)} { \pi^{\hh d} \over \Ga (\eta)}
\int_0^1 \!\!\! \d \alpha \, \alpha^{\eta_1-1}
(1-\alpha)^{\eta_2-1}   e^{i \alpha p{\cdot s}+i p{\cdot x'}} \cr
& \quad \quad \times \bigg \{ \Ga (\half d -\eta)
\big (\quar p^2\big )^{\eta - \hh d}
F_1\big (\eta - \half d + 1 ; \quar \alpha(1-\alpha) s^2 p^2 \big ) \cr
& \quad \quad \quad  + \Ga ( \eta - \half d )
\big ( \alpha(1-\alpha) s^2 \big )^{\hh d -\eta}
F_1\big (\half d -\eta + 1 ;  \quar \alpha(1-\alpha) s^2  p^2 \big )
\bigg \}  \cr
= {}& C^{\eta,\eta_1-\eta_2}(s,ip)
\int \! \d^d r \, {1\over (x'-r)^{2\eta}} \, e^{i p{\cdot r}} \cr
& {}+ \pi^{\hh d} {B(\hh d -\eta_1,\hh d -\eta_2)\over B(\eta_1,\eta_2)}
{\Ga (\eta - \hh d)\over \Ga (\eta)}
\big (s^2 \big )^{\hh d -\eta} C^{d-\eta,\eta_2-\eta_1}(s,ip) \,
e^{i p{\cdot x'}} \, , \cr}
\eqno (C.2) $$
where $s=x-x'$ and $\eta = \eta_1+\eta_2$ and $C^{a,b}(s,ip)$ is defined by
(5.8). It is easy to see that this is then equivalent to (5.11).
\bigskip
\leftline{\bigbf{Appendix D}}
\medskip

For the calculations in section 5 it is necessary to evaluate conformally
invariant integrals over $\bR_+^d$. Techniques for dealing with conformal
integrals on all $\bR^d$ are well known [29]. Here we show how the methods
of section 4 can be used to calculate the form of the integral in (5.26)
although a general discussion will be given elsewhere [30]. For $x=(y,\bx)$
and $x'=(y',\bx')$ the integrals to be considered are of the basic form
$$
F(\xi) = \int_0^\infty\!\!\! \d z\int \! \d^{d-1}\br \,
{1\over (2z)^d} F_1(\txi) F_2(\txi')\, , \quad \txi = {(x-r)^2\over 4yz}\, ,
\quad \txi' = {(x'-r)^2\over 4y'z} \, , \quad r=(z,\br) \, ,
\eqno (D.1) $$
where restricted conformal invariance guarantees that the integral is a
function
of just the invariant $\xi=(x-x')^2/4yy'$. This integral can be simplified
by integrating over $\bx$. Hence, considering the sequence of transformations
$F(\xi)\to{\hat F}(\rho)\to {\tilde {\hat F}}(k)$ defined in (4.18) and (4.25),
we obtain
$$ \eqalign {
{\hat F}(\rho) = {}& \int_0^\infty\!\!\! \d z \, {1\over 2z}
{\hat F}_1({\tilde \rho}){\hat F}_2({\tilde \rho}') \, , \quad
{\tilde \rho} = {(y-z)^2\over 4yz}   \, , \quad
{\tilde \rho}' = {(y'-z)^2\over 4y'z} \, , \cr
&{\hat F}(\sinh^2 \theta) =  {1\over 2\pi} \int \! \d k \, e^{-ik\theta}
{\tilde {\hat \F}}_1(k) {\tilde {\hat F}}_2(k) \, . \cr}
\eqno (D.2) $$
If ${\hat F}(\rho)$ can be determined then $F(\xi)$ can be recovered by
using (4.19). For our purposes it is necessary to extend this method
to deal with integrals which transform as conformal tensors. For illustration
we first consider
$$
F(\xi)\Big ( X_\mu X_\nu - {1\over d} \delta_{\mu\nu} \Big )
= \int_0^\infty\!\!\! \d z\int \! \d^{d-1}\br \,{1\over (2z)^d}
\Bigl ( \tX_\mu\tX_\nu - {1\over d} \delta_{\mu\nu} \Big )
F_1(\txi) F_2(\txi')\, ,
\eqno (D.3) $$
for $\tX_\mu =y \big[\txi(1+\txi)\big]^{-\hh}\pr_\mu \txi$ a conformal vector
of scale $0$ at $x$.
Again the form of the integral is dictated by conformal invariance to be
given in terms of the single function $F(\xi)$.
To reduce (D.3) to the previous case we introduce the differential operator
$$
{\tilde \D}_{\mu\nu} = \pr_\mu \pr_\nu + {1\over y}(n_\mu\pr_\nu +
n_\nu \pr_\mu ) - {1\over d} \delta_{\mu\nu}\Big ( \pr^2 + {2\over y} n{\cdot
\pr} \Big ) \, ,
\eqno (D.4) $$
which is constructed to give
$$
{\tilde \D}_{\mu\nu} \F(\xi) = {1\over y^2} \Big ( X_\mu X_\nu
- {1\over d} \delta_{\mu\nu} \Big ) \xi (1+\xi) \F''(\xi) \, .
\eqno (D.5) $$
If we now set
$$
F(\xi) = 4 \xi (1+\xi) \F''(\xi)
\eqno (D.6) $$
then to evaluate (D.3) it is sufficient to calculate
$$
\F(\xi) = \int_0^\infty\!\!\! \d z\int \! \d^{d-1}\br \, {1\over (2z)^d}
\F_1(\txi) F_2(\txi') \, ,
\eqno (D.7) $$
where $\F_1$ is given in terms of $F_1$ similarly to (D.6). From (D.6) and
(4.18) we can compute the transform ${\hat \F}$ without explicitly finding
$\F$ by solving (D.6) since
$$
{\hat \F}(\rho)= {\pi^\lambda \over \Gamma (\lambda+2)}
\int_0^\infty \!\!\!\! \d u \, u^{\lambda+1}\F''(u+\rho) \, .
\eqno (D.8) $$
For integrals involving the energy momentum tensor we are interested in
integrals for which $F_1\to F_T$ where
$$
F_T(\xi) =  \big [ \xi(1+\xi)\big ]^{-\hh d}\, .
\eqno (D.9) $$
In this case the required transforms are particularly simple
$$
{\hat \F}_T(\sinh^2\theta) = \half S_d \, {1\over d(d+1)} e^{-(d+1)|\theta|}
\, , \quad {\tilde {\hat \F}}_T(k) = {1\over d} S_d {1\over k^2 + (d+1)^2} \, .
\eqno (D.10) $$

It remains to treat integrals of the form given by (5.26) for which we
consider the following expression
$$
\G_{\mu\nu\si\rho} = \int_0^\infty\!\!\! \d z\int \! \d^{d-1}\br \,
{1\over (2z)^d} \Bigl ( \tX_\mu\tX_\nu - {1\over d} \delta_{\mu\nu} \Big )
\Big ( \tX'{}_{\!\si} \tX'{}_{\! \rho} - {1\over d} \de_{\si\rho} \Big )
F_T(\txi)H(\txi') \, .
\eqno (D.11) $$
We now write using (D.5)
$$
\G_{\mu\nu\si\rho} = (4yy')^2{\tilde \D}_{\mu\nu} {\tilde \D}'{}_{\!\si\rho}
\G(\xi)
\eqno (D.12) $$
where
$$
\G(\xi) = \int_0^\infty\!\!\! \d z\int \! \d^{d-1}\br \, {1\over (2z)^d}
\F_T(\txi) \H(\txi') \, , \quad 4 \xi(1+\xi) \H''(\xi) = H(\xi) \, .
\eqno (D.13) $$
Applying the transformations as described earlier we easily find
$$
{\hat \G}(\sinh^2 \theta) = {1\over 2\pi} \int \! \d k \, e^{-ik\theta}
{\tilde {\hat \F}}_T(k) {\tilde {\hat \H}}(k) \, .
\eqno (D.14) $$
The result in (D.10) for ${\tilde {\hat \F}}_T(k)$ then allows us to write
$$
\Big ( (d+1)^2 - {\d^2\over \d \theta^2} \Big ) {\hat \G}(\sinh^2\theta)
= {1\over d}S_d{\hat \H}(\sinh^2\theta) \, ,
\eqno (D.15) $$
which in turn translates into
$$
\Big ( \xi(1+\xi){\d^2\over \d\xi^2} + d(\xi+\half){\d \over \d\xi}
- d \Big ) \G(\xi) = - {1\over 4d}S_d \H(\xi) \, .
\eqno (D.16) $$
This equation is more conveniently written as
$$
\Big ( \xi(1+\xi){\d^2\over \d\xi^2} + (d+4)(\xi+\half){\d \over \d\xi}
+ d+2 \Big ) \G''(\xi) = - {1\over d}S_d {1\over 16\xi(1+\xi)}H(\xi) \, .
\eqno (D.17) $$

To evaluate (D.13) we use (2.17) and
$$ \eqalign{
{\tilde \D}_{\mu\nu}  X'{}_{\! \si} = {}& {1\over 4y^2} \, {2\xi+1\over
\xi(1+\xi)} \Big(  X_\mu I_{\nu\si}(s)
+ X_\nu I_{\mu\si}(s)  - {2\over d} \delta_{\mu\nu} X'{}_{\! \si} \Big ) \cr
& - {1\over 4y^2} {4\xi+1\over \xi(1+\xi)}  \Bigl (
X_\mu X_\nu - {1\over d} \delta_{\mu \nu} \Bigl)  X'{}_{\! \si} \cr}
\eqno (D.18) $$
to obtain
$$ \eqalign {\!\!\!\!\!\!
\G_{\mu\nu\si\rho}= {}& \I_{\mu\nu,\si\rho}(s)\, c(\xi) +  \Bigl (
X_\mu X_\nu - {1\over d} \delta_{\mu \nu} \Bigl)  \Bigl (
X'{}_{\!\si} X'{}_{\!\rho} - {1\over d} \delta_{\si\rho} \Bigl) a(\xi) \cr
& {} + \Bigl ( X_\mu X'{}_{\! \si} I_{\nu\rho}(s)
+ \mu \leftrightarrow \nu , \si \leftrightarrow \rho \cr
& {}\quad\quad  - {4\over d}
\delta_{\mu\nu} X'{}_{\! \si} X'{}_{\!\rho} - {4\over d} \delta_{\si\rho}
X_\mu X_\nu + {4\over d^2} \delta_{\mu\nu} \delta_{\si\rho} \Bigl ) b(\xi)
\, , \cr}
\eqno (D.19) $$
where the three invariant functions are given by
$$ \eqalign {
c(\xi)={}& 8\G''(\xi) \, , \qquad b(\xi) = -8(1+\xi)\Big ( 1+\xi{\d\over \d\xi}
\Big ) \G''(\xi) \, , \cr
a(\xi)={}& 16(1+\xi)^2\Big ( 2 + 4\xi{\d\over \d\xi} + \xi^2{\d^2\over \d\xi^2}
\Big ) \G''(\xi) \, . \cr}
\eqno (D.20) $$
An important test on the results (D.19,20) is provided by the conservation
equation which may be obtained by using (2.25) (with the arbitrary constant
$c=0$) giving
$$ \eqalign { \!\!\!\!\!
\pr_\mu {1\over (2y)^d} \G_{\mu\nu\si\rho} =  - {d-1\over d^2}S_d
{1\over (2y)^{d+1}} {v\over \xi}
\bigg \{& {- \Big (} I_{\nu\si}(s) X'{}_{\! \rho} +
I_{\nu\rho}(s) X'{}_{\! \si} - {2\over d} X_\nu \de_{\si\rho} \Big ) H(\xi) \cr
{}+ 2X_\nu &\Big ( X'{}_{\!\si} X'{}_{\!\rho} - {1\over d} \delta_{\si\rho}
\Big ) \big ( H(\xi) + \xi(1+\xi) H'(\xi) \big ) \bigg \} \, . \cr}
\eqno (D.21) $$
This gives two differential equations relating $a,b,c$ in (D.19) to $H$
which, with the results (D.20), are equivalent to (D.17).

For presenting the results it is more convenient to define from (2.37,38)
$$\eqalign {
\C(\xi)={}& - \half c(\xi) - b(\xi) = 4(1+2\xi) \G''(\xi) + 8(1+\xi)\xi
{\d\over \d\xi} \G''(\xi)\, , \cr
\A(\xi)={}& {1\over d^2}(d-1) \big ( (d-1)(a(\xi) + 4b(\xi)) + dc(\xi)\big )\cr
= {}& - {8\over d}(d-1)^2 (1+\xi)\xi \Big ( ( 2\xi+1){\d\over \d\xi}
+ 2 \Big ) \G''(\xi) \cr
&  + {8\over d}(d-1) \G''(\xi) - {1\over d^3}(d-1)^2 S_d H(\xi) \, . \cr}
\eqno (D.22) $$

For our applications we consider for $H(\xi)$ functions of the form
$$
H(\xi)={1\over \big [\xi(1+\xi)\big ]^{\alpha}} \, .
\eqno (D.23) $$
{}From the definition of $\H(\xi)$ in terms of $H(\xi)$ in (D.13) and using
(D.8) we find
$$
{\tilde{\hat \H}}(k)=\pi^{\hh d}{\Ga(\lambda +{\ts{3\over 2}} -\alpha) \over
\Ga(1+\alpha)}4^{\alpha-\lambda - {5\over 2}} {\tilde{\hat g}}_{\alpha-
\hh\lambda ,\hh(\lambda+3)}(k) \,.
\eqno (D.24) $$
With this result and (D.10) it is sufficient to
find the inverse transform of functions of the form
$$
{\tilde{\hat I}}(k)={{\tilde{\hat g}}_{a,b}(k) \over {\ts{1\over 16}}
k^2+\mu^2} \, , \qquad
\tilde{\hat g}_{a,b}(k) = {\Gamma (a-{\ts{i\over 4}}k)
\Gamma (a+{\ts{i\over 4}}k) \over \Gamma (b-{\ts{i\over 4}}k)
\Gamma (b+{\ts{i\over 4}}k) } \, ,
\eqno(D.25)
$$
where we are ultimately interested in taking $\mu=\quar(d+1)$,
$a=\alpha-\half \lambda$ and $b=\half (\lambda + 3)$.
By contour integration it is straightforward to obtain
$$\eqalign {
{\hat I}(\sinh^2\theta)={}& {1\over 2\pi} \int \! \d k \, e^{-ik\theta}
{\tilde{\hat I}}(k)  \cr
= {}& {4\Ga(2a) \over\Ga(b-a)\Ga(b+a)(\mu^2-a^2)}\cr
{}& \times \sum_{n=0} {1\over n!}{(2a)_n(1+a-b)_n
(a-\mu)_n(a+\mu)_n \over (a+b)_n(1+a-\mu)_n(1+a+\mu)_n}
e^{-4(a+n){|\theta|}} \cr
{}& + {\Ga(a- \mu)\Ga(a+\mu) \over \Ga(b- \mu)
\Ga(b+\mu)}{2\over\mu}e^{-4\mu{|\theta|}}  \, . \cr}
\eqno(D.26) $$
For the final inverse transform, we observe that it is only
necessary to find $I''(\xi)$ which is given by inverting (D.8)
$$
I''(\xi)={1\over \pi^\lambda \Ga(-\lambda-2)} \int_0^\infty \!\! \d\rho\,
\rho^{-\lambda-3} {\hat I}(\rho+\xi)\, .
\eqno(D.27) $$
This simplifies the calculation significantly.
{}From (B.17) and (B.18) it follows that
$$\eqalign { \!\!\!\!
I''(\xi)={}&{1\over \pi^\lambda 4^{2a-1}} {\Ga(2a+\lambda+2) \over  \Ga(b-a)
\Ga(b+a)(\mu^2-a^2)} {1 \over [\xi(1+\xi)]^{a+\hh\lambda+1}}\sum_{N=0}{h_N
\over N!} {(-1)^N \over [4 \xi(1+\xi)]^N}\cr
{}& + {1\over \pi^\lambda 4^{2\mu-1}}
{\Ga(a-\mu)\Ga(a+\mu) \over \Ga(b-\mu)\Ga(b+\mu)}
{\Ga(2 \mu + \lambda+2) \over \Ga(2\mu+1)} \cr
{}&\ \ \times {1\over \big [ \xi(1+\xi)\big ]^{\mu+\hh\lambda+1}}
F\big(\mu+\half\lambda+1,\mu-\half\lambda-\half; 2\mu+1;
-{1\over4\xi(1+\xi)}\big ) \, ,\cr
h_N ={}&{(a-\half\lambda-\half)_N(a+\half\lambda+1)_N \over (1+ 2a)_N} \cr
{}& \times \sum_{n=0}^{N} {1\over n!}{(2a)_n(1+a-b)_n (a-\mu)_n
(a+\mu)_n (a+\half \lambda+{\ts{3\over 2}})_n(-N)_n \over
(a+b)_n(1+a-\mu)_n(1+a+\mu)_n (a- \half\lambda-\half)_n(1+2a+N)_n} \, .\cr}
\eqno(D.28) $$
Now if we use the fact that $b=\half(\lambda+3)$ then the series in
$h_N$ simplifies to a terminating ${}_5F_4(\, . \, ;\, .\,; 1)$  series
and may be summed exactly once again by Dougall's theorem [27] giving
$$
h_N={(1)_N(a-\half\lambda-\half)_N(a+\half\lambda+1)_N \over
(1+a-\mu)_N(1+a+\mu)_N }\, .
\eqno(D.29) $$
With this result the  series in (D.28) can be written exactly as
a ${}_3F_2$ hypergeometric  function.
Putting it all together we obtain for $H$ given by (D.23) with general $\alpha$
$$\eqalign { \!\!
\G''(\xi)={}& {S_d \over 32 d}{1\over (d-2\alpha)(\half + \alpha)}
{1\over [\xi(1+\xi)]^{1+\alpha}} \cr
{}& \times \bigg \{
{}_3F_2\big (1,1+\alpha,\alpha-\half d;
{\ts {3\over 2}} +\alpha,1+\alpha-\half d;-{{1\over 4\xi (1+\xi) }}\big )\cr
{}&\quad \ - {\Ga(1+\half d-\alpha)\Ga(1+\alpha-\half d)\Ga(1+\half d)
\Ga({\ts {3\over 2}} +\alpha) \over \Ga(1+\alpha)\Ga({\ts {3\over 2}}+\half d)}
\big [4\xi(1+\xi)\big ]^{\alpha-\hh d} \bigg \} \, . \cr }
\eqno(D.30) $$
Note that when $\alpha = \half d$ the two terms cancel so that there is no
pole. Also from (B.28) $\G''(\xi)$ has a leading behaviour for $\xi\to 0$
with terms $\propto \xi,\xi^{1+\alpha}$. We have checked that (D.30) satisfies
(D.17) with the last term representing a solution of the homogeneous equation.

In order to evaluate the integral in (5.26) we need to take, for the
ordinary case corresponding to (5.29a),
$$
H(\xi)_{\rm ord} = {N\over S_d^{\, 3}}{4d A_{\lambda,{\rm ord}}\over (d-1)^2}\,
{1\over \big [\xi(1+\xi)\big ]^{d-2}} \, .
\eqno (D.31) $$
Applying (D.30) gives then
$$\eqalign {
\G''(\xi)_{\rm ord}=-{N\over S_d^{\, 2}}&
{d-2 \over 8(d-1)^2(d-{\ts{3\over 2}})}
{1\over \big [\xi(1+\xi) \big ]^{d-1}}  \cr
\times &  \bigg \{
{}_3F_2\big (1,d-1,\half d-2;d-\half,\half d-1; -{1\over 4\xi(1+\xi)} \big )
\cr
{}& - {\Ga(\half d -1)\Ga(3-\half d)\Ga(\half d +1)\Ga(d-\half) \over
\Ga(d-1) \Ga(\half d + {\textstyle {3\over 2}} )}
\big [4\xi(1+\xi)\big ]^{\hh d-2}\bigg \} \, .\cr }
\eqno (D.32) $$
For $\C,\A$ defined as in (D.22) we have
$$\eqalign {
\C(\xi)_{\rm ord} = {N\over S_d^{\, 2}}& {d-2 \over (d-1)^2}
{2\xi+1\over \big [\xi(1+\xi) \big ]^{d-1}}  \cr
\times &  \bigg \{
{}_3F_2\big (1,d-1,\half d-2;d-{\textstyle {3\over 2}},\half d-1
; -{1\over 4\xi(1+\xi)} \big ) \cr
& -{\Ga(\half d -1)\Ga(3-\half d)\Ga(\half d +1)\Ga(d-{\textstyle {3\over 2}})
\over \Ga(d-1) \Ga(\half d + \half )}
\big [4\xi(1+\xi)\big ]^{\hh d-2}\bigg \} \cr }
\eqno (D.33) $$
and
$$\eqalign { \!\!\!\!\!
\A(\xi)_{\rm ord} =-& {N\over S_d^{\, 2}} {8\over d^2}(d-2)^2
{1\over \big [\xi(1+\xi) \big ]^{d-2}}\cr
 - & {N\over S_d^{\, 2}} {2 \over d-1}
{1\over \big [\xi(1+\xi) \big ]^{d-1}}  \cr
\times &  \bigg \{
{}_3F_2\big (1,d-1,\half d-2;d-{\textstyle {3\over 2}},\half d
; -{1\over 4\xi(1+\xi)} \big ) \cr
& -{\Ga(\half d)^2\Ga(3-\half d)\Ga(d-{\textstyle {3\over 2}})
\over \Ga(d-1) \Ga(\half d - \half )}
\big [4\xi(1+\xi)\big ]^{\hh d-1}\Big ( 1 + {d\over d-1}{1\over4\xi(1+\xi)}
\Big ) \bigg \} \, .\cr }
\eqno (D.34) $$
\bigskip
\leftline{\bigbf{Appendix E}}
\medskip

The crucial step in deriving (7.20) is to obtain, with
${\hat D}_{\si\rho}(y,\pr)$ defined
as in (7.16), the result for arbitrary $H(\xi)$
$$ \eqalignno {
{\hat D}_{\si\rho}(y',\pr')
\Big ({1\over (4yy')^d} X_\mu & X_\nu H(\xi) \Big ) \cr
= {1\over(d^2-1)}{1\over (4yy')^d}
\bigg \{ & \I_{\mu\nu,\si\rho}(s) {1\over 2\xi(\xi+1)}H(\xi) \cr
{} + {}& \half\Big ( X_\mu X'{}_{\! \si} I_{\nu\rho}(s)
+ \mu \leftrightarrow \nu , \si \leftrightarrow \rho
- {4\over d} \delta_{\si\rho} X_\mu X_\nu \Big )
\Big ( {1\over \xi+1} H(\xi) - H'(\xi) \Big ) \cr
{} + {}& X_\mu X_\nu \Big (
X'{}_{\!\si} X'{}_{\!\rho} - {1\over d} \delta_{\si\rho} \Big )\Big (
\xi(\xi+1) H''(\xi) + 2 H'(\xi) - {2\over \xi + 1} H(\xi) \Big ) \cr
{} - {}& X_\mu X_\nu \Big (n_{\si} n_{\rho} - {1\over d} \delta_{\si\rho} \Big
)
\Big ( \xi(\xi+1) H''(\xi) + d(\xi+\half)H'(\xi) \cr
& \qquad \qquad \qquad \qquad \qquad \qquad - d H(\xi) - \half d
{1\over \xi(\xi+1)} H(\xi) \Big ) \cr
{} - {}& \de_{\mu\nu} \Big (n_{\si} n_{\rho} - {1\over d}\delta_{\si\rho}
\Big ) {1\over 2\xi(\xi+1)}H(\xi) \bigg \} \, . & (E.1) \cr}
$$
The terms dependent on $n_\si n_\rho$ violate conformal invariance but they
disappear on removing the trace in $\mu\nu$ and
$H(\xi) \to [\xi (\xi+1)]^{-\hh d}$ as appropriate for application in (7.19).
\vfill \eject
\leftline {\bigbf References}
\parskip 0pt
\vskip 15pt
\item {[1]} J. Cardy, {\it in} ``Phase Transitions and Critical Phenomena'',
vol. 11, p. 55 (C. Domb and J.L. Lebowitz eds.) Academic Press, London (1987);
\item{} J. Cardy, {\it in} ``Champs, Cordes et Ph\'enom\`enes Critiques",
(E. Br\'ezin and J. Zinn-Justin eds.) North Holland, Amsterdam 1989.
\vskip 6pt
\item {[2]} P. Ginsparg, {\it in} ``Champs, Cordes et Ph\'enom\`enes
Critiques", (E. Br\'ezin and J. Zinn-Justin eds.) North Holland, Amsterdam
1989.
\vskip 6pt
\item{[3]} H. Osborn and A. Petkou, {\it Ann. Phys. (N.Y.)}, {\bf 231} (1994)
311.
\vskip 6pt
\item{[4]} A.N. Vasil'ev, Yu.M. Pis'mak and Yu.R. Khonkenen, {\it Theoretical
and Mathematical Physics}, {\bf 46} (1981) 104; {\bf 47} (1981) 465;
{\bf 50} (1982) 127;
\item{}
A.N. Vasil'ev and M.Yu. Nalimov, {\it Theoretical
and Mathematical Physics}, {\bf 55} (1983) 423.
\vskip 6pt
\item{[5]} H.W. Diehl, in {\it Phase Transitions and Critical Phenomena},
vol 10, p. 75, (C. Domb and J.L. Lebowitz eds.) Academic Press,
London (1986).
\vskip 6pt
\item{[6]} J.L. Cardy, {\it Nucl. Phys.}, {\bf B240} [FS12] (1984) 514.
\vskip 6pt
\item{[7]} E. Eisenriegler, ``Polymers Near Surfaces'', World Scientific,
Singapore (1993).
\vskip 6pt
\item{[8]} J.L. Cardy, {\it Phys. Rev. Lett.}, {\bf 65} (1990) 1443.
\vskip 6pt
\item{[9]} D.M. McAvity and H. Osborn, {\it Nucl. Phys.}, {\bf B406} [FS]
(1993)
655.
\vskip 6pt
\item{[10]} A.J. Bray and M.A. Moore, {\it Phys. Rev. Lett.}, {\bf 38}
(1977) 735; {\it J. Phys. A}, (1977) 1927.
\vskip 6pt
\item{[11]} K. Ohno and Y. Okabe, {\it Phys. Lett. A}, {\bf 95}, (1983) 41;
{\bf 99} (1983) 54; {\it Prog. Theor. Phys.}, {\bf 70} (1983) 1226.
\vskip 6pt
\item{[12]} H.W. Diehl, and S. Dietrich, {\it Zeit. f. Phys. B}, {\bf 42}
(1981) 65, (E) {\bf 43} (1981) 281.
\vskip 6pt
\item{[13]} G. Mack and A. Salam, {\it Ann. Phys. (N.Y.)}, {\bf 53} (1969) 174.
\vskip 6pt
\item{[14]} A. Petkou, preprint DAMTP 94/12, hep-th/9410093.
\vskip 6pt
\item{[15]} D.Z. Freedman, K. Johnson and J.I. Latorre, {\it Nucl. Phys.},
{\bf B371} (1992) 353.
\vskip 6pt
\item{[16]} G. Gompper and H. Wagner, {\it Zeit. f. Phys. B}, {\bf 59} (1985)
193.
\vskip 6pt
\item{[17]} E. Br\'ezin and S.R. Wadia (eds.), ``The Large N Expansion in
Quantum Field Theory and Statistical Mechanics'', World Scientific, Singapore
(1993).
\vskip 6pt
\item{[18]}
K. Lang and W. R\"uhl, {\it Z. Phys. C}, {\bf 50} (1991) 285; {\bf 51} (1991)
127; {\it Nucl. Phys.}, {\bf B377} (1992) 371; {\bf B400}[FS] (19920 597;
{\bf B402} (1993) 573; {\it Z. Phys. C}, {\bf 61} (1994) 495; {\bf 63} (1994)
531;
\item{} A.N. Vasil'ev, and A.S. Stepanenko, {\it Theoretical
and Mathematical Physics}, {\bf 95} (1993) 160.
\vskip 6pt
\item{[19]} J.S. Reeve and A.J. Guttmann. {\it Phys. Rev. Lett.}, {\bf 45}
(1980) 1581.
\vskip 6pt
\item{[20]} H.W. Diehl, and S. Dietrich, {\it Zeit. f. Phys. B},
{\bf 50} (1983) 117.
\vskip 6pt
\item{[21]} D.M. McAvity and H. Osborn, {\it Nucl. Phys.},
{\bf B394} (1993) 728.
\vskip 6pt
\item{[22]} I.M. Gel'fand, M.I. Graev and N.Ya. Vilenkin, ``Generalised
Functions'', vol. 5,
Academic Press, New York and London (1966).
\vskip 6pt
\item{[23]} E. Eisenriegler, M. Krech and S. Dietrich,
{\it Phys. Rev. Lett.}, {\bf 70} (1993) 619, E, 2051.
\vskip 6pt
\item{[24]} J.L. Cardy, {\it Nucl. Phys.}, {\bf B275} [FS17] (1986) 200;
{\bf B324} (1989) 581;
\item{} J.L. Cardy and D.C. Lewellen, {\it Phys. Lett.}, {\bf 259B} (1991)
274;
\item{} D.C. Lewellen, {\it Nucl. Phys.}, {\bf B372} (1992) 654.
\vskip 6pt
\item{[25]} T.W. Burkhardt, E. Eisenriegler and I. Guim, {\it Nucl. Phys.},
{\bf B316} (1989) 559;
\item{} T.W. Burkhardt and T.Xue, {\it Phys. Rev. Lett.}, {\bf 66} (1991)
895; {\it Nucl. Phys.}, {\bf B354} (1991) 653;
\item{} T.W. Burkhardt and E. Eisenriegler, {\it Nucl. Phys.}, {\bf B424}
[FS] (1994) 487.
\vskip 6pt
\item{[26]} E. Eisenriegler and M. Stapper, {\it Phys. Rev.}, {\bf B50} (1994)
10009.
\vskip 6pt
\item{[27]} L.J. Slater, ``Generalised Hypergeometric Functions'', Cambridge
University Press, Cambridge (1966).
\vskip 6pt
\item{[28]} J. Thomae, {\it Math. Ann.}, {\bf 2} 427 (1870).
\vskip 6pt
\item{[29]} M. D'Eramo, L. Peliti and G. Parisi, {\it Lett. al Nuovo Cimento
ser. II}, {\bf 2} 878 (1970);
\item{} K. Symanzik, {\it Lett. al Nuovo Cimento ser. II}, {\bf 3} 734 (1972).
\vskip 6pt
\item{[30]} D.M. McAvity, in preparation.
\bye